\documentclass[article,onecolumn,superscriptaddress,showpacs,nofootinbib,showkeys,10pt]{revtex4}
\usepackage[titletoc,title,toc]{appendix}
\usepackage{graphicx}  
\usepackage{amsmath}   
\usepackage{amssymb}   
\usepackage{bm} 
\usepackage{dcolumn}
\usepackage{color}
\usepackage{mathrsfs}
\usepackage{amsfonts}
\usepackage{longtable}
\usepackage{varioref}
\RequirePackage[colorlinks,citecolor=blue,urlcolor=magenta,linkcolor=blue]{hyperref}

\def\gr{general relativity}

\def\KK{Kaluza-Klein }
\def\qnm{quasi-normal mode }
\labelformat{section}{Section #1} 
\labelformat{subsection}{Section #1} 
\labelformat{subsubsection}{Section #1}
\labelformat{subsubsubsection}{Section #1}
\labelformat{equation}{Eq.~(#1)} 
\labelformat{figure}{Fig.~#1} 
\labelformat{subfigure}{Fig.~\thefigure#1} 
\labelformat{table}{Table~#1} 
\labelformat{appendix}{Appendix #1}
\begin{document}
\title{Signatures of extra dimensions in gravitational waves from black hole quasi-normal modes}

\author{Sumanta Chakraborty}
\email{sumantac.physics@gmail.com}
\affiliation{Theoretical Physics Department, Indian Association for the Cultivation of Science, Kolkata 700032, India}
\author{Kabir Chakravarti}
\email{kabir@iucaa.in}
\affiliation{IUCAA, Ganeshkhind, Post Bag 4, Pune University Campus, Pune 411007, India}
\author{Sukanta Bose}
\email{sukanta@iucaa.in}
\affiliation{IUCAA, Ganeshkhind, Post Bag 4, Pune University Campus, Pune 411007, India}
\affiliation{Department of Physics \& Astronomy, Washington State University, 1245 Webster, Pullman, WA 99164-2814, U.S.A}
\author{Soumitra SenGupta}
\email{tpssg@iacs.res.in}
\affiliation{Theoretical Physics Department, Indian Association for the Cultivation of Science, Kolkata 700032, India}
\date{\today}  
\begin{abstract}
In this work we have derived the evolution equation for gravitational perturbation in four dimensional spacetime in presence of a spatial extra dimension. The evolution equation is derived by perturbing the \emph{effective} gravitational field equations on the four dimensional spacetime, which inherits non-trivial higher dimensional effects. Note that this is \emph{different} from the perturbation of the five dimensional gravitational field equations, existing in literature, and possess quantitatively new features. The gravitational perturbation has further been decomposed into a purely four dimensional part and another piece that depends on extra dimensions. The four dimensional gravitational perturbation now admits massive propagating degrees of freedom, owing to the existence of higher dimensions. We have also studied the influence of these massive propagating modes on the quasi-normal mode frequencies, signaling the higher dimensional nature of the spacetime, and have contrasted these massive modes 
with the massless modes in general relativity. Surprisingly, it turns out that the massive modes experience much smaller damping compared to the massless modes in general relativity and may even \emph{dominate} over and above the general relativity contribution if one observes the ringdown phase of a black hole merger event at sufficiently late times. Furthermore, the whole analytical framework has been supplemented by the fully numerical Cauchy evolution problem as well. In this context we have shown that except for minute details the overall features of the gravitational perturbations are captured in both the Cauchy evolution as well as in the analysis of quasi-normal modes. The implications on observations of black holes with LIGO and proposed space missions like LISA are also discussed.   
\end{abstract}
\maketitle
\section{Introduction}

Unification of forces has been the most challenging task the science community has ever faced. So far that quest has successfully brought the electromagnetic, strong and weak forces under one roof. However the unification scheme hits a wall when one tries to incorporate in it the only other fundamental force, namely, gravity. There have been numerous attempts, so far, to incorporate gravity in the above picture as well, leading to a unified quantum theory of nature. This has resulted in a large number of candidate theories for quantum gravity, \emph{but} without much success. This issue, broadly speaking, originates from the peculiar fact that the energy scale associated with grand unified theories is $\sim \mathcal{O}(10^{3})\textrm{GeV}$, while the natural energy scale for gravity is the Planck scale $\sim \mathcal{O}(10^{18}) \textrm{GeV}$. This huge difference between the respective energy scales manifests itself into unnatural fine tunings in various physical parameters of the model, e.g., in the mass 
of the Higgs Boson. Thus it seems legitimate to understand the origin of this fine tuning problem (known as the \textit{gauge hierarchy problem}) before delving into quantization of gravity 
\cite{PerezLorenzana:2005iv,Csaki:2004ay,Sundrum:2005jf,Polchinski:1998rq,Polchinski:1998rr,
Rovelli:2004tv,Chakraborty:2017s}.

One such natural candidate for resolving the gauge hierarchy problem in this regard corresponds to extra spatial dimensions, which can bring down the Planck scale to the realm of grand unified theories. Such a possibility was considered in 
\cite{ArkaniHamed:1998rs,Antoniadis:1998ig,Antoniadis:1990ew,Rubakov:1983bz,Rubakov:1983bb} where the extra dimensions were large enough, such that the volume spanned by them could suppress the Planck scale of the higher dimensional spacetime (known as \textit{bulk}) to the TeV scale. However this proposal harbours two conceptual drawbacks: Firstly, it seems that the problem of energy scale hierarchy has merely been transferred to another form, the volume hierarchy, e.g., if one wants to reduce the energy scale to $1~\textrm{TeV}$ the size of the extra dimensions would be $\sim 10^{11}~\textrm{m}$; and more importantly it treats the higher dimensional spacetime to be flat \cite{PerezLorenzana:2005iv}. The second one is indeed a serious issue, as gravity cannot be shielded and hence if it is present in four dimensions gravity is bound to propagate in higher spacetime dimensions as well. In order to cure this problem, Randall and Sundrum proposed a very natural solution to the hierarchy problem with warped 
extra dimensions, where presence of gravity in higher dimensions forces the effective Planck scale to reduce to TeV scale in the four dimensional hypersurface we live in (known as \textit{brane}) 
\cite{Randall:1999ee,Garriga:1999yh}. This scenario has been extensively studied in the literature in the past years in various contexts, starting from black holes 
\cite{Chamblin:1999by,Berti:2015itd,Konoplya:2008ix,Konoplya:2007jv,Berti:2009kk,Gregory:2008rf,Emparan:1999wa,Cook:2017fec,Bhattacharya:2016naa} and cosmology 
\cite{Csaki:1999mp,Csaki:1999jh,Binetruy:1999ut,Ida:1999ui,Nojiri:2001ae,Nojiri:2002hz,Charmousis:2002rc,Germani:2002pt,Gravanis:2002wy,Brax:2004xh} 
to particle phenomenology as well as possible signatures in Large Hadron Collider 
(LHC) \cite{Goldberger:1999un,Davoudiasl:1999tf,Davoudiasl:2000wi}. A lot of attention has also been devoted to the higher curvature generalization of this scenario, obtained by introducing terms like $R^{2}$, $R_{abcd}R^{abcd}$ in the gravitational action, as well as to the stabilization of these extra dimensions 
\cite{Goldberger:1999uk,Csaki:1999mp,Chacko:1999eb,Chakraborty:2014zya,Chakraborty:2013ipa,
Chakraborty:2016gpg}. 

Even though LHC provides us an observational window for the existence of extra spatial dimensions, it is important to know if there exists any other observational tests that can either prove or disprove their existence independently. It is obvious that in order to probe these effects, one has to investigate high energy/high curvature regime, which can originate from either high energy collisions like in LHC or from physics near black holes. The second possibility opens up a few interesting observational avenues --- (a) the black hole continuum spectrum, originating from accretion disc around a supermassive black hole, (b) strong gravitational lensing around supermassive black holes and finally, (c) gravitational waves from collision of two massive black holes. We have already elaborated on the continuum spectrum from supermassive black holes and their implications regarding the presence of extra dimensions in \cite{Banerjee:2017hzw}, while strong gravitational lensing has been discussed in detail in \cite{
Chakraborty:2016lxo}. In this work we aim to address the third possibility, i.e., the effect of higher dimensions on gravitational waves, in light of the recent detections 
\cite{Abbott:2017vtc,TheLIGOScientific:2016pea,Abbott:2016nmj,TheLIGOScientific:2016src,Abbott:2016blz} of the same in Advanced Laser Interferometer Gravitational-Wave Observatory (aLIGO). The whole process of collision between two black holes can broadly be divided into three categories --- inspiral phase, merger phase and ringdown phase. The first two phases are best described by a combination of post-Newtonian and numerical approaches 
\cite{Asada:1996ya,Damour:2008qf,Damour:2001tu,Nagar:2011fx,Damour:2008gu,Baiotti:2010xh,Blanchet:2006zz,
Flanagan:2005yc,Centrella:2010mx,Ajith:2007kx,Buonanno:2000ef,Delsate:2014hba,Pai:2000zt,Bose:2005fm,Pretorius:2005gq,
Arun:2006yw}, which we leave for the future, concentrating here on the ringdown phase only. In this situation the \qnm frequencies play a very fundamental role in determining the ring down phase and in this work we will concentrate on deriving the \qnm frequencies for this higher dimensional scenario 
\cite{Kanti:2005xa,Berti:2009kk,Konoplya:2011qq,Berti:2015itd,Toshmatov:2016bsb,Andriot:2017oaz}. 

To understand the behaviour of \qnm frequencies in the context of higher spacetime dimensions, one can follow two possible approaches --- (a) One starts from the gravitational field equations in the bulk and then consider its perturbation around a bulk solution, which manifests itself as a black hole on the brane. This one we refer to as the \emph{bulk based approach}. (b) Otherwise, one projects the bulk gravitational field equations on the brane hypersurface resulting in an effective description of the brane dynamics inherited from the bulk, referred to as the \emph{brane based approach}. In this case as well one perturbs the effective gravitational field equations on the brane, around a given bulk solution representing again a brane black hole.  Some aspects of this problem along the first line of attack has already been elaborated and explored in \cite{Seahra:2004fg,Clarkson:2005eb,Seahra:2009gw,Clarkson:2006pq,Clarkson:2005mg,Seahra:2009zz,Witek:2013ora}, while to our knowledge the second avenue is 
hitherto unexplored. In this work, we wish to fill this gap by providing a thorough analysis of the second approach in relation to the black hole perturbation theory and possible discords with the bulk based approach. In particular, we will try to understand whether the results derived in \cite{Seahra:2004fg} using Cauchy evolution of initial data matches with our \qnm frequency analysis. Further for completeness we will present the Cauchy evolution for the brane based approach as well. This will not only help to contrast these two approaches but will also depict whether the \qnm analysis and the Cauchy evolution are compatible with each other. Besides providing yet another independent route towards understandings of higher dimensions, this will also be of significant interest to the gravitational wave community. 

The paper is organized as follows: We start in \ref{Perturb_Effective} with a brief introduction of the effective equation formalism in the context of higher spatial dimensions and then we build up our gravitational perturbation equation based on the above. This has been applied in \ref{Sph_Symm} to derive the evolution equations for the master variables associated with spherically symmetric brane and possible effects from higher dimensions. In \ref{qnm_analysis} we have studied these perturbation equations in Fourier space and have derived the \qnm frequencies using the continued fraction method as well as the direct integration scheme. Using these \qnm frequencies the time evolution of the master variable has been determined for both the bulk and the brane based approach in \ref{Num_qnm}. \ref{consistency_Cauchy} deals with Cauchy evolution of the initial data and its possible harmony with the \qnm analysis. We conclude with a discussion and implications of the results obtained, in \ref{conclusion}. Some 
detailed calculations pertaining to derivation of gravitational perturbation equation on the brane have been presented in \ref{App_A}, while those associated with continued fraction method have been elaborated in \ref{App_B}.  

\emph{Notations and Conventions:} We will set the fundamental constant $c$ as well as the combination $GM$ to unity, where $M$ is the mass of the black hole. Indices running over all the bulk coordinates are denoted by uppercase Latin letters, while all the brane indices are denoted by Greek letters. Any geometrical quantity associated with the brane hypersurface alone is being denoted with a superscript $(4)$. Further, all the matrix valued quantities will be denoted by boldfaced letters. Finally the signature convention adopted in this work is the mostly positive one.
\section{Perturbing effective gravitational field equations on the brane}\label{Perturb_Effective}

We start this section by providing a very brief introduction to the effective gravitational field equations on the brane, which will be necessary for our later purposes. Since we are interested in signatures of higher dimensions only, it will be sufficient to work within the context of Einstein gravity in five spacetime dimensions, in which case the gravitational Lagrangian density is the five dimensional Ricci scalar $R$. Thus the five dimensional gravitational field equations will read $G_{AB}=8\pi G_{(5)}T_{AB}$, where $T_{AB}$ stands for the matter energy momentum tensor, which may be present in the bulk and $G_{(5)}$ is the five dimensional gravitational constant. In the specific context when the bulk energy momentum tensor is originating from a negative cosmological constant $\Lambda$, one arrives at the following static and spherically symmetric solution on the brane,
\begin{align}\label{GW_Eq01}
d&s_{\rm unperturbed}^{2}
\nonumber
\\
&=e^{-2ky}\left(-f(r)dt^{2}+\frac{dr^{2}}{f(r)}+r^{2}d\Omega ^{2}\right)+dy^{2}~,
\end{align}
with $f(r)=1-(2/r)$ and $k\propto \sqrt{-\Lambda}$. Note that from the perspective of a brane observer located on a $y=\textrm{constant}$ hypersurface, the spacetime structure on the brane is given by the Schwarzschild solution. 

This raises the following interesting question: What happens to the gravitational field equations on the brane, given the gravitational field equations on the bulk? It has been answered for Einstein gravity in \cite{Shiromizu:1999wj} and has been extended recently to various other scenarios involving alternative gravity theories \cite{Chakraborty:2014xla,Chakraborty:2015bja,Chakraborty:2015taq}. The derivation goes as follows, one first chooses the brane hypersurface, say $y=0$, and determines the normal $n_{A}=\nabla _{A}y$, yielding the induced metric on the brane hypersurface to be $h_{AB}=g_{AB}-n_{A}n_{B}$, such that $n_{A}h^{A}_{B}=0$. Given the induced metric, one can introduce the notion of covariant derivative on a brane hypersurface and hence a notion of brane curvature using commutator between the brane covariant derivatives. This enables one to express the bulk curvature in terms of the brane curvature and extrinsic curvatures associated with the brane hypersurface. Further contractions will 
enable one to relate the bulk Einstein's equations with curvatures on the brane, referred to as the effective gravitational field equations on the brane. The effective equations in vacuum brane differ from four dimensional Einstein's equations by an additional term inherited from the bulk Weyl tensor and takes the following form,
\begin{equation}\label{GW_Eq02}
~^{(4)}G_{\mu \nu}+E_{\mu \nu}=0~.
\end{equation}
Here $E_{\mu \nu}$ stands for a particular projection of the bulk Weyl tensor $C_{ABCD}$ on the brane hypersurface (commonly known as the electric part) given by,
\begin{equation}\label{GW_Eq03}
E_{\mu \nu}=C_{ABCD}e^{A}_{\mu}n^{B}e^{C}_{\nu}n^{D}~,
\end{equation}
where $n_{A}$ is the normalized normal introduced earlier and $e^{A}_{\mu}=\partial x^{A}/\partial y^{\mu}$ is the bulk to brane projector, with $x^{A}$ being the bulk coordinates and $y^{\mu}$ are the brane coordinates \cite{gravitation,Poisson}.

At this stage it is worth mentioning that in order to arrive at the above relation we have assumed that the bulk cosmological constant and the brane tension cancels each other, leading to a vanishing effective cosmological constant on the brane hypersurface \cite{Shiromizu:1999wj,Randall:1999ee}. The above cancellation has its origin in the fact that in the effective field equation the effective cosmological constant is the difference between bulk cosmological constant and brane tension, and this difference has to be zero for the stability of the background spacetime. Further, note that even though \ref{GW_Eq02} acts as the effective field equations on the brane, to solve it explicitly one does require information of the bulk, hidden in $E_{\mu \nu}$ through the bulk Weyl tensor. 

There are two ways to solve this equation --- (a) Assume certain bulk geometry as ansatz (which for our case corresponds to \ref{GW_Eq01}) and then try to see what sort of brane configuration solves \ref{GW_Eq02}. (b) Take $E_{\mu \nu}$ as an arbitrary tensor and try to solve \ref{GW_Eq02} with $E_{\mu \nu}$ treated as a source: e.g., in the context of spherical symmetry one often divides $E_{\mu \nu}$ into an energy density (known as \emph{dark radiation}) and pressure (known as \emph{dark pressure}). Even though one can have very interesting results emerging from the second scenario \cite{Maartens:2001jx}, it has the drawback that the bulk metric remains unknown and in general it is not even clear whether there exists a bulk metric that would satisfy Einstein's equations in the bulk. Thus we will adopt the first scenario and shall take \ref{GW_Eq01} as the background metric which indeed satisfies \ref{GW_Eq02} as well 
\cite{Dadhich:2000am,Maartens:2001jx,Harko:2004ui,Maartens:2010ar,Chakraborty:2014xla}. 

This procedure must be contrasted with the perturbation of bulk Einstein's equations around the solution presented in \ref{GW_Eq01}, since in the case of effective field equations, the perturbation of bulk Weyl tensor will play a crucial role. Thus it is not at all clear a priori how the perturbed equations in the brane based approach will behave in contrast to the bulk based approach, even though they are being perturbed around the same solution. With this motivation in the backdrop, let us concentrate on perturbation of \ref{GW_Eq03} around the bulk metric $g_{AB}$ given in \ref{GW_Eq01}, such that,
\begin{align}\label{GW_Eq04}
g_{AB}\rightarrow g_{AB}+h_{AB}~.
\end{align}
Here $h_{AB}$ is the perturbed metric around $g_{AB}$ and all the expressions to follow will be evaluated to the first order in the perturbed metric $h_{AB}$ \footnote{In principle one should write down $g_{AB}\rightarrow g_{AB}+\epsilon~ h_{AB}$, with small $\epsilon$ and then keeping only terms linear in $\epsilon$.}. 

It is also well known that not all the components of $h_{AB}$ are dynamical, there are redundant gauge degrees of freedom. These gauge choices must be made according to convenience of calculations. In this particular situation the following gauge conditions will turn out to be useful later on,
\begin{align}\label{GW_Eq05}
\nabla _{A}h^{A}_{B}=0;\qquad h^{A}_{A}=0;\qquad h_{AB}=h_{\alpha \beta}e^{\alpha}_{A}e^{\beta}_{B}~,
\end{align}
known as the Randall-Sundrum gauge. The usefulness of this gauge condition can also be anticipated from the fact that these imply $h_{AB}n^{A}=0$ and hence the perturbed bulk metric takes the following form,
\begin{equation}\label{GW_Eq06}
ds^{2}_{\rm perturbed}=\Big[q_{\alpha \beta}(y,x^{\mu})+h_{\alpha \beta}(y,x^{\mu})\Big]dx^{\alpha}dx^{\beta}+dy^{2}~,
\end{equation}
where $q_{\alpha \beta}$ solves \ref{GW_Eq02} and is given by 
\begin{align}\label{GW_N01}
&q_{\alpha \beta}dx^{\alpha}dx^{\beta}
\nonumber
\\
&=e^{-2ky}\left(-f(r)dt^{2}+\frac{dr^{2}}{f(r)}+r^{2}d\theta ^{2}+r^{2}\sin ^{2}\theta d\phi ^{2} \right)~.
\end{align}
Even though \ref{GW_Eq01} opts for $f(r)=1-(2/r)$, in the rest of the analysis we will keep $f(r)$ as general as possible. Then to linear order in the perturbed metric $h_{\alpha \beta}$ one can expand the four dimensional Einstein tensor as,
\begin{align}\label{GW_Eq07}
~^{(4)}G_{\mu \nu}\simeq ~^{(4)}G_{\mu \nu}^{(q)}&+~^{(4)}R^{(h)}_{\mu \nu}
\nonumber
\\
&-\frac{1}{2}q_{\mu \nu}~^{(4)}R^{(h)}-\frac{1}{2}h_{\mu \nu}~^{(4)}R^{(q)}~,
\end{align}
where terms with superscript $(q)$ denote that they have to be evaluated for the brane background metric $q_{\alpha \beta}$ given in \ref{GW_N01} and superscript $(h)$ implies that it has been evaluated for the perturbed metric $h_{\mu \nu}$. The index $(4)$ implies that these are all four dimensional geometrical quantities. 

Another ingredient in the perturbation of effective brane based approach is the perturbation of the bulk Weyl tensor. For that one has to write down the bulk Weyl tensor in terms of the bulk Riemann, Ricci tensor and Ricci scalar and expand all of them to leading order in the gravitational perturbation $h_{\alpha \beta}$. The above procedure leads to,
\begin{widetext}
\begin{align}\label{GW_Eq08}
C_{ABCD}&=R_{ABCD}-\frac{1}{3}R_{AC}g_{BD}+\frac{1}{3}R_{AD}g_{BC}+\frac{1}{3}R_{BC}g_{AD}-\frac{1}{3}R_{BD}g_{AC}
+\frac{1}{12}R\left(g_{AC}g_{BD}-g_{AD}g_{BC}\right)
\nonumber
\\
&\simeq C^{(g)}_{ABCD}+\Bigg\lbrace R^{(h)}_{ABCD}-\frac{1}{3}R^{(g)}_{AC}h_{BD}-\frac{1}{3}R^{(h)}_{AC}g_{BD} 
+\frac{1}{3}R^{(h)}_{AD}g_{BC}+\frac{1}{3}R^{(g)}_{AD}h_{BC}-\frac{1}{3}R^{(h)}_{BD}g_{AC}
\nonumber
\\
&-\frac{1}{3}R^{(g)}_{BD}h_{AC}+\frac{1}{3}R^{(h)}_{BC}g_{AD}+\frac{1}{3}R^{(g)}_{BC}h_{AD}
+\frac{1}{12}R^{(h)}\Big(g_{AC}g_{BD}-g_{AD}g_{BC}\Big)
\nonumber
\\
&+\frac{1}{12}R^{(g)}\Big(g_{AC}h_{BD}+h_{AC}g_{BD}-g_{AD}h_{BC}-h_{AD}g_{BC}\Big)\Bigg\rbrace~.
\end{align}
\end{widetext}
Here superscript $(g)$ denotes that the respective quantity is evaluated for the bulk background metric $g_{AB}$. Note that due to dependence of $q_{\alpha \beta}$ on extra dimensional coordinate $y$, quantities evaluated for the bulk metric will inherit $y$-derivatives of $q_{\alpha \beta}$ and hence will differ from their four-dimensional counterparts. Given the perturbation of bulk Weyl tensor the corresponding projection of the perturbed bulk Weyl tensor onto the brane hypersurface results in,
\begin{widetext}
\begin{align}\label{GW_Eq09}
E_{\mu \nu}&\simeq E_{\mu \nu}^{(g)}+\Bigg\lbrace R^{(h)}_{ABCD}e^{A}_{\mu}n^{B}e^{C}_{\nu}n^{D}-\frac{1}{3}R^{(h)}_{AC} e^{A}_{\mu}e^{C}_{\nu}
\nonumber
\\
&-\frac{1}{3}R^{(h)}_{BD}n^{B}n^{D}q_{\mu \nu}-\frac{1}{3}R^{(g)}_{BD}n^{B}n^{D}h_{\mu \nu}+\frac{1}{12}R^{(h)} q_{\mu \nu}+\frac{1}{12}R^{(g)}h_{\mu \nu}\Bigg\rbrace ~.
\end{align}
\end{widetext}
Note that in the above perturbation equation for the projected bulk Weyl tensor, the first order corrections to bulk Riemann, Ricci tensor and Ricci scalar appears. One can decompose all these perturbed quantities evaluated for the bulk metric in terms of the respective brane metric and extra dimensional contributions. This has been explicitly carried out in \ref{App_A} and ultimately leads to the following expression for the projected bulk Weyl tensor,
\begin{align}\label{GW_Eq18}
E_{\mu \nu}^{(h)}&=\frac{1}{6}~^{(4)}\square h_{\mu \nu}-\frac{1}{3}\partial _{y}^{2}h_{\mu \nu}-k\partial _{y}h_{\mu \nu}+\frac{1}{3}k^{2}h_{\mu \nu}
\nonumber
\\
&+\frac{1}{3}h^{\alpha}_{\beta}~^{(4)}R^{(q)\beta}_{~~~~\mu \alpha \nu}-\frac{1}{6}h^{\alpha}_{\mu}~^{(4)}R^{(q)}_{\alpha \nu}
\nonumber
\\
&-\frac{1}{6}h^{\alpha}_{\nu}~^{(4)}R^{(q)}_{\alpha \mu}+\frac{1}{12}~^{(4)}R^{(q)}h_{\mu \nu}~.
\end{align}
At this stage it is worth emphasizing that the gauge conditions elaborated in \ref{GW_Eq05}, take a simpler form in this context. In particular, the spatial part of the differential condition $\nabla _{A}h^{A}_{\mu}=0$, when expanded in terms of four dimensional quantities immediately yields $\nabla _{\nu}h^{\nu}_{\mu}=0$. Use of this relation and commutator of four dimensional covariant derivative results into $\nabla ^{\mu}E^{(h)}_{\mu \nu}=0$, as is evident from \ref{GW_Eq18} in the context of vacuum solutions. 

One can also try to understand this result from a different perspective. Since we are perturbing around vacuum solutions, it follows from \ref{GW_Eq02} that  $E^{(h)}_{\mu \nu}\propto ~^{(4)}G^{(h)}_{\mu \nu}$. Thus it immediately implies that $\nabla _{\mu}E^{(h)~\mu \nu}=0$ as it should be, by virtue of Bianchi identity. Finally collecting all the pieces from perturbation of bulk Weyl tensor elaborated in \ref{GW_Eq18} as well as perturbation of original Einstein tensor as in \ref{GW_Eq07} we obtain,
\begin{align}\label{GW_Eq19}
e^{2ky}&\Big\lbrace ~^{(4)}\square h_{\mu \nu}+2h_{\alpha \beta}~^{(4)}R^{\beta~\alpha}_{~\mu ~\nu}\Big\rbrace 
\nonumber
\\
&+\left\lbrace -k^{2}h_{\mu \nu}+3k\partial _{y}h_{\mu \nu}+\partial _{y}^{2}h_{\mu \nu}\right\rbrace=0~.
\end{align}
In order to arrive at the above relation we have used the fact that $q_{\alpha \beta}=\exp(-2ky)g_{\alpha \beta}$, where in this particular situation $g_{\alpha \beta}$ is the Schwarzschild metric. Note that we have not used this fact explicitly anywhere in this section, except for assuming that $g_{\alpha \beta}$ must satisfy vacuum Einstein's equations on the brane. Further, all the geometrical quantities present in the above equation are evaluated for the brane metric $g_{\alpha \beta}$. 

At this stage it is instructive to split the perturbation equations into parts depending on four dimensional spacetime and those depending on extra dimensions, such that, $h_{\alpha \beta}(y,x^{\mu})=h_{\alpha \beta}(x^{\mu})\chi (y)$. Following the separability of the perturbed metric, the above equations can also be decomposed into two parts, which for vacuum brane solution reduce to,
\begin{align}
e^{-2ky}\left\lbrace-k^{2}\chi+3k\partial _{y}\chi+\partial _{y}^{2}\chi\right\rbrace&=-\mathcal{M}^{2}\chi (y)~,
\label{GW_Eq20a}
\\
~^{(4)}\square h_{\mu \nu}+2h_{\alpha \beta}~^{(4)}R^{\alpha~\beta}_{~\mu ~\nu}-\mathcal{M}^{2}h_{\mu \nu}&=0~.
\label{GW_Eq20b}
\end{align}
Remarkably, the effect of the whole analysis is just the emergence of a massive gravitational perturbation modes. With $\mathcal{M}=0$, one immediately recovers the dynamical equation governing gravitational perturbation in a non-trivial background. As we will see later, \ref{GW_Eq20a} will lead to a series of masses denoted by $m_{n}$ and is called the $n$th \KK mode mass of gravitational perturbation. For each \KK mode, say of order $n$, there will be a solution $h_{\mu \nu}^{(n)}$ to \ref{GW_Eq20b}. When all these $n$ values are summed over one ends up with the full solution of the gravitational perturbation. 

To summarize, we have started from the effective gravitational field equations on the 3-brane, which depends on the bulk Weyl tensor and hence on the bulk geometry. The main problem of this approach being, not all the components of the projected bulk Weyl tensor $E_{\mu \nu}$ are determined in terms of quantities defined on the brane. In particular, the transverse-traceless part of the projected bulk Weyl tensor, representing the graviton modes in the bulk spacetime, can not be determined. This is intimately related to the fact that the effective field equations on the brane are \emph{not} closed \cite{Shiromizu:1999wj}. In this work, we have circumvented this problem by using the gauge freedom for the gravitational perturbation. We have started with the Schwarzschild anti-de Sitter spacetime (as in \ref{GW_Eq01}) which identically satisfies the effective gravitational field equations on the brane. We then consider perturbation around this background, which certainly involves graviton modes propagating in 
the bulk spacetime. However the use of Randall-Sundrum gauge (presented in \ref{GW_Eq05}) enables one to reduce the  number of propagating degrees of freedom and hence the effective field equations (at least in the perturbative regime) becomes closed. Finally the method of separation of variables enables one to separate a four dimensional part from the extra dimensional one and arrives at \ref{GW_Eq20a} and \ref{GW_Eq20b} respectively. The presence of extra dimension essentially translates into the infinite tower of \KK modes as far as the propagation of gravitational wave in four dimension is considered.

Let us now emphasise the key differences between our approach and the bulk based one. Interestingly, \ref{GW_Eq20b} governing the evolution of gravitational perturbation of the four dimensional brane is identical to that of bulk based approach, while the eigenvalue equation, i.e., \ref{GW_Eq20a} determining the mass of graviton is different. Hence the Kaluza-Klein mass modes of graviton in the brane based approach will be different from that in the bulk based approach and hence will have interesting observational consequences in both high energy collision experiments as well as in propagation of gravitational waves. In this work we will mainly be interested in the effect of the mass term originating from \ref{GW_Eq20b}, in particular how it modifies the behaviour of perturbations in contrast to \gr\ and also how the brane and bulk based approach differs. This is what we will concentrate on in the next sections.    
\section{Specialising to spherically symmetric vacuum brane}\label{Sph_Symm}

We have described a general method for deriving the dynamical equations pertaining to gravitational perturbation, starting from the effective gravitational field equations on the brane in the previous section. We would now like to apply the above scenario in the context of black holes on the brane. In particular we are interested in perturbations around the background given by \ref{GW_Eq01}. Thus in this section, with the above scenario in the backdrop, we specialise to vacuum and spherically symmetric solution on the brane, such that, $g_{\alpha \beta}=\textrm{diag}(-f(r),f^{-1}(r),r^{2},r^{2}\sin ^{2}\theta)$. For the moment we concentrate on situations with arbitrary choices for $f(r)$, while later on we will choose a specific form for $f(r)$, namely, $f(r)=1-(2/r)$. Further being a vacuum solution, the Ricci tensor and Ricci scalar identically vanishes. 

The main focus now will be understanding the evolution equation of the gravitational perturbation $h_{\mu \nu}$ before discussing the \KK modes. In general the perturbation $h_{\mu \nu}$ can depend on all the spacetime coordinates $(t,r,\theta,\phi)$. The spherical symmetry associated with this problem demands a separation between $(t,r)$ and $(\theta,\phi)$ part, which results into decomposition of the angular part into spherical harmonics. In particular, for the gravitational perturbation we obtain,
\begin{align}\label{GW_Eq21}
h_{\alpha \beta}^{(n)}=\sum _{l=0}^{\infty}\sum _{m=-l}^{l}\sum _{i=1}^{10}h_{i}^{(nlm)}(t,r)\left\{Y^{(i)}_{lm}\right\}_{\alpha \beta}(\theta,\phi)~,
\end{align}
where the perturbation $h_{\alpha \beta}$ have been broken up into ten independent parts, separated into $h_{i}^{(nlm)}$ depending on $(t,r)$ and the rest depending on the angular coordinates. Further, $n$ stands for the \KK mode index, while $l$ is the angular momentum and $m$ being its $z$-component. The quantities $\{Y_{lm}\}_{\alpha \beta}$ are the tensorial spherical harmonics in four spacetime dimensions. In order to define these tensor harmonics one should introduce the following normalised basis vectors,
\begin{align}\label{GW_Eq22}
t^{\alpha}&=\frac{1}{\sqrt{f(r)}}\left(\partial _{t}\right)^{\alpha};\qquad r^{\alpha}=\sqrt{f(r)}\left(\partial _{r}\right)^{\alpha}; 
\nonumber
\\
\theta ^{\alpha}&=\frac{1}{r}\left(\partial _{\theta}\right)^{\alpha};\qquad \phi ^{\alpha}=\frac{1}{r\sin \theta}\left(\partial _{\phi}\right)^{\alpha}~.
\end{align}
It is clear that they are orthogonal to each other, while the factors in the front ensures that they are normalised as well. Given this structure one can introduce an induced metric on the $(\theta,\phi)$ plane such that, $\mu_{\alpha \beta}=g_{\alpha \beta}+t_{\alpha}t_{\beta}-r_{\alpha}r_{\beta}$, leading to, $t^{\alpha}\mu _{\alpha \beta}=0=r^{\alpha}\mu_{\alpha \beta}$. One can also define an antisymmetric tensor $\epsilon _{\alpha \beta}=\theta _{\alpha}\phi _{\beta}-\phi_{\alpha}\theta _{\beta}$. Given this one can construct ten such irreducible representations, which include $t_{\alpha}t_{\beta}Y_{lm}$, $\mu_{\alpha \beta}Y_{lm}$ and so on involving no derivatives of $Y_{lm}$, as well as terms like $r_{(\alpha}\mu_{\beta) \rho}\nabla ^{\rho}Y_{lm}$, $t_{(\alpha}\mu_{\beta)\rho}\nabla ^{\rho}Y_{lm}$ etc. depending on derivatives of $Y_{lm}$. Among all these choices, three terms among the ten will depend on the antisymmetric combination $\epsilon _{\alpha \beta}$ and will pick up a term $(-1)^{l+1}$ 
under parity. These we will refer to as \emph{axial perturbations}. On the other hand, the remaining seven components will inherit an extra factor of $(-1)^{l}$ under parity transformation and are referred to as \emph{polar perturbations}. Thus the spherical harmonic decomposition of $h^{(n)}_{\alpha \beta}$ in \ref{GW_Eq21} can be further subdivided into axial and polar parts.

The above decomposition is useful in simplifying the algebra further. It is evident that the operators acting on $h_{\alpha \beta}$ in \ref{GW_Eq20b} are invariant under parity. Thus the solutions to \ref{GW_Eq20b} which are eigenfunctions of parity with different eigenvalues decouple from each other. Hence in the present scenario, the polar and axial perturbations differ from each other in parity eigenvalue and hence evolves independently of one another. Further two axial (or, polar) modes having different $l$ and $m$ values also have different eigenvalues under parity and hence they also decouple. Thus one can solve for the evolution of a given $l$ mode for axial (or, polar) perturbation separately. 

Due to complicated nature of the polar perturbations, we content ourselves with the axial perturbations only. The angular part of the axial perturbations contain essentially three terms, two depending on single derivative of $Y_{lm}$, while the third one depends on double derivatives of $Y_{lm}$. Thus for $l=0$ all the axial modes identically vanishes and for $l=1$, the term involving double derivatives of $Y_{lm}$ does not contribute. Hence in what follows we will concentrate on the $l\geq 2$ scenario. In this case there are two master variables, which we will denote by $u_{n,l}$ and $v_{n,l}$ respectively and their evolution equations read,
\begin{align}
\mathcal{D}u_{n,l}+f(r)\Big\{m_{n}^{2}&+\frac{l(l+1)}{r^{2}}-\frac{6}{r^{3}} \Big\}u_{n,l}
\nonumber
\\
&+f(r)\frac{m_{n}^{2}}{r^{3}}v_{n,l}=0~,
\label{GW_Eq23a}
\\
\mathcal{D}v_{n,l}+f(r)\Big\{m_{n}^{2}&+\frac{l(l+1)}{r^{2}} \Big\}v_{n,l}+4f(r)u_{n,l}=0~.
\label{GW_Eq23b}
\end{align}
Here, $\mathcal{D}$ is the differential operator $\partial _{t}^{2}-\partial _{r_{*}}^{2}$, where $r_{*}$ is the tortoise coordinate defined using $f(r)$ as $dr_{*}=dr/f(r)$. Note that these two differential equations are coupled to each other and provides a complete set. The massless limit also turns out to be interesting. As far as $u_{n,l}$ is concerned \ref{GW_Eq23a} decouples and the corresponding potential reduces to the well known Regge-Wheeler form. The potential for $v_{n,l}$ resembles to that of an electromagnetic field. Note that an identical form for the equations were derived in \cite{Seahra:2004fg}, however  from a different perspective. This is due to the fact explained in \ref{Perturb_Effective}, i.e, the evolution of gravitational perturbation equation is identical to \cite{Seahra:2004fg} modulo the \KK decomposition and hence the mass term.  

\begin{table*}
\begin{center}
\caption{Numerical estimates of the first ten \KK mass modes correct to second decimal place for two possible choices of the inter-brane separation $d$ and bulk curvature scale $\ell$ have been presented for brane based approach. First \ref{GW_Eq29} has been solved for $z_{n}$ and the result has been presented in the second column. Incidentally, the solution for $z_{n}$ is insensitive to choices of $d/\ell$ as far as solutions accurate to second decimal places are considered. To avoid any instability present in the problems the inverse of bulk curvature scale has been chosen such that, the mass of lowest lying \KK mode is greater than or equal to $0.43$ in geometrised units.}\label{Table_01}       
%
%
\begin{tabular}{p{3.5cm}p{1.5cm}p{4cm}p{4.5cm}}
\hline\noalign{\smallskip}
\hline\noalign{\smallskip}
\KK Modes    & ~~$z_{n}$ & ~~~~~~Associated Mass         &  ~~~~~~Associated Mass             \\
             &           &($d/\ell=20; 1/\ell=6\times 10^{7}$) & ($d/\ell=30; 1/\ell=1.3\times 10^{12}$)  \\
\hline \noalign{\smallskip}
\hline \noalign{\smallskip}
~~~~~~n=1    & 3.56      & ~~~~~~~~~ 0.44                & ~~~~~~~~~~ 0.43                    \\
~~~~~~n=2    & 6.74      & ~~~~~~~~~ 0.83                & ~~~~~~~~~~ 0.82                    \\
~~~~~~n=3    & 9.88      & ~~~~~~~~~ 1.22                & ~~~~~~~~~~ 1.20                    \\
~~~~~~n=4    & 13.03     & ~~~~~~~~~ 1.61                & ~~~~~~~~~~ 1.58                    \\
~~~~~~n=5    & 16.17     & ~~~~~~~~~ 2.00                & ~~~~~~~~~~ 1.98                    \\
~~~~~~n=6    & 19.32     & ~~~~~~~~~ 2.39                & ~~~~~~~~~~ 2.35                    \\
~~~~~~n=7    & 22.48     & ~~~~~~~~~ 2.78                & ~~~~~~~~~~ 2.73                    \\
~~~~~~n=8    & 25.60     & ~~~~~~~~~ 3.17                & ~~~~~~~~~~ 3.11                    \\
~~~~~~n=9    & 28.75     & ~~~~~~~~~ 3.56                & ~~~~~~~~~~ 3.50                    \\
~~~~~~n=10   & 31.89     & ~~~~~~~~~ 3.94                & ~~~~~~~~~~ 3.88                    \\
\noalign{\smallskip}
\hline\noalign{\smallskip}
\hline \noalign{\smallskip}
\end{tabular}
\end{center}
\end{table*}

Having discussed the scenario for gravitational perturbation, let us explore the higher dimensional effects, i.e., determination of the mass term by solving \ref{GW_Eq20a}. We will be concerned with the even parity eigenfunctions of \ref{GW_Eq20a}, as the derivation of effective field equations assume existence of $Z_{2}$ symmetry. Further, \ref{GW_Eq20a} being a second order differential equation will require two boundary conditions to uniquely arrive at the solution. Rather than imposing boundary conditions on $\chi(y)$ we will impose boundary conditions in $\partial _{y}\chi(y)$. Before engaging with the boundary conditions let us solve \ref{GW_Eq20a}, which on introduction of the new variable, $\zeta =e^{ky}$, becomes
\begin{align}\label{GW_Eq24}
\zeta ^{-2}\left\lbrace -k^{2}\chi+k^{2}\zeta ^{2}\frac{d^{2}\chi}{d\zeta ^{2}}+4k^{2}\zeta \frac{d\chi}{d\zeta}\right\rbrace +m^{2}\chi =0~,
\end{align}
where the following results have been used,
\begin{align}\label{GW_Eq25}
\frac{d\chi}{dy}=k\zeta \frac{d\chi}{d\zeta};
\qquad
\frac{d^{2}\chi}{dy^{2}}=k^{2}\zeta ^{2}\frac{d^{2}\chi}{d\zeta ^{2}}+k^{2}\zeta \frac{d\chi}{d\zeta}~.
\end{align}
One can further transform the above equation to a more manageable form by introducing yet another variable $\xi$, replacing $\zeta$, such that $m\zeta=\xi$ and the transformed version of \ref{GW_Eq24} takes the following form,
\begin{align}\label{GW_Eq26}
k^{2}\xi ^{2}\frac{d^{2}\chi}{d\xi ^{2}}+4k^{2}\xi \frac{d\chi}{d\xi}+\left(\xi ^{2}-k^{2}\right)\chi=0~.
\end{align}
The above equation is essentially Bessel's differential equation and hence it's two independent solutions in terms of modified Bessel functions of first and second kinds are 
\begin{align}\label{GW_Eq27}
\chi (y)=e^{-\frac{3}{2}ky}\left[C_{1}J_{\nu}\left(\frac{me^{ky}}{k}\right)
+C_{2}Y_{\nu}\left(\frac{me^{ky}}{k}\right)\right]~,
\end{align}
with $\nu=\sqrt{13}/2$. The departure from bulk based approach should now be evident from the above analysis. The effect of higher dimensions is through the extra dimensional part of the gravitational perturbation, namely $\chi(y)$. This is certainly a discriminating feature between the bulk and the brane based approach, since the order of the Bessel functions appearing in these two approaches to determine the \KK mode masses are different \cite{Seahra:2004fg}. Thus it is clear that the mass spectrum of our model will be different when compared to the bulk based approach. 

Let us briefly point out the reason behind the difference between \KK mode masses when one follows the brane-based approach, on the one hand, and the bulk-based approach, on the other hand. This is basically due to the difference in the gravitational field equations. For example, when perturbing the bulk gravitational field equations, the Weyl tensor plays no role. By contrast, the perturbation of the Weyl tensor plays a central role in the brane-based approach. Therefore, the basic field equations governing dynamics of gravity in the two approaches differ, but the Schwarzschild AdS spacetime is still a solution of both the field equations. Hence, even though the background solution is the same in both cases, the perturbations follow different dynamics pertaining to the fact that field equations themselves are different. This is why the \KK mode masses are also different. An analogy may be helpful here. For instance, the Schwarzschild solution is a solution of both Einstein gravity as well as 
$f(R)
$ gravity. However, the field equations of both these theories are widely different. Thus, the perturbations about the Schwarzschild background will satisfy different evolution equations in these theories (see, for example \cite{Bhattacharyya:2017tyc,Pratten:2015rqa,Berti:2015itd}), like the scenario we are considering in this work. The fact that the field equations in the bulk and brane based approaches are different is known and is manifested in the fact that there exist solutions to the field equations in the brane-based approach, with no bulk correspondence whatsoever \cite{Dadhich:2000am,Chamblin:2000ra,Maartens:2001jx,Maartens:2010ar,Dadhich:2001ry}. This explains the difference in the masses of the \KK modes associated with the brane and the bulk based approaches respectively.
\begin{table*}
\begin{center}
\caption{Numerical estimates of the mass of first ten \KK modes have been presented for the bulk based approach, by solving \ref{GW_Eq29} for $\nu=2$. It is clear from \ref{Table_01} that the solution $z_{n}$ of \ref{GW_Eq29} is different in the bulk based approach in comparison to the brane based one. Among the two sets of choices for the inter-brane separation $d$ and bulk curvature scale $\ell$, one is identical to that of brane based approach, while the other slightly differs. Both these situations clearly depict the differences of the \KK mass modes in the brane and the bulk based approach.}
\label{Table_02}       
%
%
\begin{tabular}{p{3.5cm}p{1.5cm}p{4cm}p{4.5cm}}
\hline\noalign{\smallskip}
\hline\noalign{\smallskip}
\KK Modes             & ~~$z_{n}$ & ~~~~~~Associated Mass             & ~~~~~~Associated Mass                \\
                      &           &  ($d/\ell=20$; $1/\ell=6\times 10^{7}$) & ($d/\ell=30$; $1/\ell=1.2\times 10^{12}$)  \\
\hline \noalign{\smallskip}
\hline \noalign{\smallskip}
~~~~~~~n=1            & 3.83      & ~~~~~~~~~ 0.47                    & ~~~~~~~~~~~ 0.43                     \\
~~~~~~~n=2            & 7.01      & ~~~~~~~~~ 0.87                    & ~~~~~~~~~~~ 0.79                     \\
~~~~~~~n=3            & 10.18     & ~~~~~~~~~ 1.26                    & ~~~~~~~~~~~ 1.14                     \\
~~~~~~~n=4            & 13.33     & ~~~~~~~~~ 1.65                    & ~~~~~~~~~~~ 1.50                     \\
~~~~~~~n=5            & 16.46     & ~~~~~~~~~ 2.03                    & ~~~~~~~~~~~ 1.85                     \\
~~~~~~~n=6            & 19.61     & ~~~~~~~~~ 2.42                    & ~~~~~~~~~~~ 2.20                     \\
~~~~~~~n=7            & 22.76     & ~~~~~~~~~ 2.81                    & ~~~~~~~~~~~ 2.56                     \\
~~~~~~~n=8            & 25.91     & ~~~~~~~~~ 3.20                    & ~~~~~~~~~~~ 2.91                     \\
~~~~~~~n=9            & 29.05     & ~~~~~~~~~ 3.59                    & ~~~~~~~~~~~ 3.26                     \\
~~~~~~~n=10           & 32.19     & ~~~~~~~~~ 3.98                    & ~~~~~~~~~~~ 3.61                     \\
\noalign{\smallskip}
\hline\noalign{\smallskip}
\hline \noalign{\smallskip}
\end{tabular}
\end{center}
\end{table*}
 
To find the unknown coefficients $C_{1}$ and $C_{2}$ we need to impose boundary conditions and as emphasised earlier these conditions will be on derivatives of $\chi(y)$. To make the analysis at par with possible resolutions of the hierarchy problem, we assume the existence of another brane located at some $y=d$. Incidentally, the distance $d$ need not be constant but varying, known as radion field, whose stabilisation would lead to a non-zero inter-brane separation $d$ \cite{Goldberger:1999uk}. We have also neglected effects of brane bending, if any, by assuming that $d$ is a pure constant. Hence the boundary conditions imposed are given by, $[\partial _{y}+(\nu+(3/2)) k]\chi =0$ at $y=0$ and also on the other brane hypersurface at $y=d$. This leads to the following two algebraic equations satisfied by the two unknown coefficients $C_{1}$ and $C_{2}$ as 
\begin{align}
C_{1}J_{\nu-1}(m/k)&+C_{2}Y_{\nu-1}(m/k)=0~,
\label{GW_Eq28a}
\\
C_{1}J_{\nu-1}(\{m/k\}e^{kd})&+C_{2}Y_{\nu -1}(\{m/k\}e^{kd})=0~.
\label{GW_Eq28b}
\end{align}
Using the first relation one can determine the ratio $C_{1}/C_{2}$ and hence the solution for $\chi(y)$ gets determined except for an overall normalisation. On the other hand, substitution of the same in \ref{GW_Eq28b} results into the algebraic equation 
\begin{align}\label{GW_Eq29}
Y_{\nu-1}(m_{n}/k)J_{\nu-1}(z_{n})-J_{\nu-1}(m_{n}/k)Y_{\nu-1}(z_{n})=0~,
\end{align}
where $m_{n}=\{z_{n}k\}e^{-kd}$ yields an infinite series of solutions for the mass, where $n$ stands for a particular \KK mode. The masses for the first ten \KK modes have been presented in \ref{Table_01} for two different sets of choices of inter-brane separation $d$ and bulk curvature scale $\ell=1/k$. This has been achieved by first solving for $z_{n}$ using \ref{GW_Eq29} and then obtaining the \KK mass $m_{n}$. 

To see clearly the difference between brane and bulk based approach, we have presented the masses of the first ten lowest lying \KK modes in the context of bulk based approach as well. This requires solving \ref{GW_Eq29} for $z_{n}$ with $\nu=2$. It is evident from \ref{Table_02} that the solutions for $z_{n}$ are completely different in the two scenarios. In particular the numerical values of $z_{n}$ in the brane based approach are lower than the corresponding numerical values in the bulk based approach. This results in lowering of the masses of \KK modes in the brane based approach, as evident from \ref{Table_01} and \ref{Table_02} for the choices $d/\ell=20$ and $\ell^{-1}=6\times 10^{7}$ in geometrised units. The numerical values are so chosen that they are in agreement with other constraints already present in this framework. For example, $d/\ell \geq 13$ is necessary to arrive at the desired warping required to get around the gauge hierarchy problem, while the table top experiment of Newton's law would 
demand $1/\ell \geq 10^{7}(M/M_{\odot})$ (or, $\ell\leq 0.1~\textrm{mm}$), where $M_{\odot}$ is the solar mass \cite{Hoyle:2004cw,Long:2002wn,Smullin:2005iv}. This explains the choices of $d/\ell$ as well as that of $1/\ell$. Numerical estimates of the masses of the \KK modes for two such choices of $d/\ell$ and $1/\ell$ values have been presented in \ref{Table_01} and \ref{Table_02} respectively. Masses of these \KK modes will be used in the next section for determination of the quasi-normal modes for the brane black hole. 

At this stage it is worth mentioning about the Gregory-Laflamme instability, which originates due to instability of the bulk metric under perturbation pertaining to long wavelength modes \cite{Gregory:2008rf,Gregory:2000gf,Gregory:2011kh,Lehner:2011wc,Frolov:2009jr}. The fact that there exist another brane at $y=d$ helps to evade the instability by providing a cutoff on the long wavelength modes. The separation $d$ between the two branes as well as the bulk curvature scale $\ell\sim 1/k$ (see \ref{GW_Eq01}) are also bounded by the fact that we have not seen any influence of the extra dimension on the gravitational interaction in our observable universe. The above instability essentially translates through $d$ and $\ell$ into the mass of the \KK modes and for $m_{n}\gtrsim 0.43$ the above instability can be avoided, which is also reflected in both the tables depicting masses of the \KK modes (see also \cite{Konoplya:2008yy}).   

Finally, given a particular \KK mode $n$, one can determine the extra dimensional part of the gravitational perturbation as 
\begin{align}\label{GW_Eq30}
\chi _{n}(y)=N_{n}\Bigg[&Y_{\nu-1}(m_{n}/k)J_{\nu}(\{m_{n}/k\} e^{ky})
\nonumber
\\
&-J_{\nu-1}(m_{n}/k)Y_{\nu}(\{m_{n}/k\} e^{ky})\Bigg]~.
\end{align}
Here $N_{n}$ is the overall normalisation factor and $\nu=\sqrt{13}/2$ is the order of the Bessel functions. Thus the complete solution to the gravitational perturbation can be written in the following form,
\begin{align}
h_{\alpha \beta}(t,r,\theta,\phi;y)&=\sum _{n=0}^{\infty}N_{n}\Big\{Y_{\nu-1}(m_{n}/k)J_{\nu}(\{m_{n}/k\} e^{ky})
\nonumber
\\
&-J_{\nu-1}(m_{n}/k)Y_{\nu}(\{m_{n}/k\} e^{ky})\Big\}
\nonumber
\\
&\times \sum _{l=0}^{\infty}\sum _{m=-l}^{l} \Bigg\{\sum _{i=1}^{7}P_{i}^{(nlm)}(t,r)\mathcal{P}^{(i)lm}_{\alpha \beta}(\theta,\phi)
\nonumber
\\
&+\sum _{i=1}^{3}A_{i}^{(nlm)}(t,r)\mathcal{A}^{(i)lm}_{\alpha \beta}(\theta,\phi)\Bigg\}~.
\end{align}
Here the first part is the contribution from extra dimensions, while the four dimensional effects have been divided into polar and axial perturbations respectively. The first seven are the polar perturbations, while the last three are the axial ones. As already emphasised earlier, these two contributions do not mix and hence one can treat them separately. We have already provided the evolution equations for the master variables associated with the axial perturbation in \ref{GW_Eq23a} and \ref{GW_Eq23b}, which we will solve next. The solution (or, evolution) can be obtained in two ways, by the calculation of quasi-normal modes, or, performing a fully numerical Cauchy evolution of the initial data. We have performed both these analysis in this work and shall present the calculation of quasi-normal modes in the next section before taking up the Cauchy evolution of initial data.
\section{The spectrum of associated quasi-normal modes}\label{qnm_analysis}

\begin{table*}
\begin{center}
\caption{Real and Imaginary parts of the \qnm frequencies have been presented. These are obtained from the \emph{brane based approach} with the following choice of parameters associated with the extra dimensions: $d/\ell=20; 1/\ell=6\times 10^{7}$. In particular, results for the first two \KK mass modes have been presented for four different choices of the angular momentum.}
\label{Table_03}       
%
%
\begin{tabular}{p{2cm}p{3cm}p{3cm}p{3cm}p{3cm}}
\hline\noalign{\smallskip}
\hline\noalign{\smallskip}
        &  $m=0.44, l=2$  &               & $m=0.83, l=2$  &                   \\
\hline\noalign{\smallskip}
\hline\noalign{\smallskip}
Mode     & Real             &   Imaginary   &  Real             &  Imaginary        \\
\hline \noalign{\smallskip}
\hline \noalign{\smallskip}
j=1     & 0.467             &  -0.051       & 0.396             & -0.038            \\
j=2     & 0.530             &  -0.071       & 0.543             & -0.104            \\
j=3     & 0.378             &  -0.197       & 0.183             & -0.168            \\
j=4     & 0.473             &  -0.239       & 0.243             & -0.369            \\
\noalign{\smallskip}
\hline\noalign{\smallskip}
\hline \noalign{\smallskip}
        & $m=0.44, l=3$  &               & $m=0.83, l=3$  &                   \\
\hline\noalign{\smallskip}
\hline \noalign{\smallskip}
j=1     & 0.653             &  -0.078       & 0.843             & -0.051            \\
j=2     & 0.708             &  -0.084       & 0.708             & -0.134            \\
j=3     & 0.618             &  -0.244       & 0.773             & -0.176            \\
j=4     & 0.562             &  -0.437       & 0.576             & -0.327            \\
\noalign{\smallskip}
\hline\noalign{\smallskip}
\hline \noalign{\smallskip}
        & $m=0.44, l=4$   &              & $m=0.83, l=4$  &                   \\
\hline\noalign{\smallskip}
\hline \noalign{\smallskip}
j=1     & 0.847             &  -0.086       & 0.951             & -0.064            \\
j=2     & 0.827             &  -0.263       & 0.960             & -0.219            \\
j=3     & 0.791             &  -0.451       & 0.896             & -0.393            \\
\noalign{\smallskip}
\hline\noalign{\smallskip}
\hline \noalign{\smallskip}
        & $m=0.44, l=5$   &              & $m=0.83, l=5$  &                   \\
\hline\noalign{\smallskip}
\hline \noalign{\smallskip}
j=1     & 1.043             &  -0.090       & 1.123             & -0.076            \\
j=2     & 1.084             &  -0.272       & -1.098            & -0.233            \\
j=3     & 1.002             &  -0.460       & 1.051             & -0.404            \\
\noalign{\smallskip}
\hline\noalign{\smallskip}
\hline \noalign{\smallskip}
\end{tabular}
\end{center}
\end{table*}

\begin{table*}
\begin{center}
\caption{Real and Imaginary parts of the first few \qnm frequencies have been depicted. These values are obtained starting from the \emph{brane based approach}, with the following choices of the extra dimensional parameters: $d/\ell=30; 1/\ell=1.3\times 10^{12}$. Results have been presented for two lowest lying \KK mass modes and for four choices of angular momentum associated with each modes.}
\label{Table_04}       
%
%
\begin{tabular}{p{2cm}p{3cm}p{3cm}p{3cm}p{3cm}}
\hline\noalign{\smallskip}
\hline\noalign{\smallskip}
        &  $m=0.43, l=2$  &                & $m=0.82, l=2$    &                 \\
\hline\noalign{\smallskip}
\hline\noalign{\smallskip}
Mode     & Real             &   Imaginary    & Real		    &  Imaginary      \\
\hline \noalign{\smallskip}
\hline \noalign{\smallskip}
j=1     & 0.462             &  -0.053        & 0.702              & -0.006          \\
j=2     & 0.527             &  -0.072        & 0.385              & -0.041          \\
j=3     & 0.377             &  -0.201        & 0.541              & -0.109          \\
j=4     & 0.471             &  -0.241        & 0.253              & -0.326          \\
\noalign{\smallskip}
\hline\noalign{\smallskip}
\hline \noalign{\smallskip}
        & $m=0.43, l=3$  &                & $m=0.82, l=3$    &                 \\
\hline\noalign{\smallskip}
\hline \noalign{\smallskip}
j=1     & 0.650             &  -0.079        & 0.796              & -0.036          \\
j=2     & 0.616             &  -0.246        & 0.705              & -0.138          \\
j=3     & 0.679             &  -0.261        & 0.770              & -0.179          \\
j=4     & 0.562             &  -0.439        & 0.576              & -0.331          \\
\noalign{\smallskip} 
\hline\noalign{\smallskip}
\hline \noalign{\smallskip}
        & $m=0.43, l=4$   &               & $m=0.82, l=4$    &                 \\
\hline\noalign{\smallskip}
\hline \noalign{\smallskip}
j=1     & 0.846             &  -0.086        & 0.948              & -0.065          \\
j=2     & 0.826             &  -0.264        & 0.906              & -0.204          \\
j=3     & 0.790             &  -0.452        & 0.833              & -0.373          \\
\noalign{\smallskip}
\hline\noalign{\smallskip}
\hline \noalign{\smallskip}
        & $m=0.43, l=5$   &               & $m=0.82, l=5$    &                 \\
\hline\noalign{\smallskip}
\hline \noalign{\smallskip}
j=1     & 1.041             &  -0.090        & 1.120              & -0.076          \\
j=2     & 1.027             &  -0.272        & 1.095              & -0.234          \\
j=3     & 1.001             &  -0.461        & 1.050              & -0.406          \\
\noalign{\smallskip}
\hline\noalign{\smallskip}
\hline \noalign{\smallskip}
\end{tabular}
\end{center}
\end{table*}

\begin{table*}
\begin{center}
\caption{In this table we have presented numerical estimates for real and imaginary parts of the \qnm frequencies obtained from the \emph{bulk based approach}. The parameters characterising the bulk spacetime corresponds to: $d/\ell=20$; $1/\ell=6\times 10^{7}$. In this situation as well we have presented the \qnm frequencies for four possible choices of angular momentum given the two lowermost \KK mode masses.}
\label{Table_05}       
%
%
\begin{tabular}{p{2cm}p{3cm}p{3cm}p{3cm}p{3cm}}
\hline\noalign{\smallskip}
\hline\noalign{\smallskip}
        &  $m=0.47, l=2$  &                & $m=0.87, l=2$    &                 \\
\hline\noalign{\smallskip}
\hline\noalign{\smallskip}
Mode     & Real             &   Imaginary    & Real               & Imaginary       \\
\hline \noalign{\smallskip}
\hline \noalign{\smallskip}
j=1     & 0.480             &  -0.046        & 0.437              & -0.015          \\
j=2     & 0.540             &  -0.067        & 0.542              & -0.087          \\
j=3     & 0.381             &  -0.185        & 0.119              & -0.128          \\
j=4     & 0.477             &  -0.231        & 0.242              & -0.318          \\
\noalign{\smallskip}
\hline\noalign{\smallskip}
\hline \noalign{\smallskip}
        & $m=0.47, l=3$  &                & $m=0.87, l=3$    &                 \\
\hline\noalign{\smallskip}
\hline \noalign{\smallskip}
j=1     & 0.660             &  -0.076        & 0.862              & -0.045          \\
j=2     & 0.716             &  -0.082        & 0.719              & -0.117          \\
j=3     & 0.623             &  -0.239        & 0.785              & -0.163          \\
j=4     & 0.564             &  -0.431        & 0.576              & -0.313          \\
\noalign{\smallskip} 
\hline\noalign{\smallskip}
\hline \noalign{\smallskip}
        & $m=0.47, l=4$   &               & $m=0.87, l=4$    &                 \\
\hline\noalign{\smallskip}
\hline \noalign{\smallskip}
j=1     & 0.853             &  -0.085        & 1.010              & -0.067          \\
j=2     & 0.831             &  -0.259        & 0.920              & -0.193          \\
j=3     & 0.793             &  -0.447        & 0.840              & -0.359          \\
\noalign{\smallskip}
\hline\noalign{\smallskip}
\hline \noalign{\smallskip}
        & $m=0.47, l=5$   &               & $m=0.87, l=5$    &                 \\
\hline\noalign{\smallskip}
\hline \noalign{\smallskip}
j=1     & 1.047             &  -0.089        & 1.176              & -0.077          \\
j=2     & 1.032             &  -0.269        & 1.108              & -0.227          \\
j=3     & 1.004             &  -0.457        & 1.058              & -0.396          \\
\noalign{\smallskip}
\hline\noalign{\smallskip}
\hline \noalign{\smallskip}
\end{tabular}
\end{center}
\end{table*}

\begin{table*}
\begin{center}
\caption{Real and Imaginary parts of the \qnm frequencies have been depicted in a bulk spacetime with the following set of parameters: $d/\ell=30$; $1/\ell=1.2\times 10^{12}$ in the \emph{bulk based approach}. The values have been presented for four choices of angular momentum, given the two lowest lying \KK modes.}
\label{Table_06}       
%
%
\begin{tabular}{p{2cm}p{3cm}p{3cm}p{3cm}p{3cm}}
\hline\noalign{\smallskip}
\hline\noalign{\smallskip}
        &  $m=0.43, l=2$  &                & $m=0.79, l=2$    &                 \\
\hline\noalign{\smallskip}
\hline\noalign{\smallskip}
Mode     & Real             &   Imaginary    & Real               & Imaginary       \\
\hline \noalign{\smallskip}
\hline \noalign{\smallskip}
j=1     & 0.462             &  -0.053        & 0.672              & -0.006          \\
j=2     & 0.527             &  -0.072        & 0.456              & -0.014          \\
j=3     & 0.377             &  -0.201        & 0.542              & -0.087          \\
j=4     & 0.471             &  -0.241        & 0.534              & -0.123          \\
\noalign{\smallskip}
\hline\noalign{\smallskip}
\hline \noalign{\smallskip}
        & $m=0.43, l=3$  &                & $m=0.79, l=3$    &                 \\
\hline\noalign{\smallskip}
\hline \noalign{\smallskip}
j=1     & 0.650             &  -0.079        & 0.825              & -0.055          \\
j=2     & 0.616             &  -0.246        & 0.696              & -0.149          \\
j=3     & 0.678             &  -0.261        & 0.761              & -0.187          \\
j=4     & 0.562             &  -0.439        & 0.666              & -0.377          \\
\noalign{\smallskip} 
\hline\noalign{\smallskip}
\hline \noalign{\smallskip}
        & $m=0.43, l=4$   &               & $m=0.79, l=4$    &                 \\
\hline\noalign{\smallskip}
\hline \noalign{\smallskip}
j=1     & 0.846             &  -0.086        & 0.937              & -0.067          \\
j=2     & 0.826             &  -0.264        & 0.898              & -0.210          \\
j=3     & 0.790             &  -0.452        & 0.829              & -0.382          \\
\noalign{\smallskip}
\hline\noalign{\smallskip}
\hline \noalign{\smallskip}
        & $m=0.43, l=5$   &               & $m=0.79, l=5$    &                 \\
\hline\noalign{\smallskip}
\hline \noalign{\smallskip}
j=1     & 1.041             &  -0.090        & 1.112              & -0.078          \\
j=2     & 1.027             &  -0.272        & -1.089             & -0.238          \\
j=3     & 1.001             &  -0.461        & 1.045              & -0.411          \\
\noalign{\smallskip}
\hline\noalign{\smallskip}
\hline \noalign{\smallskip}
\end{tabular}
\end{center}
\end{table*}

In this section we will investigate the characteristic frequencies, namely the quasi-normal modes associated with the propagation of massive \KK modes in the Schwarzschild geometry induced on the brane hypersurface. This is usually performed by going over to the frequency space, such that,
\begin{align}
u_{n,l}(t,r)&=\int d\omega~e^{-i\omega t} \psi_{n,l}(\omega,r)~,
\label{Eq_qnm_01a}
\\
v_{n,l}(t,r)&=\int d\omega~e^{-i\omega t} \phi_{n,l}(\omega,r)~.
\label{Eq_qnm_01b}
\end{align}
At this stage all possible frequencies are allowed, but as we will see later on, this is not the case. Only some specific set of frequencies are allowed, known as the \qnm frequencies and hence the above integral will be converted to a sum over all the \qnm frequencies. The single most important fact about this expansion is that the \qnm frequencies can be imaginary. Since we do not expect any runaway situations associated with this problem, thus $\textrm{Im}(\omega)<0$ are the allowed \qnm frequencies \cite{Leaver:1985ax,Chandrasekhar:1985kt,Nollert:1999ji,Kokkotas:1999bd,Pani:2013pma,Cardoso:2011xi,
Rosa:2011my,Konoplya:2008rq,Konoplya:2006br,Kanti:2005xa}. Before getting into the details of obtaining the \qnm frequencies in this context, let us briefly discuss about another prediction of \ref{GW_Eq23a} and \ref{Eq_qnm_01a}, namely late time wave tails. Since \qnm frequencies have a real as well as imaginary parts, it is exponentially suppressed and at late times ($t\rightarrow \infty$) it produces vanishing contribution. Therefore the wave tail, originating from existence of branch cut in the frequency integral of \ref{Eq_qnm_01a} dominates the late time behaviour of the gravitational perturbation $u_{n,l}(t)$. It turns out that the power law scaling of the perturbation modes have a universal behaviour. In particular, for massive gravitational modes, which includes the scenario presented in this work, the late time behaviour essentially corresponds to the following universal power law behaviour $u_{n,l}(t)\sim t^{-5/6}\sin (\omega t)$. Here the oscillation frequency $\omega$ depends on the mass of the 
perturbation mode linearly. Thus the late time behaviour is essentially governed by the $t^{-5/6}$ universal factor. We will need this fact in the later parts of this work. For the moment being we will exclusively concentrate on the \qnm analysis.

In order to determine the \qnm frequencies one also needs to impose suitable boundary conditions on the solution space. These are --- (a) the \qnm must be ingoing at the black hole horizon and (b) these modes must be outgoing in the asymptotic regions. These conditions are best suited in terms of the tortoise coordinate $r_{*}$, defined as integral of $\{dr/f(r)\}$, in which the horizon corresponds to $r_{*}\rightarrow -\infty$, while the asymptotic region implies $r_{*}\rightarrow \infty$. Thus the condition that quasi-normal modes are ingoing at the horizon implies that $u_{n,l}(\omega,r_{*})$ as well as $v_{n,l}(\omega,r_{*})$ behave as $\exp(-i\omega r_{*})$ in the near horizon regime. A similar situation will exist for the asymptotic region as well. These boundary conditions will dictate the discrete values of the frequencies associated with the quasi-normal modes. These values will have three indices, the \KK mode index $n$, the angular momentum index $l$ and the \qnm index $p$. Having obtained the 
corresponding quasi-normal modes one can substitute them back to \ref{Eq_qnm_01a} and \ref{Eq_qnm_01b} respectively and thus obtain the time evolution of the both $u_{n,l}(t,r_{*})$ and $v_{n,l}(t,r_{*})$. These estimates can then be compared with the Cauchy evolution problem and a match between the two will ensure correctness of our method presented here. Thus for completeness and consistency we will also present results for Cauchy evolution in the next section. We will mainly content ourselves with the continued fraction method but will briefly discuss the forward integration scheme as well. 

\subsection{Continued fraction method}

The frequency spectrum associated with the quasi-normal modes can be obtained by starting with a suitable ansatz for $u_{n,l}(t,r)$ and $v_{n,l}(t,r)$ respectively. Given this ansatz one can try to obtain a power series solution associated with the differential equations presented in \ref{GW_Eq23a} and \ref{GW_Eq23b}, resulting in recursion relation between the coefficients of various terms in the power series. This recursion relation will be satisfied provided the \qnm frequencies are discrete. For this purpose we start with the following general form of the coupled differential equations,
\begin{align}
-\frac{\partial ^{2}u_{n,l}}{\partial t^{2}}+\frac{\partial ^{2}u_{n,l}}{\partial r_{*}^{2}}-f(r)&\left(m_{n}^{2}+\frac{l(l+1)r-6}{r^{3}}\right)u_{n,l}
\nonumber
\\
&-f(r)\frac{m_{n}^{2}}{r^{3}}v_{n,l}=0~,
\\
-\frac{\partial ^{2}v_{n,l}}{\partial t^{2}}+\frac{\partial ^{2}v_{n,l}}{\partial r_{*}^{2}}-f(r)&\left(m_{n}^{2}+\frac{l(l+1)}{r^{2}} \right)v_{n,l}
\nonumber
\\
&-4f(r)u_{n,l}=0~,
\end{align}
where $f(r)=1-(2/r)$. Subsequently eliminating derivatives with respect to $r_{*}$ in favour of $r$ and writing down the two master variables $u_{n,l}(t,r)$ and $v_{n,l}(t,r)$ as in \ref{Eq_qnm_01a} and \ref{Eq_qnm_01b} we obtain after simplifications,
\begin{align}
r(r&-2)\frac{d^{2}\psi_{n,l}}{dr^{2}}+2\frac{d\psi _{n,l}}{dr}+\frac{\omega ^{2}r^{3}}{r-2}\psi _{n,l}
\nonumber
\\
&-\left[m_{n}^{2}r^{2}+l(l+1)-\frac{6}{r}\right]\psi _{n,l}-\frac{m_{n}^{2}}{r}\phi _{n,l}=0~,
\label{Eq_qnm_02a}
\\
r(r&-2)\frac{d^{2}\phi_{n,l}}{dr^{2}}+2\frac{d\phi _{n,l}}{dr}+\frac{\omega ^{2}r^{3}}{r-2}\phi _{n,l}
\nonumber
\\
&-\left[m_{n}^{2}r^{2}+l(l+1)\right]\phi _{n,l}-4r^{2}\psi _{n,l}=0~.
\label{Eq_qnm_02b}
\end{align}
Having derived the basic equations governing $\psi_{n,l}$ and $\phi_{n,l}$ one normally writes down both these master variables in terms of various powers of $r$ and $(r-2)$, such that the boundary conditions at horizon and at asymptotic regions can be satisfied. Subsequently the remaining pieces of $\psi_{n,l}$ and $\phi_{n,l}$ are solved by using the power series method. The resulting recursion relation between the coefficients of these power series will also be coupled and it is only helpful to combine them into a single matrix equation with off-diagonal entries illustrating the coupling between the systems. Performing the same for the master variables involved here as well, one ends up with the following matrix equation for $j>0$, with integer $j$ as,
\begin{align}
\mathbf{P}_{j}\mathbf{V}_{j+1}+\mathbf{Q}_{j}\mathbf{V}_{j}+\mathbf{R}_{j}\mathbf{V}_{j-1}=0~.
\label{Eq_qnm_07}
\end{align}
Here, the coefficients $\mathbf{P}_{j}$, $\mathbf{Q}_{j}$ and $\mathbf{R}_{j}$ depend on the details of the system, i.e., on the parameters involved. The vector $\mathbf{V}_{j}$ on the other hand corresponds to a column matrix constructed out of the power series coefficients for $\psi_{n,l}$ and $\phi_{n,l}$, such that one obtains
\begin{align}
\mathbf{P}_{j}&=\left(\begin{array}{ll}
                      \alpha _{j} & 0\\
                      0 & \alpha _{j}
                     \end{array}\right)~,\qquad
\mathbf{Q}_{j}=\left(\begin{array}{ll}
                      \beta _{j}+3 & -\frac{m_{n}^{2}}{2}\\
                      -4 & \beta _{j}
                     \end{array}\right)~,
                     \nonumber
                     \\
\mathbf{R}_{j}&=\left(\begin{array}{ll}
                      \gamma _{j}-3 & \frac{m_{n}^{2}}{2}\\
                      0 & \gamma _{j}
                     \end{array}\right)~,
\end{align}
where the unknown coefficients $\alpha_{j}$, $\beta _{j}$ and $\gamma _{j}$ can be written in terms of the \KK mode mass and the quasi-normal mode frequency $\omega$ as,
\begin{align}
\alpha _{j}&=(j+1)(j+1-4i\omega)~,
\nonumber
\\
\gamma _{j}&=\left(j-1+\frac{(\omega -i\lambda)^{2}}{\lambda}\right)\left(j+1+\frac{(\omega -i\lambda)^{2}}{\lambda}\right)~,
\nonumber
\\
\beta _{j}&=-2j^{2}+\left(-2+\frac{8i\omega \lambda -2\omega ^{2}+6\lambda ^{2}}{\lambda} \right)j
\nonumber
\\
&-l(l+1)+\frac{1}{\lambda}\Big(3\lambda ^{2}-\omega ^{2}-12i\omega \lambda^{2}-4\lambda ^{3}
\nonumber
\\
&+4i\omega \lambda +12\lambda \omega ^{2}+4i\omega ^{3} \Big)~.
\end{align}
where, $\lambda=\sqrt{m_{n}^{2}-\omega ^{2}}$. The above recursion relation must be supplemented with the zeroth order recursion relation, which simply reads, $\mathbf{P}_{0}\mathbf{V}_{1}+\mathbf{Q}_{0}\mathbf{V}_{0}=0$. Given this one can use \ref{Eq_qnm_07} to replace $\mathbf{V}_{1}$ in terms of $\mathbf{V}_{0}$ and $\mathbf{V}_{2}$. Subsequently one can again replace $\mathbf{V}_{2}$ by higher order terms using \ref{Eq_qnm_07} repeatedly. This method of solving the matrix valued recursion relation presented in \ref{Eq_qnm_07} is known as the method of continued fraction. In this method, following the procedure outlined above one ends up with an equation of the form $\mathbf{M}\mathbf{V}_{0}=0$, where the matrix $\mathbf{M}$ reads,
\begin{align}
\mathbf{M}=\mathbf{Q}_{0}-\mathbf{P}_{0}\left[\mathbf{Q}_{1}-\mathbf{P}_{1}\{\mathbf{Q}_{2}+\mathbf{P}_{2}\mathbf{M}_{2}\}\mathbf{R}_{2} \right]^{-1}\mathbf{R}_{1}~.
\end{align}
Here $\mathbf{M}_{j}$ is a matrix which can be written in terms of $\mathbf{P}_{j+1}$, $\mathbf{Q}_{j+1}$, $\mathbf{R}_{j+1}$ and most importantly also depends on $\mathbf{M}_{j+1}$. Moreover the matrix $\mathbf{M}_{j}$ when acts on $\mathbf{V}_{j}$ yields $\mathbf{V}_{j+1}$. Thus in order for the matrix equation $\mathbf{M}\mathbf{V}_{0}=0$ to have non-trivial solutions for $\mathbf{V}_{0}$, one must have 
\begin{align}\label{Eq_qnm_08}
\textrm{det}~\mathbf{M}=0~.
\end{align}
\begin{figure*}[t!]
\centering
\includegraphics[scale=0.45]{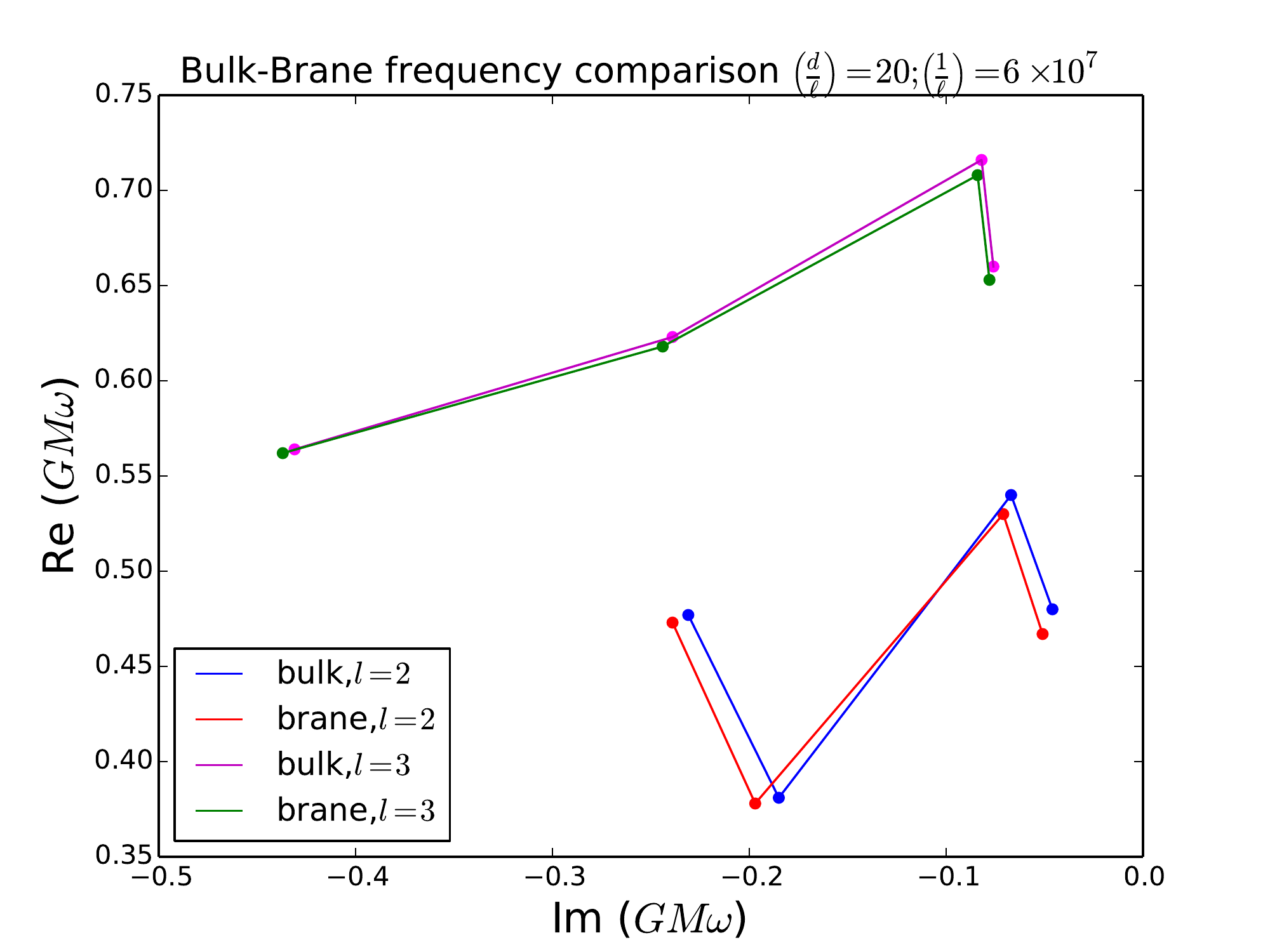}~~
\includegraphics[scale=0.45]{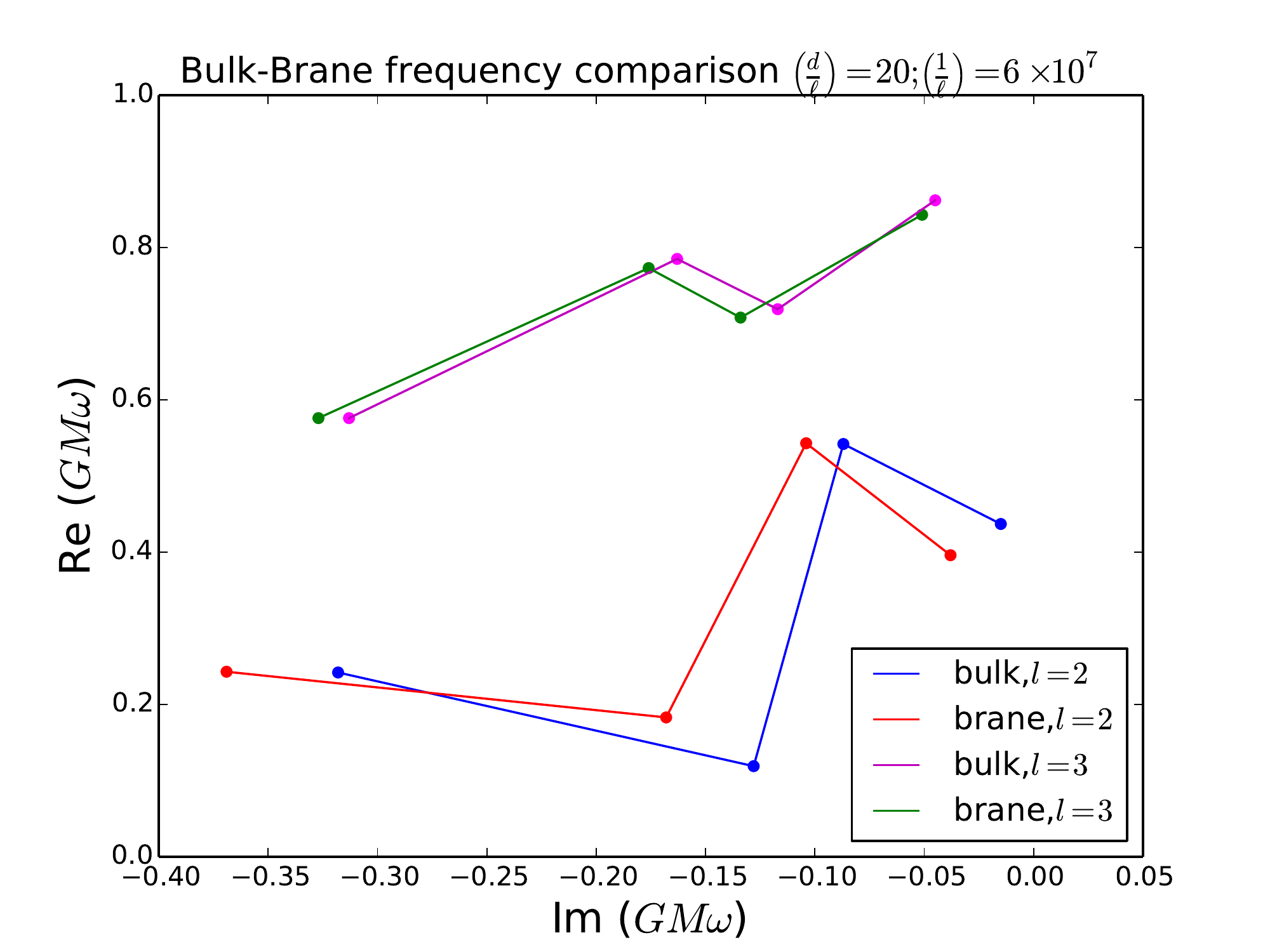}
\caption{Real and imaginary parts of the \qnm frequencies have been plotted. The figure on the left corresponds to the \qnm frequencies associated with the lowest lying \KK mass modes in both the brane and bulk based approach. The curves at the bottom shows the $l=2$ case, while the curves at the top depicts the situation when $l=3$. The figure to the right illustrates an identical situation but for the next \KK mass modes. As evident from the curves, the imaginary part of \qnm frequencies are smaller in the case of bulk based approach resulting in less damping. We will confirm this behaviour in the later sections.}
\label{Fig_QNM}
\end{figure*}
In principle one needs to take into account an infinite number of terms to solve the above equation. However in practice one truncates $\mathbf{M}_{j}$ at some order $J$, and hence obtain all the lower order matrices starting from $\mathbf{M}_{J}$. Of course, at a later stage one needs to check the independence of the solution of \ref{Eq_qnm_08} explicitly on the truncation order $J$. We have solved the above matrix valued recursion relation using the continued fraction method discussed earlier in the symbolic manipulation package MATHEMATICA and have obtained the corresponding lowest lying \qnm frequencies associated with various \KK mode masses for different values of angular momentum. These values are listed in four tables. In \ref{Table_03} we present both the real and imaginary parts of the \qnm frequencies for the two lowest lying \KK mass modes associated with the following values: $d/\ell=20; 1/\ell=6\times 10^{7}$. It is clear that as the mass increases the imaginary part of the lowest \qnm 
frequency decreases, while it increases with angular momentum. For example, when $l=2$, $\textrm{Im}~\omega=-0.05$ for $m_{1}=0.44$, while it becomes $-0.04$ as the mass increases to $m_{2}=0.83$. Hence more massive the \KK modes are, the \qnm functions are less and less damped --- a feature in complete agreement with the result of \cite{Seahra:2004fg}. While for $m=0.44$, the imaginary part of the lowest \qnm frequency will read $\textrm{Im}~\omega=-0.051$ for $l=2$, while it becomes $-0.078$ as the angular momentum increases to $l=3$. Thus with an increase of angular momentum the imaginary part of the \qnm frequency also increases. Hence among the modes with $l=2$ and $l=3$, the time evolution of the $l=3$ mode will be more damped in comparison to the $l=2$ one. This feature is also present in \ref{Table_04}, where the \qnm frequencies have been presented for a different choice of the ratio between brane separation and bulk curvature scale, namely for $d/\ell=30$ and $1/\ell=1.3\times 10^{12}$. 

These numerical values are again chosen to be consistent with previous experimental bounds on $d$ and $\ell$ as explained earlier. In this case also as the mass of the \KK mode increases the imaginary part of the \qnm frequency decreases, while the increase of angular momentum has a reverse effect. For the same choices of the bulk parameters, the \KK mode masses for the brane based and the bulk based approach differs as evident from \ref{Table_01} and \ref{Table_02}. For example, in the situation where $d/\ell=20; 1/\ell=6\times 10^{7}$, the lowest lying \KK mode mass in brane based approach is $m_{1}=0.44$, while that in the bulk based approach being $m_{1}=0.47$. Hence the imaginary part of the \qnm frequency will be lower for the bulk based approach. This has interesting implications --- the axial perturbation generated from bulk Einstein's equations will decay in a slower pace in time when compared to the corresponding perturbation mode originated from effective field equations on the brane. This 
situation has been clearly depicted in \ref{Table_05} and \ref{Table_06} respectively (see also \ref{Fig_QNM}). One can also check that the \qnm frequencies derived here indeed matches with those derived in the direct integration scheme which we will discuss next. 
\subsection{Direct integration method}

In the previous section we have discussed one particular method of determining the \qnm frequencies associated with the perturbation of brane world black hole. However for completeness we present another supplementary method of computing the \qnm frequencies, which can be used along with the continued fraction method to correctly predict the \qnm frequencies. In this method, as the name suggests, one integrates directly from the horizon to the asymptotic region given the boundary conditions mentioned earlier. In this problem we have two master variables characterising the axial gravitational perturbation and satisfying two second order coupled ordinary differential equations (see \ref{Eq_qnm_03a} and \ref{Eq_qnm_03b} respectively in \ref{App_B}). The solution in the near horizon regime will have $e^{-i\omega r_{*}}$ times a power series around the horizon, while at infinity it will behave as $e^{-k_{\infty}r_{*}}$, where $k_{\infty}$ is the wave number in the asymptotic region. The asymptotic solution will 
be characterised by a two-dimensional column vector $\{b_{\infty}^{(1)},b_{\infty}^{(2)}\}$, for which one can choose a suitable orthonormal system of basis vectors. Numerical integration of these differential equations from the horizon out to infinity will lead to a $(2\times 2)$ matrix $\mathbf{S}(m_{n},\omega)$, which can be expanded in the basis introduced above. Finally, setting the determinant of this matrix $\mathbf{S}$ to zero one can solve for the \qnm frequencies 
\cite{Pani:2013pma,Rosa:2011my}. 

Further note that, this method is particularly suited for determination of quasi-bound states, for which the leading order behaviour of the fields at infinity is well understood. However for the determination of \qnm frequencies one needs to extract additional, sub-dominant behaviour of the mode functions at infinity, which makes this approach prone to numerical errors. However if the imaginary part is small compared to the real part, one can determine the \qnm frequencies to sufficient accuracy. In practice one integrates these differential equation to some high value of radial distance and the result must be impermeable to any shift in this distance. Also one can supplement one of these methods by checking whether for a given \KK mode mass and angular momentum one obtains the same \qnm frequency from the other. We have explicitly checked that this is indeed the case, the values obtained from the continued fraction method is in good agreement with those obtained from the direct integration scheme as well. 
This depicts the internal consistency of our model in a straightforward manner.
\begin{figure*}[t!]
\centering
\includegraphics[scale=0.65]{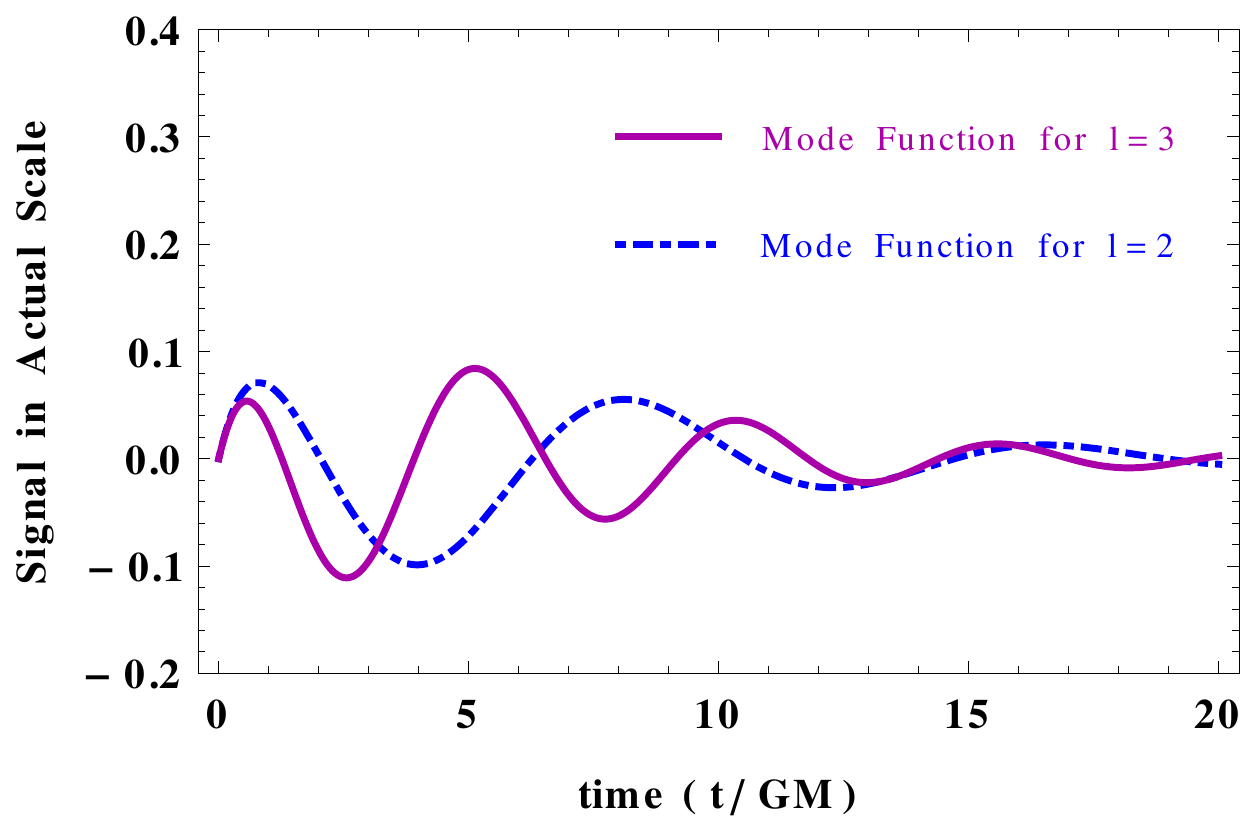}~~
\includegraphics[scale=0.65]{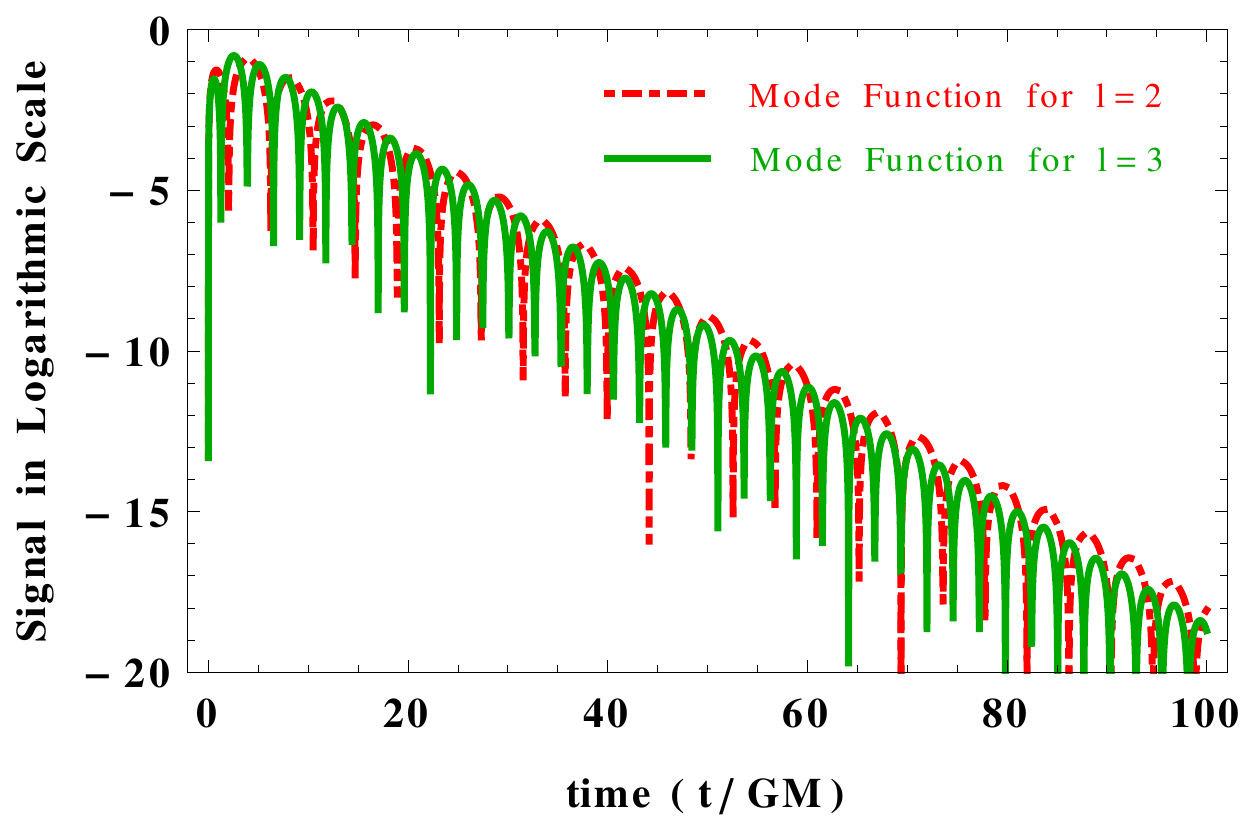}
\caption{Time evolution of the master mode function $u_{n,l}(t)$ associated with axial gravitational perturbation for two different values of angular momentum $l$ in the context of \gr\ have been depicted. The time scale has been normalised to the mass of the central hole, i.e., $t\rightarrow t/GM$. Moreover the figure in the left illustrates the actual evolution of the mode function with time, while the right one presents the same but in a Logarithmic scale. The amplitude of the mode function corresponding to $l=3$ is slightly smaller compared to the mode function having $l=2$, as evident from the right figure. In both of them the dotted one stands for mode function with $l=2$, while the continuous one is the mode function with $l=3$. We will contrast this scenario with the respective ones in the presence of extra dimensions.}
\label{Fig_01_GR}
\end{figure*}
\section{Numerical analysis of the quasi-normal modes}\label{Num_qnm}

The principal aim of this work was to determine the time evolution of the perturbations obtained from the effective gravitational field equations on the brane. Also, to contrast the same with the time evolution of perturbation derived from bulk Einstein's equations. One can achieve this by following two possible avenues --- (a) Obtaining the \qnm frequencies and hence obtaining the time evolution and (b) Solving the Cauchy evolution problem numerically and hence arrive at the evolution of the gravitational perturbation. 

In this section we will follow the first method where the time evolution of the mode function $u_{n,l}(t)$ depicting axial gravitational perturbation will be presented, using the \qnm analysis performed in \ref{qnm_analysis}. For this purpose we will use \ref{Eq_qnm_01a}, where the integral over all frequencies will now be replaced by summation over all the \qnm frequencies. Thus our strategy will be as follows, we will use the numerically computed \qnm frequencies and then sum over them in order arrive at the time evolution for the mode function $u_{n,l}(t)$. Here we would like to reiterate the fact that $n$ stands for the \KK modes and $l$ is the angular momentum associated with the gravitational perturbation. For example, $u_{0,2}$ corresponds to the axial gravitational perturbation associated with angular momentum $l=2$ around a \gr\ solution, while $u_{1,3}$ is the axial gravitational perturbation associated with the lowest lying \KK mode and with angular momentum $l=3$. In what follows using the 
numerical values of \qnm frequencies we will present time evolution of $u_{n,l}(t)$ for a few low lying \KK modes with different choices of angular momentum $l$. These will be contrasted with the mode functions $u_{0,l}$ associated with \gr.

Note that this process is inherently approximate, since in principle one should add all the \qnm frequencies in order to obtain the time evolution of the perturbation, while here we will consider a few lowest lying quasi-normal modes to perform the same. Even though this is certainly an approximate description, it will nevertheless provide the overall behaviour of the gravitational perturbation with time and the key features that will distinguish the scenario presented here from that in \gr. More refined results can be obtained using the Cauchy evolution, which we will present in the next section. This will provide another self-consistency check of our formalism and hence of the associated results.  
\begin{figure*}[t!]
\centering
\includegraphics[scale=0.65]{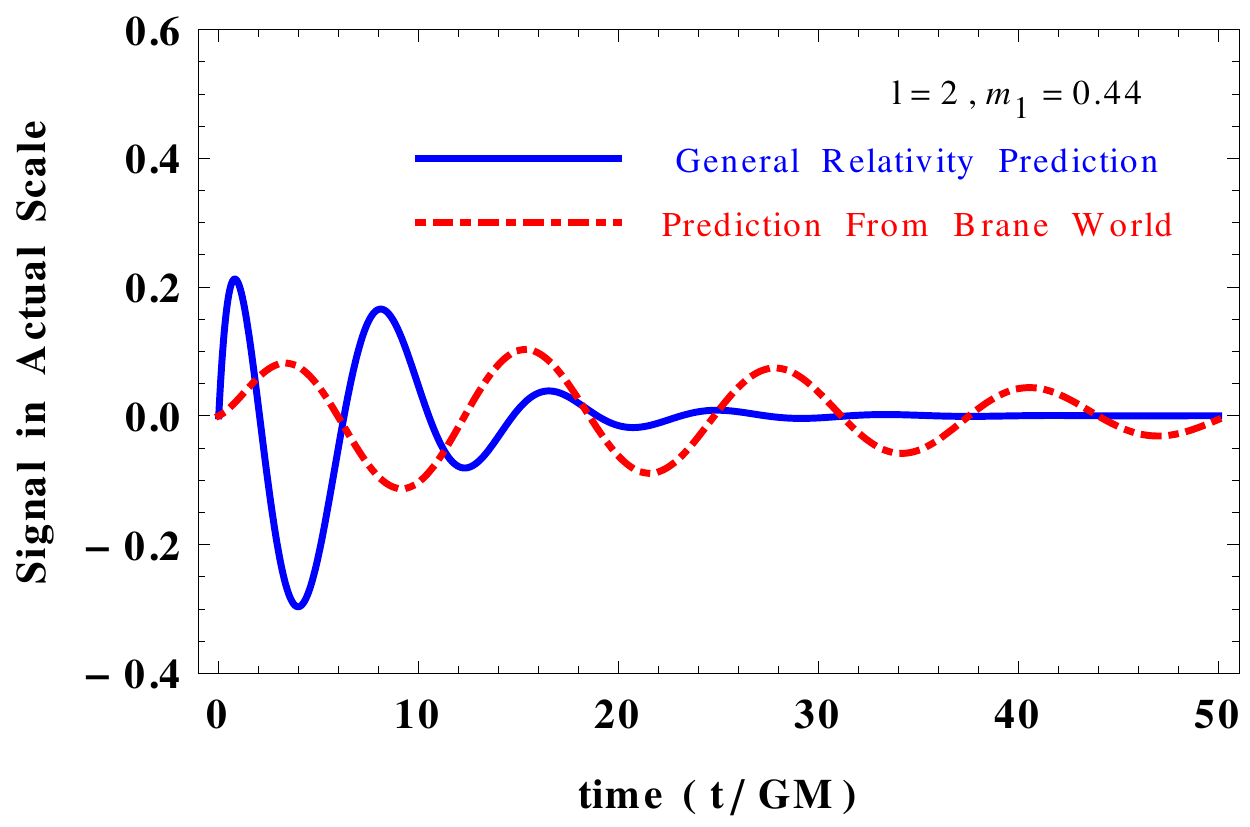}~~
\includegraphics[scale=0.65]{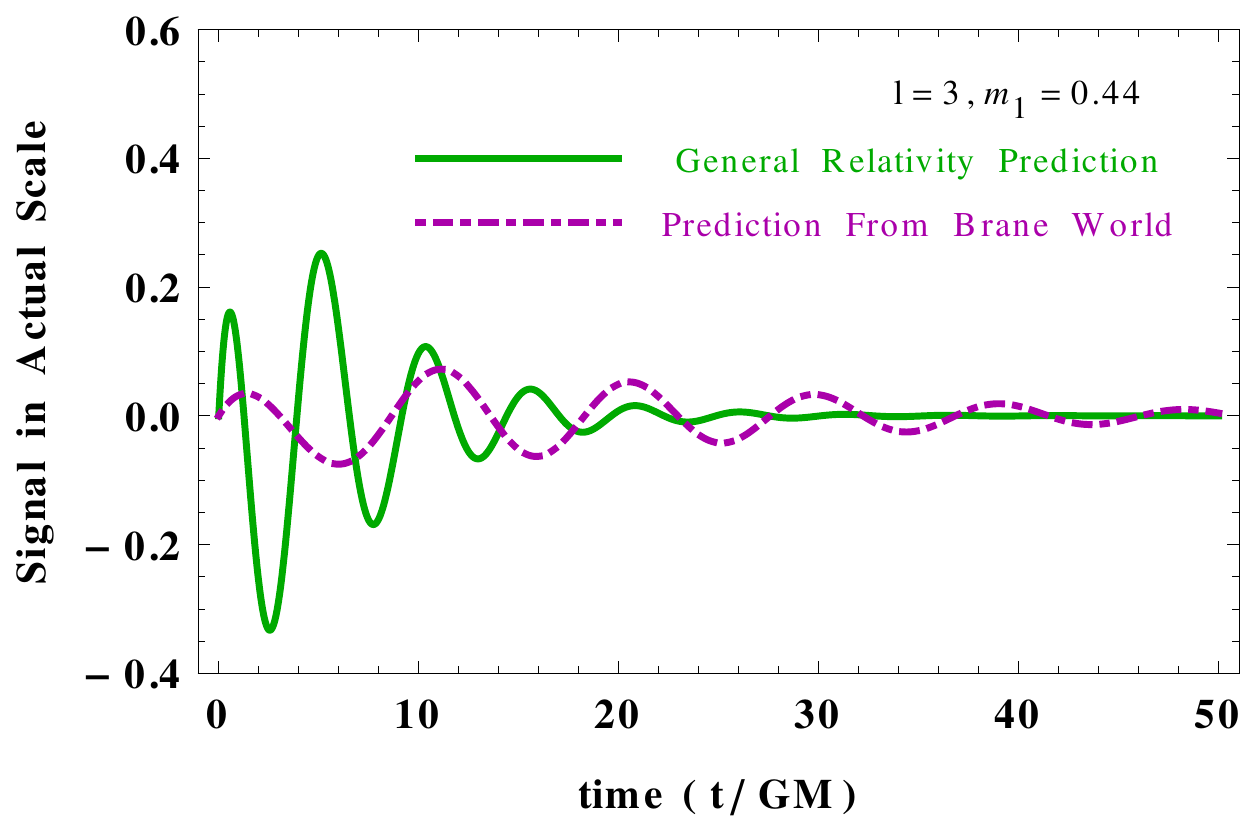}\\
\includegraphics[scale=0.65]{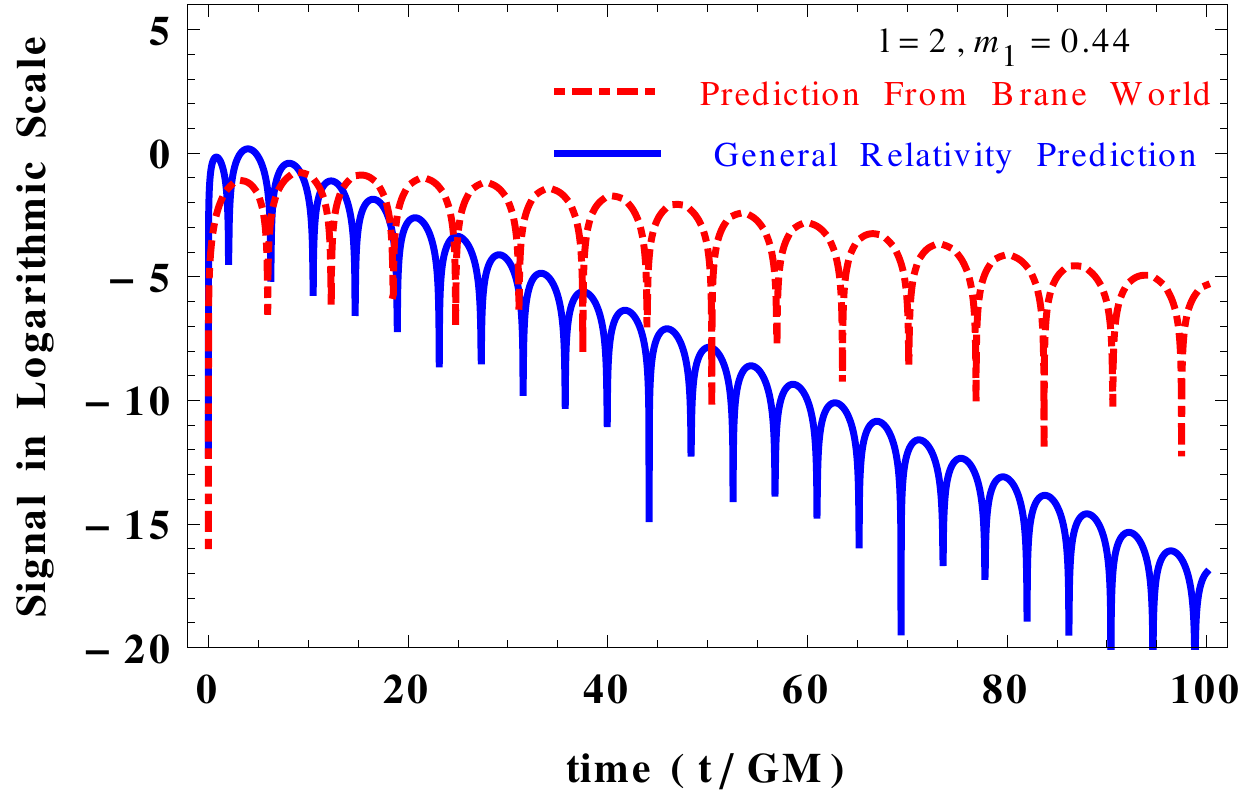}~~
\includegraphics[scale=0.65]{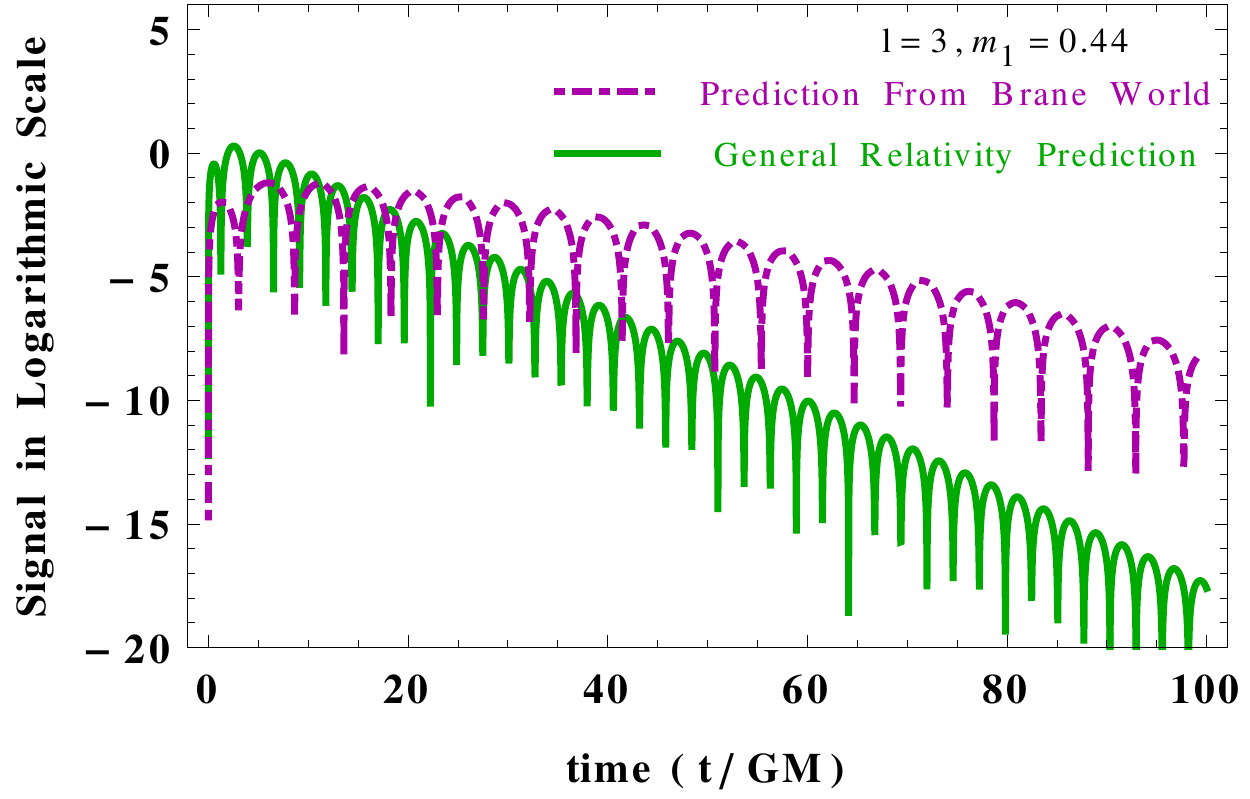}\\
\caption{Time evolution of the master mode function $u_{n,l}(t)$ for two different values of angular momentum both in the context of \gr\ ($n=0$) as well as in the \emph{brane based approach} have been presented. All the figures are associated with the lowest lying \KK mode mass $m_{1}=0.44$ (see \ref{Table_01}) but for two different choices of the angular momentum. For brevity we have presented both --- (a) the figures have been drawn in a Logarithmic scale (in the below panel) and (b) the figures in actual scale (in the top panel). All these figures clearly bring out the key differences between these two scenarios. See text for more discussions.}
\label{Fig_02_GR_vs_KK_AL_123}
\end{figure*}

As a first step towards the same we will present the time evolution of the axial perturbation in the context of \gr\ alone. This will set the stage for what to come next. This has been presented in \ref{Fig_01_GR}, where we have depicted how the mode functions evolve with time in the actual scale as well as in the Logarithmic scale. The advantage of the Logarithmic scale is that, it can enhance very tiny differences, while the disadvantage being, it will make large differences to appear as a small one. The left figure in \ref{Fig_01_GR} presents the actual time evolution of the $l=2$ and $l=3$ mode functions in \gr, i.e., $u_{0,2}(t)$ and $u_{0,3}$ respectively, while the right one presents the same in Logarithmic scale. It is clear that there is appreciable difference between the two at earlier times, which gets washed out as the modes gradually decay down. On the other hand the Logarithmic plot shows exactly the opposite nature as explained earlier.

\begin{figure*}[t!]
\centering
\includegraphics[scale=0.65]{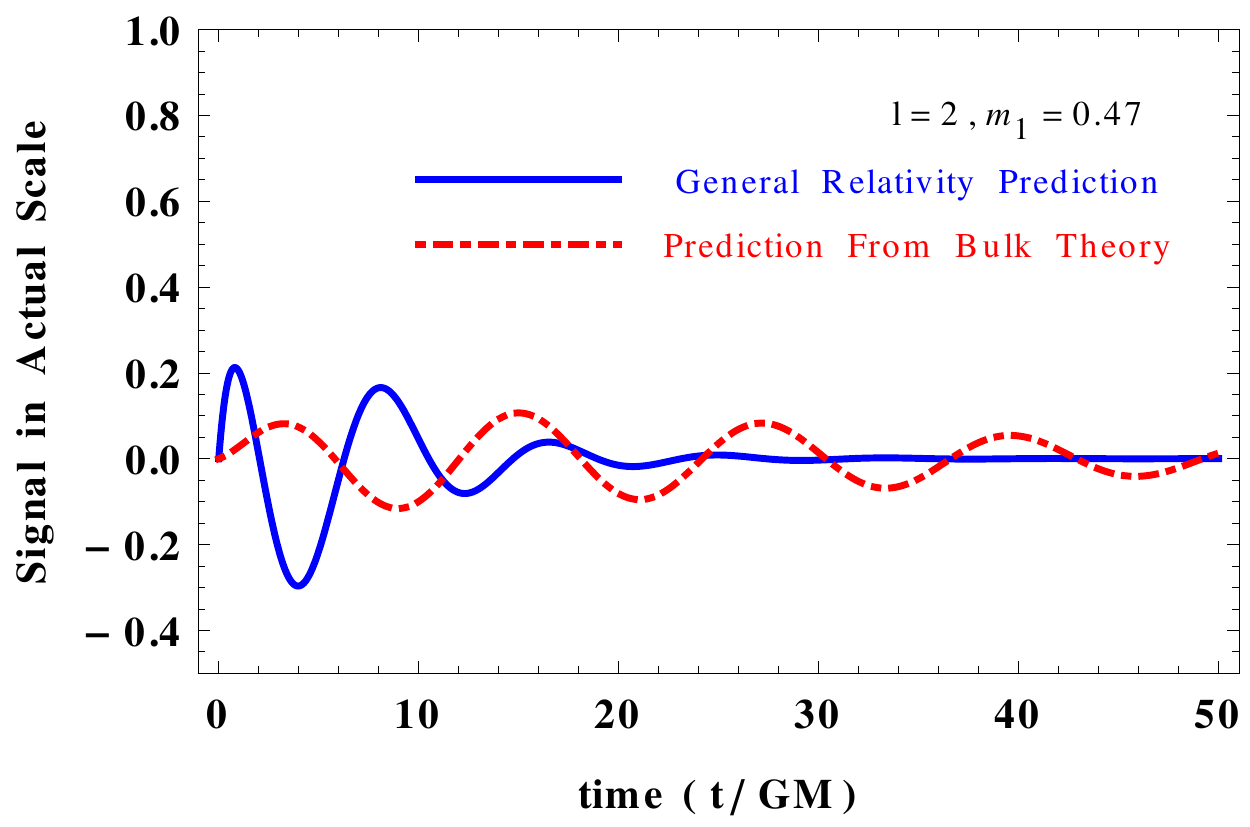}~~
\includegraphics[scale=0.65]{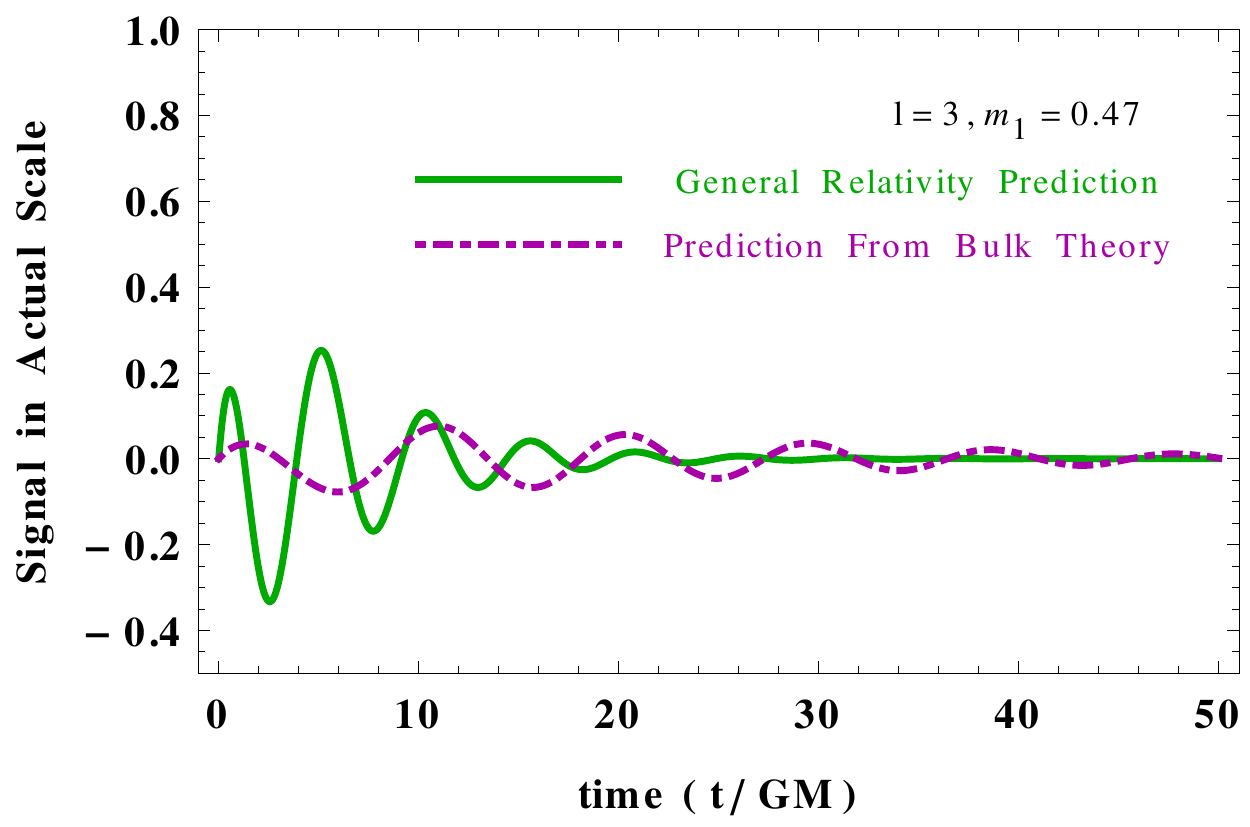}\\
\includegraphics[scale=0.65]{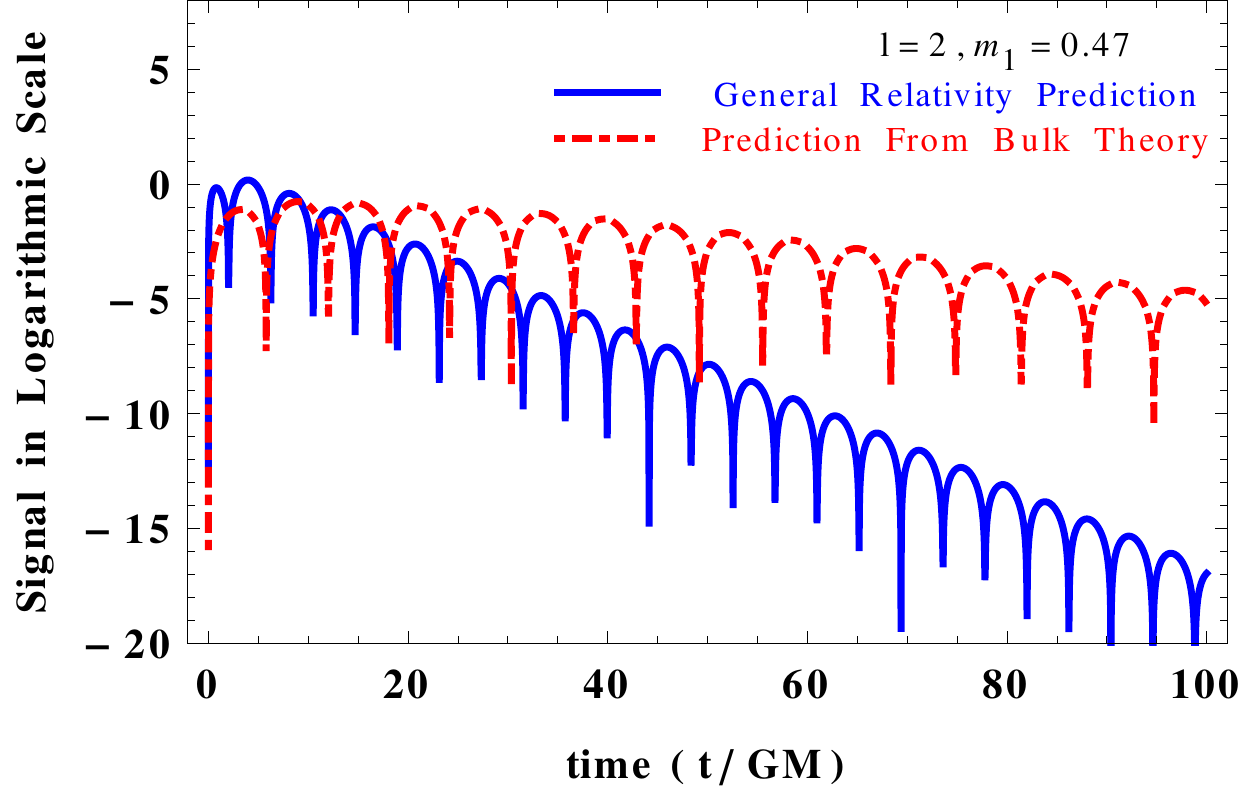}~~
\includegraphics[scale=0.65]{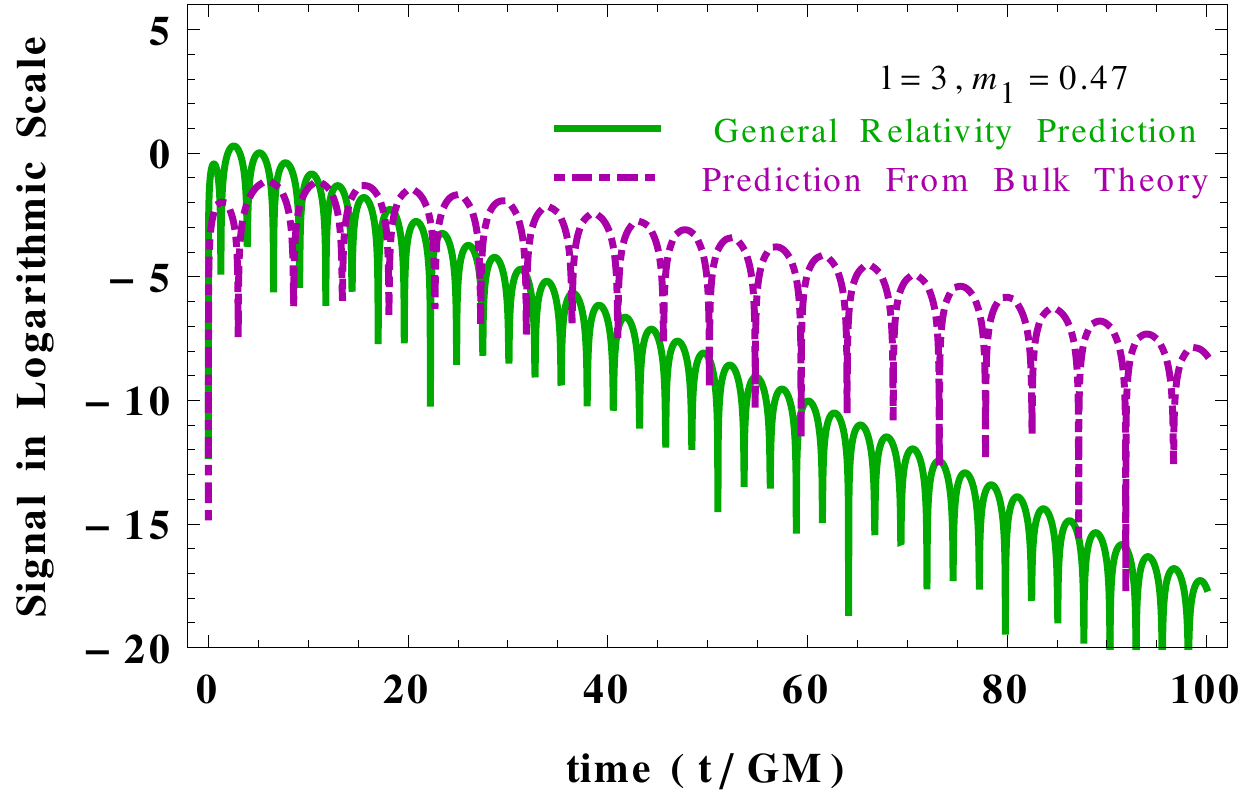}\\
\caption{The above figures depict the time evolution of the master mode function $u_{n,l}(t)$ in the \emph{bulk based approach} and has been contrasted with that in \gr\ (n=0). In this case also behaviour of the mode functions in a Logarithmic scale as well as in the actual scale have been presented. The \KK mode mass associated with the master variable presented here corresponds to the lowest one with $m_{1}=0.47$ with the following parameters: $d/\ell=20$; $1/\ell=6\times 10^{7}$.}
\label{Fig_04_GR_vs_KKB_AL_123}
\end{figure*}
\begin{figure*}[t!]
\centering
\includegraphics[scale=0.65]{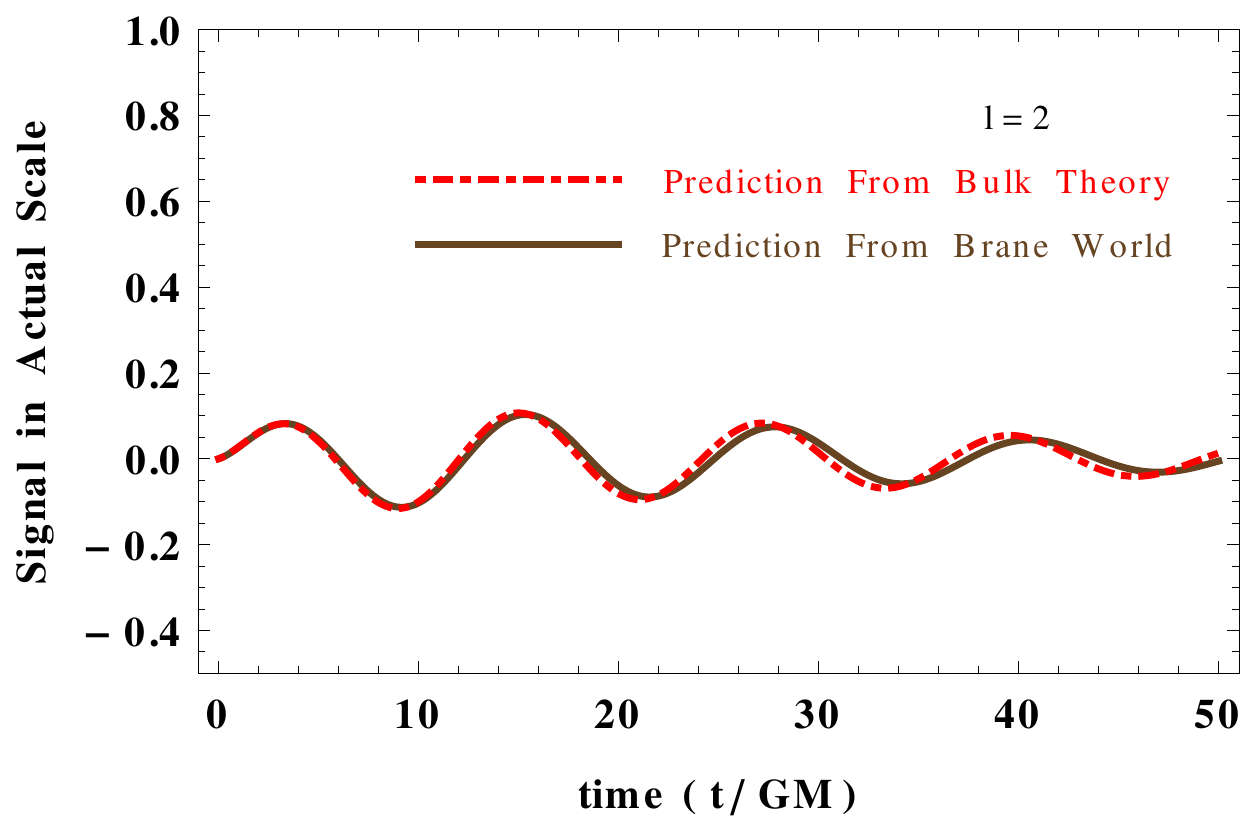}~~
\includegraphics[scale=0.65]{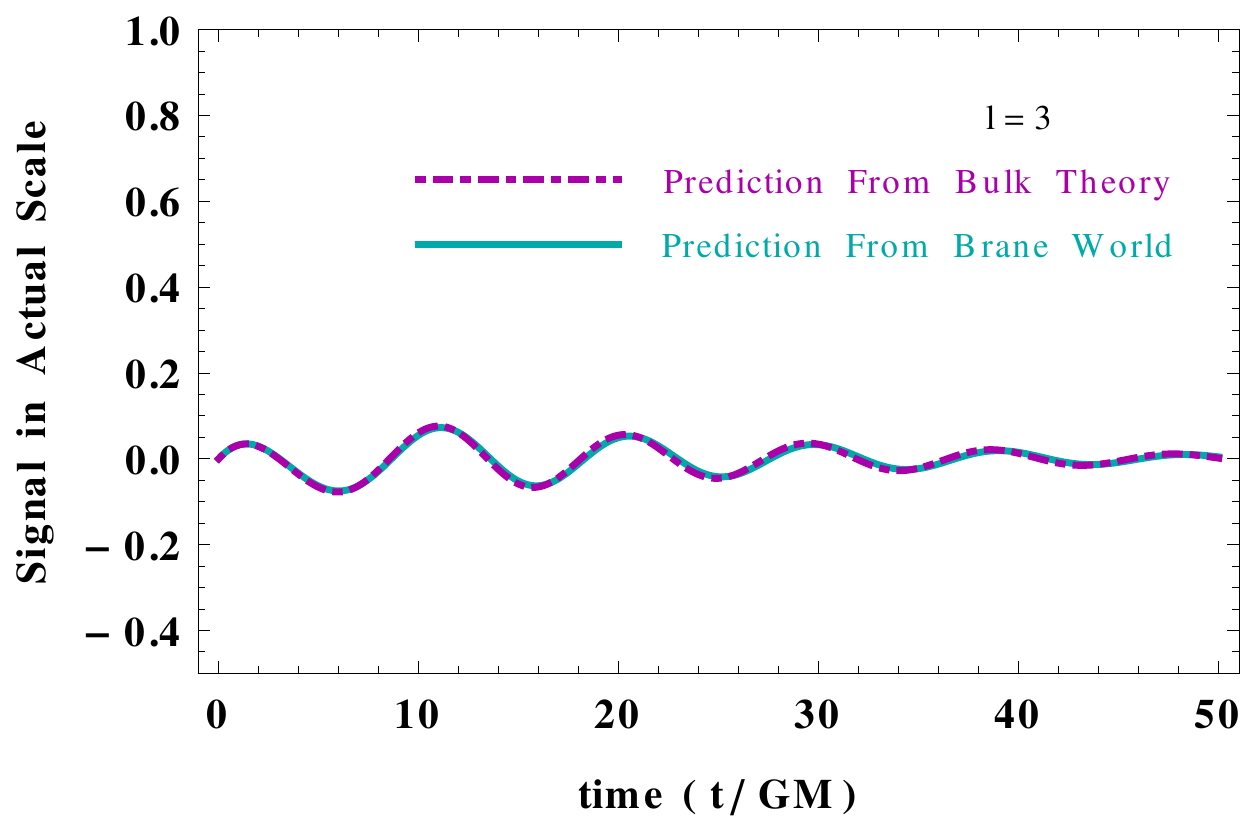}\\
\includegraphics[scale=0.65]{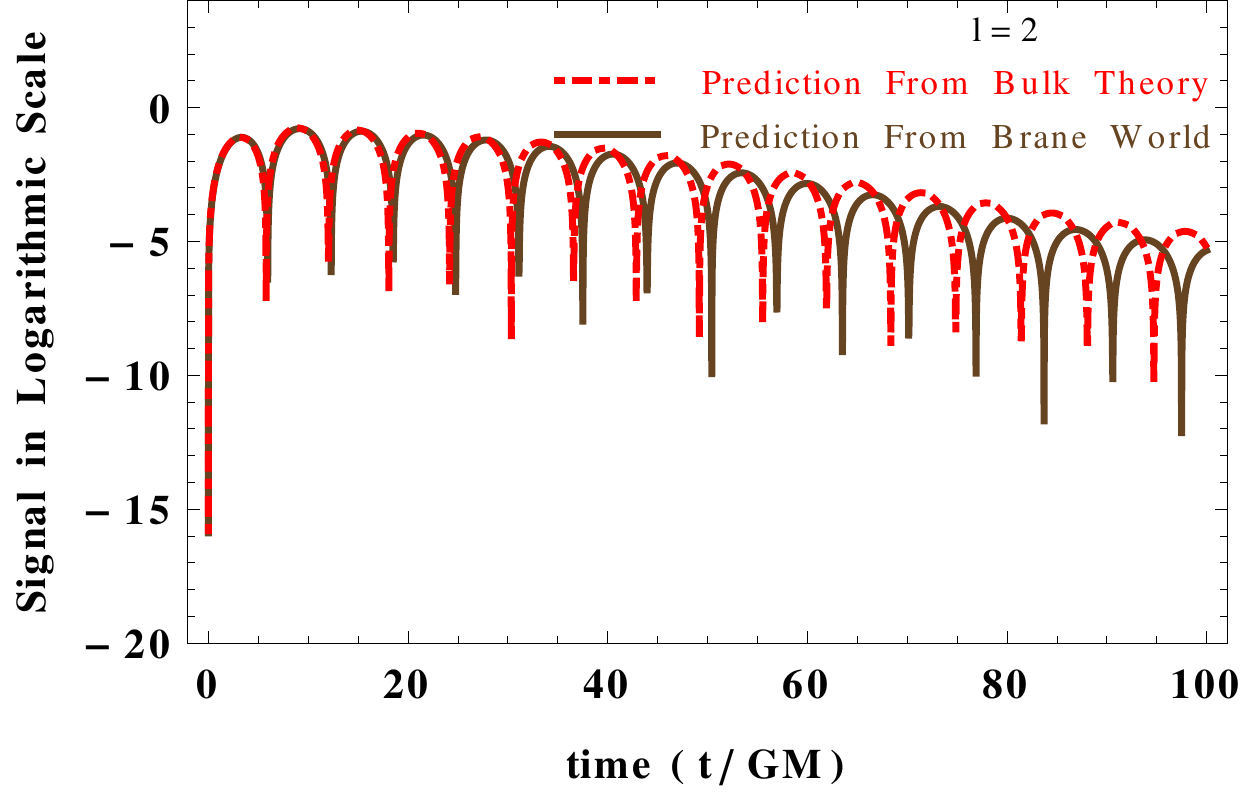}~~
\includegraphics[scale=0.65]{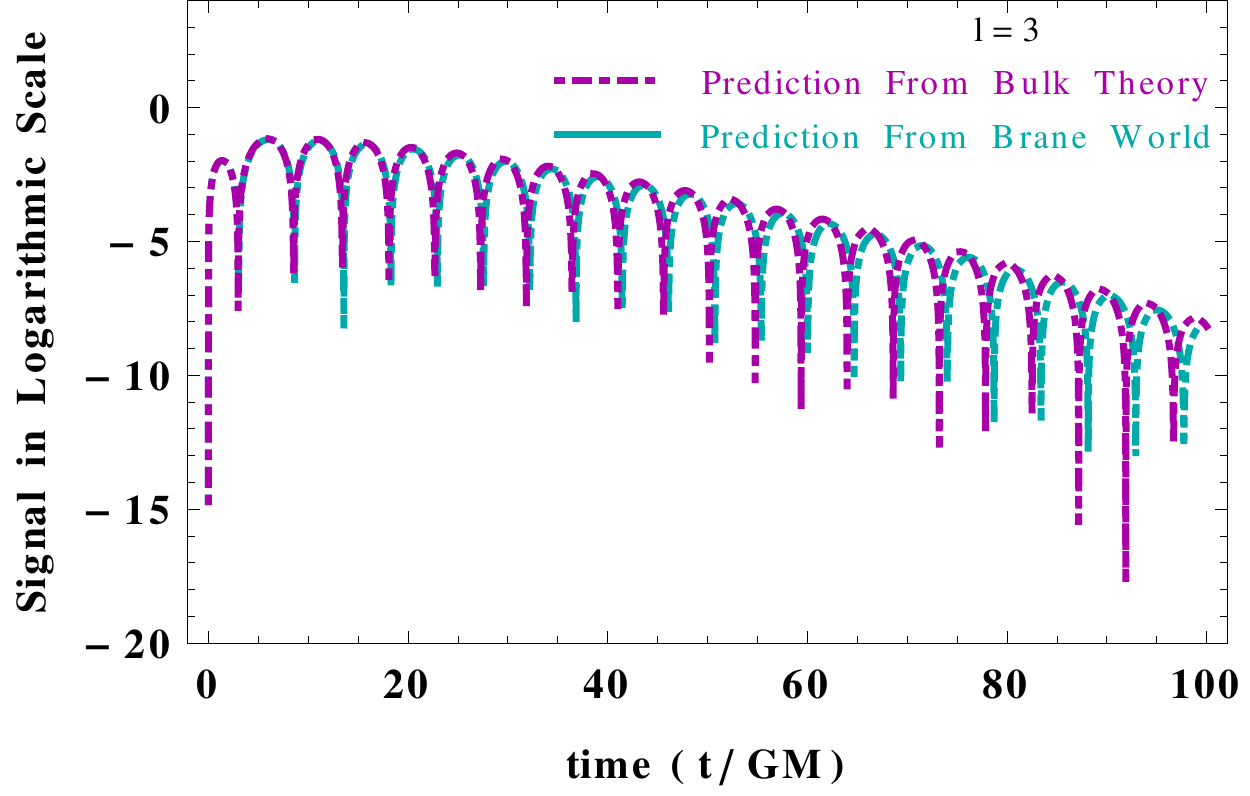}
\caption{The two scenarios presented in this work, namely the perturbation of effective four dimensional Einstein's equations or the perturbation of bulk Einstein's equations, have been illustrated in this figure. Both for the identical choice of the extra dimensional parameters, i.e., $d/\ell=20$; $1/\ell=6\times 10^{7}$. It is clear that the time evolution of the mode function $u_{n,l}(t)$ differ from each other in these two distinct approaches. This is primarily due to the difference between the \KK mode masses in these two approaches. See text for more discussions.}
\label{Fig_05_KK_vs_KKB_AL_123}
\end{figure*}

Returning back to our main goal, we have illustrated time variation of the perturbation associated with the lowest lying \KK mode having mass $m_{1}=0.44$ and have contrasted the same with \gr\ in \ref{Fig_02_GR_vs_KK_AL_123}. The figures on the left depict time variation of the perturbation for $l=2$ in both actual and Logarithmic scale, while those on the right are for $l=3$. The main difference emerging from \ref{Fig_02_GR_vs_KK_AL_123} is that the damping time scale of the massive modes are much greater compared to those in \gr. The same is true for the Logarithmic plots as well, where the fact that modes in \gr\ are heavily damped in comparison to the massive modes is much pronounced. The features of the massive modes remain identical as one considers the second lowest \KK mode having mass $m_{2}=0.83$ as well. Here also the slower decay of the massive modes with time is the key distinguishing feature between \gr\ and the higher dimensional model discussed here.  

So far we have been discussing the time evolution of the gravitational perturbation starting from the effective gravitational field equations on the brane. At this stage let us try to understand the corresponding situation when the gravitational perturbation originating from the bulk Einstein's equations is being considered. As emphasised earlier this will be similar to the brane based approach but will have an associated \KK mass mode, which will be different. For example, as evident from \ref{Table_01} and \ref{Table_02} for the same choice of bulk parameters, i.e., $d/\ell=20$; $1/\ell=6\times 10^{7}$, the \KK mass spectrum will be different in the two scenarios. Thus in \ref{Fig_04_GR_vs_KKB_AL_123} we have presented the time evolution of the gravitational perturbation derived from the bulk based approach. Here also we observe the same key features, e.g., very slow decay of the perturbation in contrast to that in \gr. Thus if the ring down phase of any black hole merger is being probed to intermediate 
times, where the evolution of gravitational perturbation is still dominated by the quasi-normal modes, any departure from the \gr\ prediction can possibly signal the existence of extra spacetime dimensions. Following the general trend, in \ref{Fig_04_GR_vs_KKB_AL_123} as well we have presented the time evolution in actual as well as in Logarithmic scale for two possible choices of angular momentum $l=2$ and $l=3$ respectively.  

This enables one to compare the bulk and the brane based approach given the same bulk parameters. The resulting discord should be attributed to the difference between the masses of the respective \KK modes. As evident from \ref{Fig_05_KK_vs_KKB_AL_123} this difference is really very small unlike the situation with \gr. Moreover since the masses of the \KK modes are higher in the bulk based approach they would decay slower. This can be clearly seen from both the Logarithmic plots in \ref{Fig_05_KK_vs_KKB_AL_123}, where the perturbation in the bulk based approach becomes larger than the brane based one at late times. Same features appear for both the angular momentums as well, however the difference is much smaller in higher angular momentum compared to the lower one.  

All these features can also be seen for the second lowest \KK mode mass, $m_{2}=0.87$, in the bulk based approach.The time evolution of the corresponding gravitational perturbation in both actual and Logarithmic scale for two choices of the angular momentum shows very similar features when compared with the lowest lying \KK mass mode. As expected, the massive modes decay much slowly in comparison with \gr. Further from the comparison of brane and bulk based approach for the second lowest lying \KK mode, one may infer that the difference only becomes sensible after a large time has elapsed and hence if the ring down phase can be probed minutely at very late times one may infer the preference of the bulk based approach over the brane based one or vice versa. However, the situation is not so simple and another subtle effect comes into play at late times, which corresponds to the wave tail. All the quasi-normal modes are inherently exponentially suppressed and hence at very late times their effects are 
negligibly small. In this situation the wave tails enter the picture and in most of the cases the late time behaviour is essentially governed by the wave tail decaying only as a power law (this is identical to the ``bulk based approach" as well, see \cite{Seahra:2009gw}). Since the scaling of the power law is mostly universal, independent of the nature of fields and mass of the fields under consideration, both the bulk and brane based approach will decay by an identical power law behaviour. This will make the detection of these two different approaches by late time measurements extremely difficult.
\section{Consistency of the approach: Comparison with Cauchy evolution}\label{consistency_Cauchy}

The previous section illustrates the methods to determine the \qnm frequencies, using which we have obtained the time evolution of the perturbation mode $u_{n,l}$. This is so because, in the limit of $m_{n}\rightarrow 0$ (i.e., \gr\ limit) this mode represents the axial gravitational perturbation, while the other essentially becomes a gauge degree of freedom. Hence we can compare the time evolution of $u_{n,l}$ with the respective one in \gr\ and see the harmony as well as possible discord among the two. We have already performed the same in the previous section. However in principle one expects the above approach to match with the Cauchy evolution of the perturbation equations presented in \ref{Eq_qnm_03a} and \ref{Eq_qnm_03b} respectively in \ref{App_B}. This is what we will explore in this section.  

For this purpose, we closely follow the analysis put forward in \cite{Konoplya:2011qq} but modifying it wherever necessary. Referring back to  \ref{GW_Eq23a} and \ref{GW_Eq23b} as the key differential equations for the master variables, one can write them in a compact manner as 
\begin{equation}\label{Eq_Cauchy_01}
\mathcal{D}\mathbf{\Psi}+\mathbf{V}(r)\mathbf{\Psi}=0~.
\end{equation}
Here, $\mathbf{\Psi}$ is a two dimensional column matrix constructed out of $u_{n,l}$ and $v_{n,l}$ respectively. Rather than working with the normal $(t,r)$ coordinates it is instructive to transform to the light cone coordinates. The transformation into the light-cone coordinates can be achieved by introduction of the null coordinates as: $u=t-r_{*}$, $v=t+r_{*}$. Use of these null coordinates modifies \ref{Eq_Cauchy_01} to
\begin{equation}
4\partial_u\partial_v\mathbf{\Psi}+\mathbf{V}(u,v)\mathbf{\Psi}=0~,
\end{equation}
where
\begin{equation}
\mathbf{\Psi}= 
\begin{pmatrix} u_{n,l} \\ v_{n,l}  \end{pmatrix}~, \qquad
\mathbf{V}=
\begin{pmatrix} V_{11} & V_{12}\\ V_{21} & V_{22} \end{pmatrix}~.
\end{equation}
Here all the matrix coefficients of $\mathbf{V}$ are dependent on the black hole solution on the brane and the \KK mode mass $m_{n}$. 

\begin{figure*}[t!]
\centering
\includegraphics[scale=0.45]{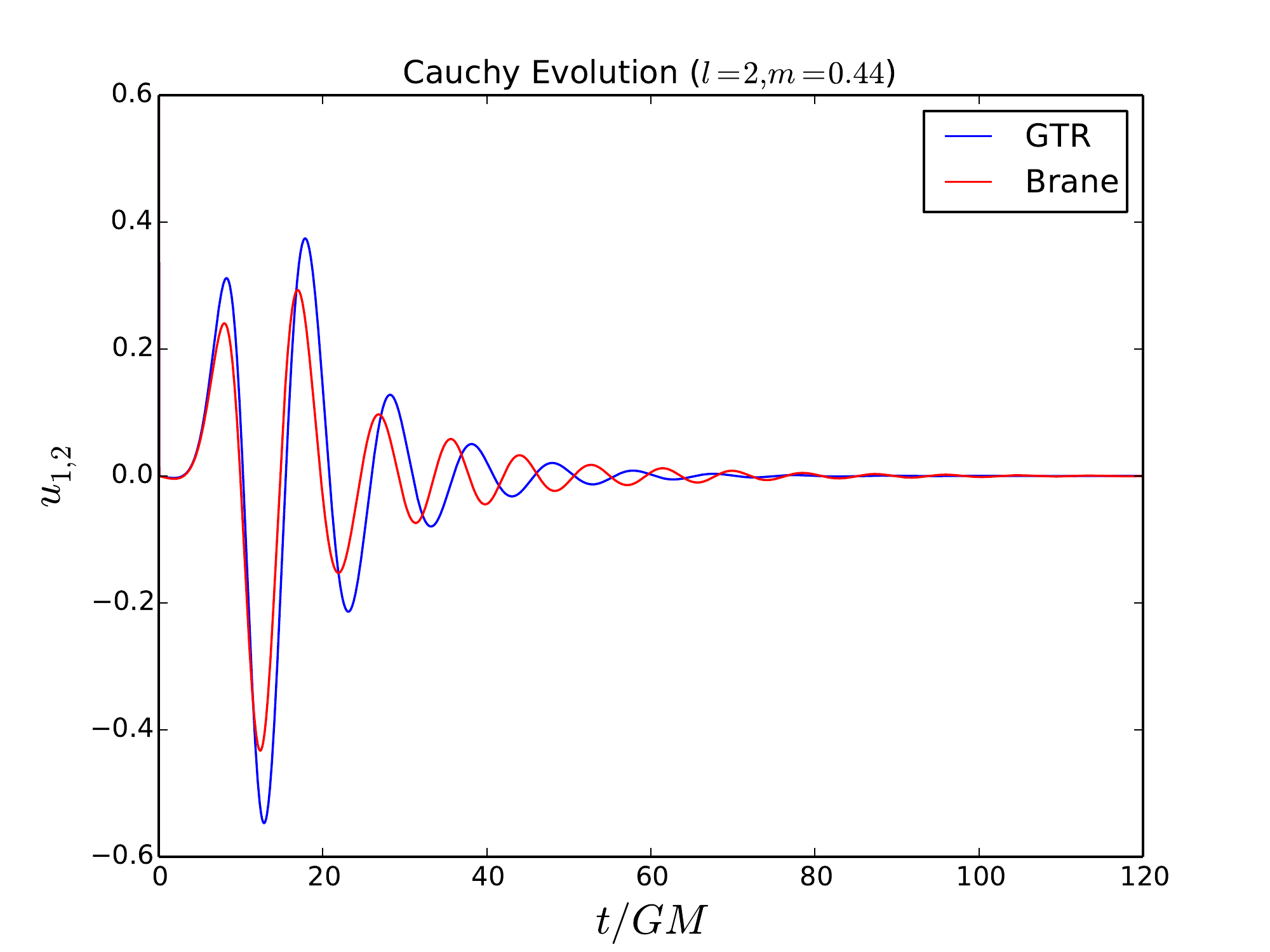}~~
\includegraphics[scale=0.45]{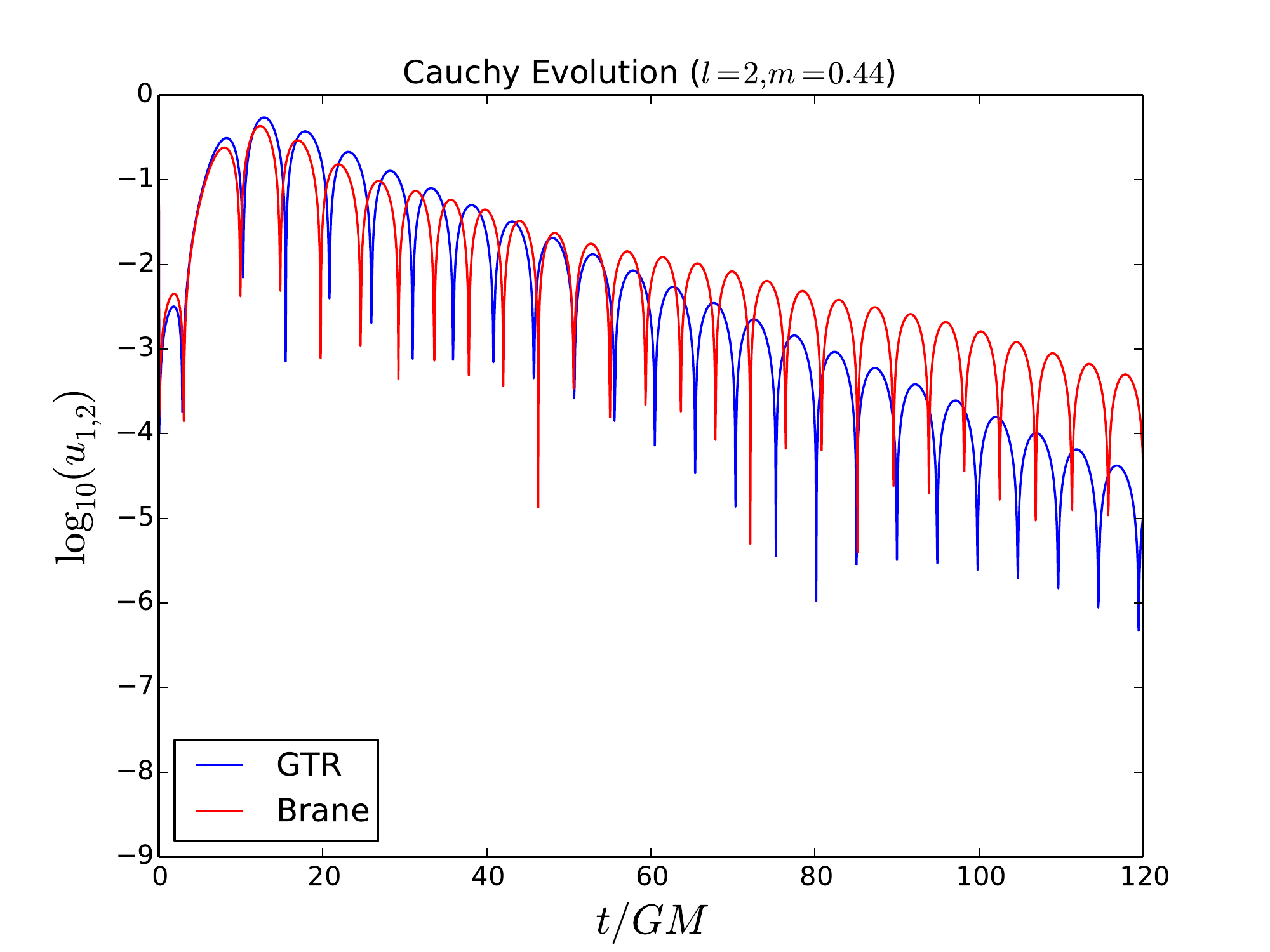}\\
\includegraphics[scale=0.45]{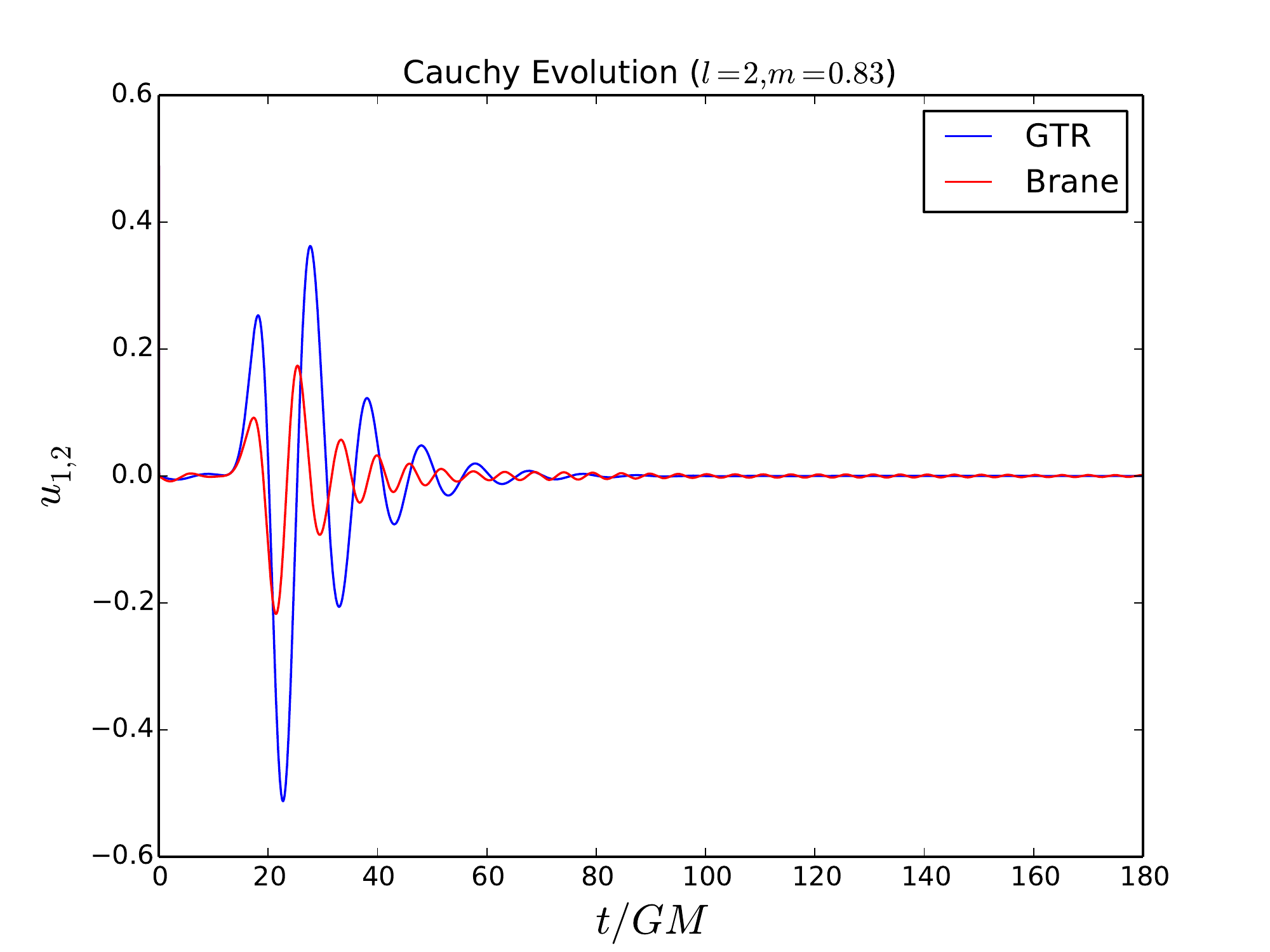}~~
\includegraphics[scale=0.45]{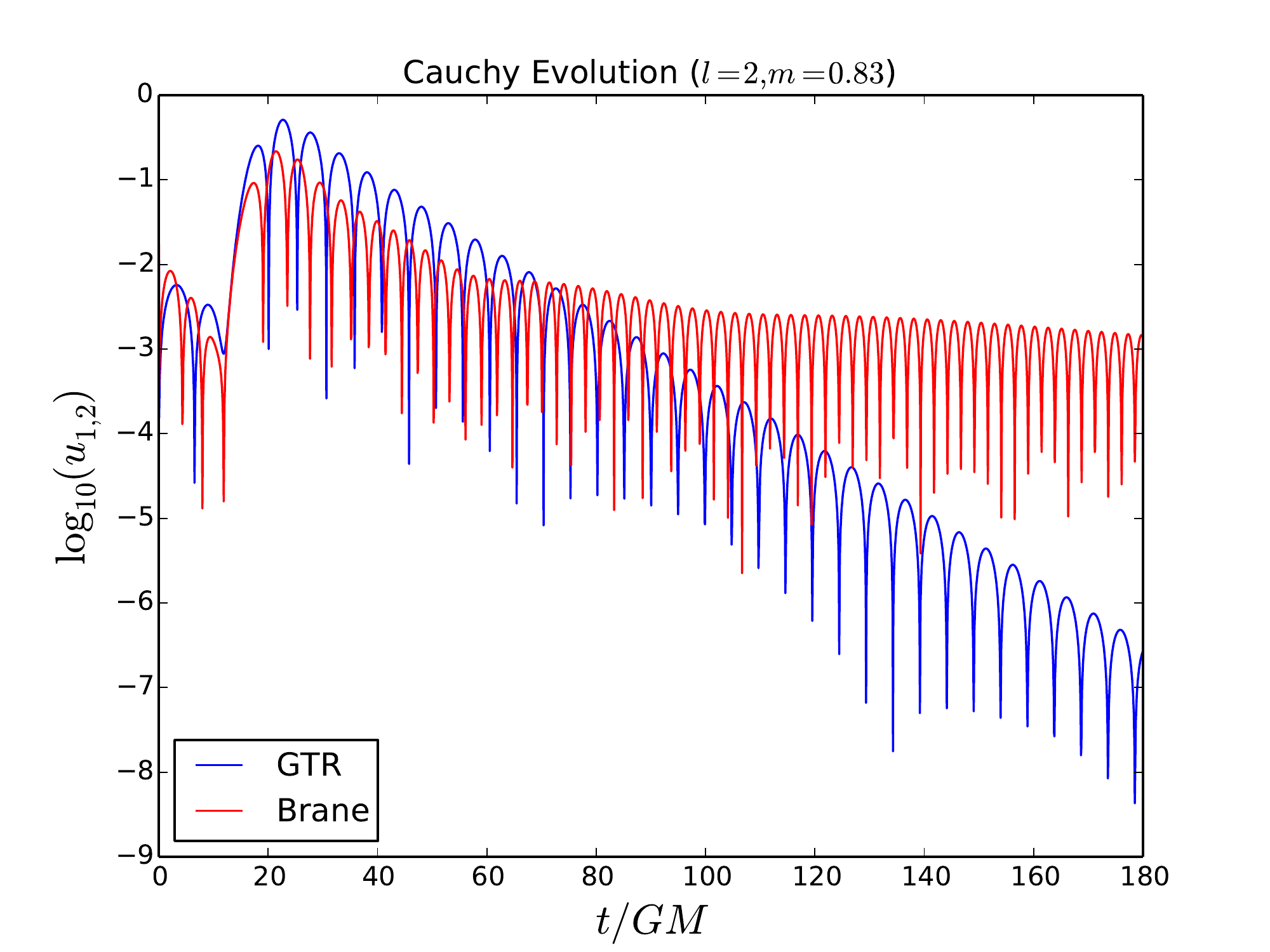}\\
\caption{The Cauchy evolution of the master variable $u_{n,l}(t)$ has been plotted for two different choices of the \KK mode masses for a given angular momentum. The figures in the top panel depicts the evolution of the master variable for $l=2$ and $m_{1}=0.44$, the lowest lying \KK mode with $d/\ell=20$ and $1/\ell=6\times 10^{7}$. The figure on the left is the actual variation of the master variable with time, while that on the right presents the same variation but in a Logarithmic scale. While the figures in the bottom panel illustrates the same, however for $l=2$ and \KK mode mass $m_{2}=0.83$. It is clear that as the mass increases the master variable becomes less and less damped in comparison with \gr. Further we clearly observe that the overall features present in the Cauchy evolution of the master variable are identical to those obtained by the \qnm analysis, illustrating the internal consistency of both the methods adapted here.}
\label{Fig_QNM_Cauchy}
\end{figure*}

For clarity, we have suppressed the functional dependence of the potential matrix $\mathbf{V}$ for the time being. To proceed further we need to introduce a notion of time evolution operator. For this purpose, we note that for an arbitrary function of time $f(t)$, the function given by $e^{h\partial_t}$ is the time evolution operator, in the sense that $e^{h\partial_t}f(t)=f(t+h)$. Thus in order to obtain the time evolution of the mode functions $\mathbf{\Psi}$, we apply the time evolution operator on $\Psi$. This yields,
\begin{equation}\label{Eq_Cauchy_02}
\mathbf{\Psi}(t+h)=e^{h\partial_t}\mathbf{\Psi}=e^{h\partial_u+h\partial_v}\mathbf{\Psi}~.
\end{equation}
\ref{Eq_Cauchy_02} can be written in a nice manner by expanding the right hand side, resulting into,
\begin{widetext}
\begin{align}\label{Eq_Cauchy_03}
&\mathbf{\Psi}(u+h,v+h)=\sum_{j=0}\frac{1}{j!}(h\partial_u)^{j}\sum_{k=0}\frac{1}{k!}(h\partial_v)^{k}\Psi(u,v)
\nonumber
\\
&=\left[e^{h\partial_u} + e^{h\partial_v} -1 +\frac{1}{2} h^2\partial_u\partial_v\left(1+\frac{h\partial_u}{2!} +\frac{{h\partial_u}^2}{3!}+\cdots\right) \left(1+\frac{h\partial_v}{2!} +\frac{{h\partial_v}^2}{3!}+\cdots \right)\right]\mathbf{\Psi}(u,v) 
\nonumber
\\
&=\Bigg[e^{h\partial_u} + e^{h\partial_v}-1+\frac{1}{2}h^2\partial_u\partial_v
\Bigg\{\left(e^{h\partial_u}-{(h\partial_u)}^2\left(\frac{1}{2!}-\frac{2}{3!}\right)-\cdots\right)
+\partial _{u}\rightarrow \partial _{v}\Bigg\}\Bigg]\mathbf{\Psi}(u,v) 
\nonumber
\\
&=\left[e^{h\partial_u} + e^{h\partial_v}-1+\frac{1}{2}h^2\partial_u\partial_v \left\{\left(e^{h\partial_u}+\mathcal{O}(h^2)\right) +\left(e^{h\partial_v}+\mathcal{O}(h^2)\right)\right\}\right]\mathbf{\Psi}(u,v)~.
\end{align}
\end{widetext}
The last expression of \ref{Eq_Cauchy_03} can be expanded immediately and hence finally we have,
\begin{align}
\mathbf{\Psi}(u+h,v+h)=&\mathbf{\Psi}(u+h,v)+\mathbf{\Psi}(u,v+h)-\mathbf{\Psi}(u,v) 
\nonumber
\\
-\frac{h^2}{8}&\Big\{\mathbf{V}(u+h,v)\mathbf{\Psi}(u+h,v)
\nonumber
\\
&+\mathbf{V}(u,v+h)\mathbf{\Psi}(u,v+h)\Big\}~.
\end{align}
This can be thought of as an evolution equation in the light-cone coordinates $u$ and $v$. The interesting aspect of this formalism is that once initial data is specified in the $u,v$ coordinates, we need no additional boundary conditions, which is unlike the physical coordinates $t,r$ (or, for that matter $t,r_{*}$). We evolve the system with Gaussian initial data in $u$ and constant data in $v$. \ref{Fig_QNM_Cauchy} illustrates the numerical evolution of $\Psi$ as a function of time for different choices of the angular momentums and \KK mode masses obtained by numerically integrating the above evolution equation in light cone coordinates. Interestingly and as expected, it illustrates all the basic properties that we have already observed from a \qnm analysis. For example, in all the cases illustrated in \ref{Fig_QNM_Cauchy} it is clear that at intermediate times (i.e., when the spectrum of quasi-normal modes dominate the evolution of gravitational perturbation) the mode functions due to massive \KK modes 
will dominate over those in \gr. This is again due to the fact that the massive modes suffer much less damping compared to the respective ones in \gr. The translation of the same in the \qnm language corresponds to the imaginary part of the \qnm frequency to be smaller for massive \KK modes compared to the modes in \gr. Further the fact that as the mass of the \KK mode increases it experiences less and less damping is also borne out by both Cauchy evolution (see \ref{Fig_QNM_Cauchy}) and the \qnm analysis. Of course, there are minute differences present between both these methods, which have their origin in the initial conditions and the fact that the Cauchy evolution is more accurate compared to the \qnm analysis. All in all, the \qnm analysis and the Cauchy evolution of initial data provides a complete and consistent description of the time evolution of gravitational perturbation in presence of extra spatial dimensions.

\begin{figure*}[t!]
\centering
\includegraphics[scale=0.45]{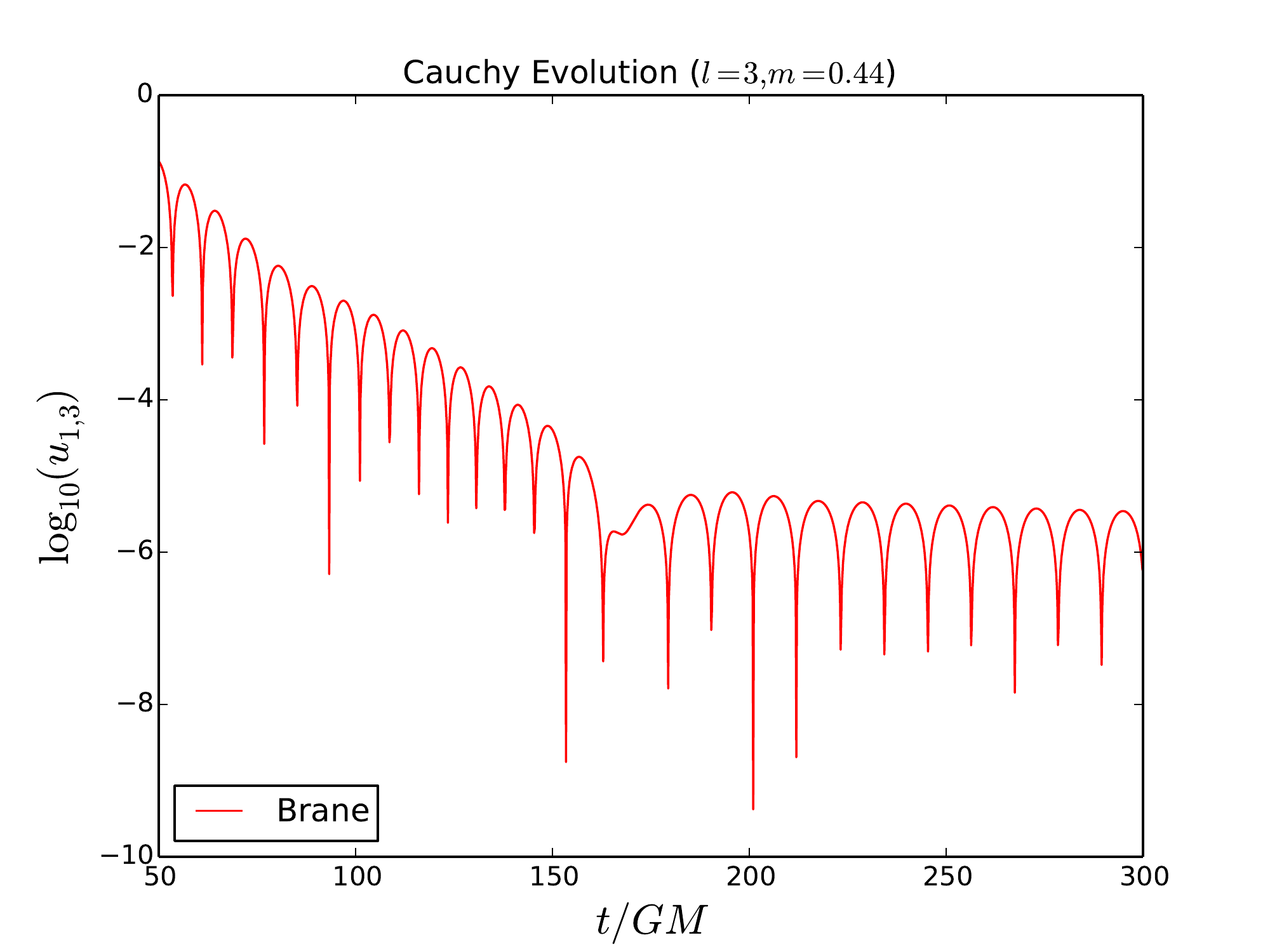}~~
\includegraphics[scale=0.45]{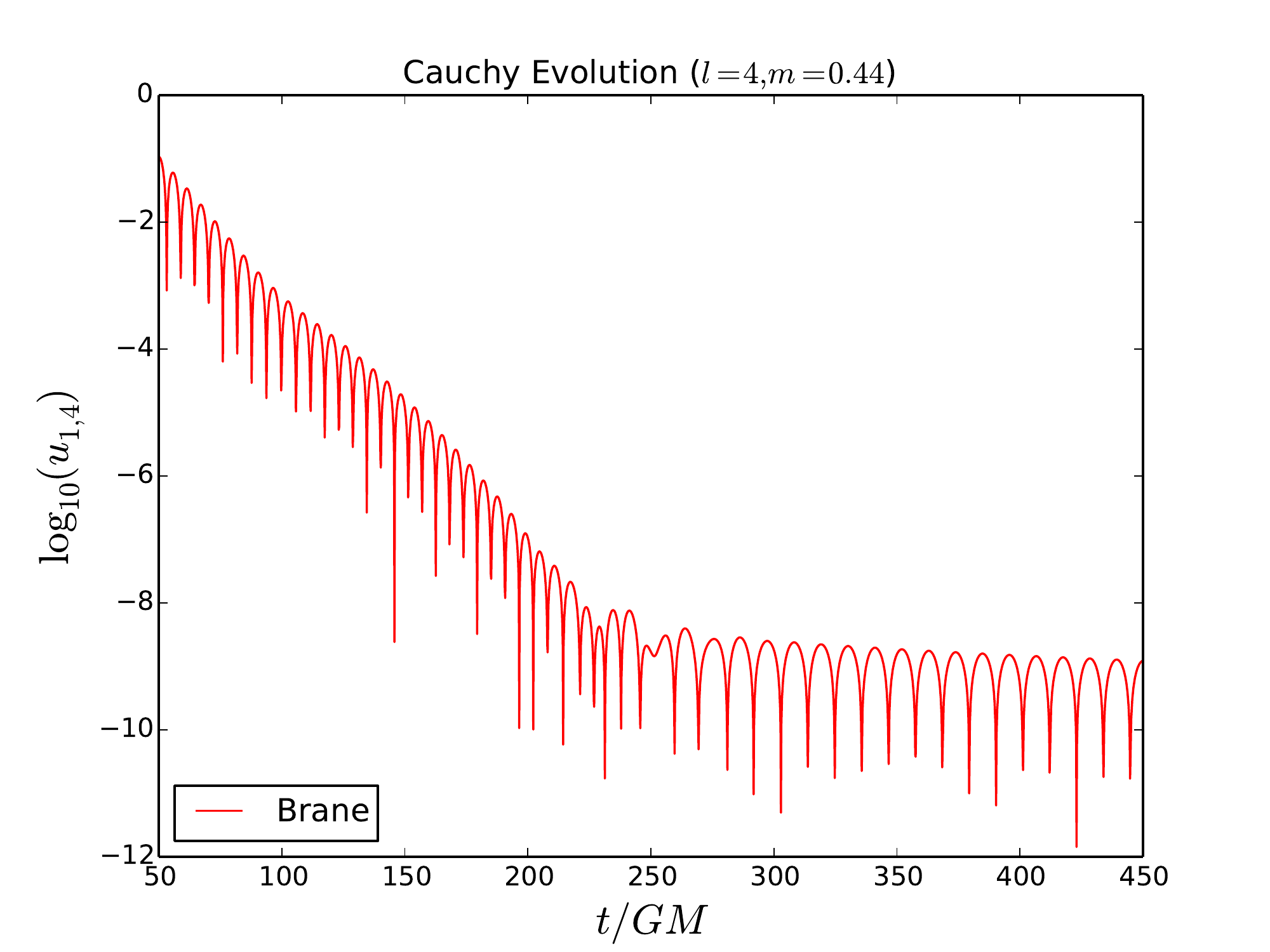}\\
\caption{The Cauchy evolution of the master variable $u_{n,l}(t)$ associated with the brane based approach has been plotted for two different choices of the angular momentum given a \KK mode mass to illustrate the late time behaviour. The figure in the left depicts the evolution of the master variable for $l=3$ and $m_{1}=0.44$, the lowest lying \KK mode with $d/\ell=20$ and $1/\ell=6\times 10^{7}$. The figure on the right is for $l=4$ and identical \KK mode mass. It is clear that as time progresses the power law tail dominates over the exponential damping due to quasi-normal modes. This once again illustrates the consistency of Cauchy evolution with the theoretical methods adapted here.}
\label{Cauchy_Tail}
\end{figure*}

Besides being consistent with the \qnm analysis, Cauchy evolution has more additional features to offer. The most important such feature is the presence of late time power law tail. Since at the intermediate stages, the contributions from \qnm dominates, the behaviour of the mode function $u_{n,l}(t)$ as presented in \ref{Fig_QNM_Cauchy} resembles those in \ref{Num_qnm}. However, if one can perform the Cauchy analysis for a sufficiently long time, gradually the contributions from quasi-normal modes become smaller compared to the late time tail. Thus the Cauchy evolution of the initial data for a longer time must result in the desired power law tail and will serve as another consistency check of our approach. Following this we have presented a long time Cauchy evolution of the perturbation equation in \ref{Cauchy_Tail}, which distinctly depicts the late time wave tail. As evident from \ref{Cauchy_Tail}, the mode function is initially dominated by the quasi-normal modes and hence decays 
linearly in the Logarithmic scale. However in the late stages of Cauchy evolution the power law takes over and dominates the quasi-normal modes, thus presenting an almost constant-in-time behaviour of the same. Hence, the numerical analysis of the Cauchy evolution of the perturbation equation is completely consistent with theoretical expectation, providing one more consistency check of our formalism.

\section{Discussion and concluding remarks}\label{conclusion}

\begin{table*}
\begin{center}
\caption{Frequencies (in Hz) of oscillation for the quasi-normal modes emanating from black holes having different masses have been depicted. Numerical estimates for the frequencies have been presented for \gr\ as well as for the two lowest lying \KK modes with masses $m_{1}=0.43$ and $m_{2}=0.83$ respectively. It is also clear that the frequency of the modes increases with an increase in the $l$ value. It is clear that as the mass of the black hole increases the frequency decreases. Thus more massive the black hole is it is more problematic to detect in aLIGO. While the \KK modes have better chance of originating if the mass of the black hole increases, this leads to lowering of the frequency and hence have less chance of getting detected in aLIGO.}
\label{Table_07}       
%
%
\begin{tabular}{p{2cm}p{3cm}p{3.5cm}p{3.5cm}}
\hline\noalign{\smallskip}
\hline\noalign{\smallskip}
                    &                      & Frequencies for $l=2$   &                          \\
\hline\noalign{\smallskip}
\hline\noalign{\smallskip}
$(M/M_{\odot})$     &   General Relativity & KK Mode ($m_{1}=0.43$)     & KK Mode ($m_{2}=0.82$)   \\
\hline \noalign{\smallskip}
\hline \noalign{\smallskip}
$1$                 & 24140                &  14930                     & 12441                    \\
$10$                & 2414                 &  1493                      & 1244.1                   \\
$10^{2}$            & 241.4                &  149.3                     & 124.4                    \\
$10^{3}$            & 24.1                 &  14.9                      & 12.4                     \\
$10^{4}$            & 2.4                  &  1.5                       & 1.2                      \\
$10^{5}$            & 0.2                  &  0.1                       & 0.1                      \\
\noalign{\smallskip}
\hline\noalign{\smallskip}
\hline \noalign{\smallskip}
                    &                      & Frequencies for $l=3$   &                          \\
\hline\noalign{\smallskip}
\hline \noalign{\smallskip}
$1$                 & 38779                &  21328                     & 27856                    \\
$10$                & 3877.9               &  2132.8                    & 2785.6                   \\
$10^2$              & 387.7                &  213.3                     & 278.6                    \\
$10^3$              & 38.7                 &  21.3                      & 27.8                     \\
$10^4$              & 3.9                  &  2.1                       & 2.8                      \\
$10^5$              & 0.3                  &  0.2                       & 0.3                      \\
\noalign{\smallskip} 
\hline\noalign{\smallskip}
\hline \noalign{\smallskip}
\end{tabular}
\end{center}
\end{table*}

In this work we set out to achieve three goals in a single framework --- (a) Effect of extra spatial dimensions on the gravitational perturbation and whether one can provide some possible observational signatures of the same in the ring down phase of black hole merger; (b) How the two possible methods to determine the gravitational perturbation on the brane, namely by either perturbing the bulk gravitational field equations or perturbing the \emph{effective} gravitational field equations on the brane, differ as far as the behaviour of the gravitational wave solution is concerned; (c) Whether the analysis using quasi-normal modes is consistent with the fully numerical Cauchy evolution of the initial data. We believe to have addressed all of them in a satisfactory manner in this work which we summarise below.

We have explicitly demonstrated that the existence of extra spatial dimensions indeed modifies the gravitational perturbation equation by essentially introducing a tower of massive perturbation modes in addition to the standard massless one. Thus the presence of massive gravitational perturbation modes is a definitive signature of the existence of higher dimensions. To see the consequences of the above we have discussed the behaviour of quasi-normal modes in this context. In particular, we have shown that for the massive modes the imaginary part of the \qnm frequencies are much small compared to those in \gr. This has resulted in the time evolution of the massive gravitational perturbations to exhibit a weak decay rate in comparison to the massless modes as in \gr. The above phenomenon opens up the observational window to probe the possible existence of higher dimensions using gravitational wave observation. If the ring down phase during the merger of two black holes is loud enough to be accurately measured for a sufficient amount of time (unlike the aLIGO-VIRGO observations to-date) it may be possible to detect any departure from the \gr\ prediction and thus may lead to concrete observational signature for the existence of higher dimensions or may provide stringent constraints on the associated parameters. This will become feasible as the sensitivity of the aLIGO detectors are further improved or the space based gravitational wave detector LISA becomes operational. We will address the detailed observational aspects of this particular signature of extra dimension in light of the recent detection of gravitational waves at aLIGO in a future work.

The evolution equation for the gravitational perturbation obtained by perturbing the bulk field equations has already been derived in \cite{Seahra:2004fg}, while in this work we have derived the evolution equation by perturbing the \emph{effective} gravitational field equations on the brane hypersurface. From the structure of the equation itself, difference between these two approaches should be evident. In both the bulk based and the brane based approach the four dimensional perturbation equation looks identical with one crucial difference, namely, the masses associated with both the approaches are different. This is because the differential equation satisfied by the extra dimensional part are different in these two scenarios. This in turn leads to difference in the \qnm frequencies as evident from \ref{Fig_QNM}, the imaginary parts of the \qnm frequencies are smaller than the bulk based approach in comparison to the brane based one. Since the difference is small there is possibly no way in foreseeable future to observationally distinguish these two effects (see e.g., \ref{Fig_05_KK_vs_KKB_AL_123}) however theoretically there does exist a difference between these approaches. Naively speaking, this is due to the fact that a solution of the \emph{effective} gravitational field equation on the brane may not have any higher dimensional embedding.  

\begin{figure*}[t!]
\centering
\includegraphics[scale=0.80]{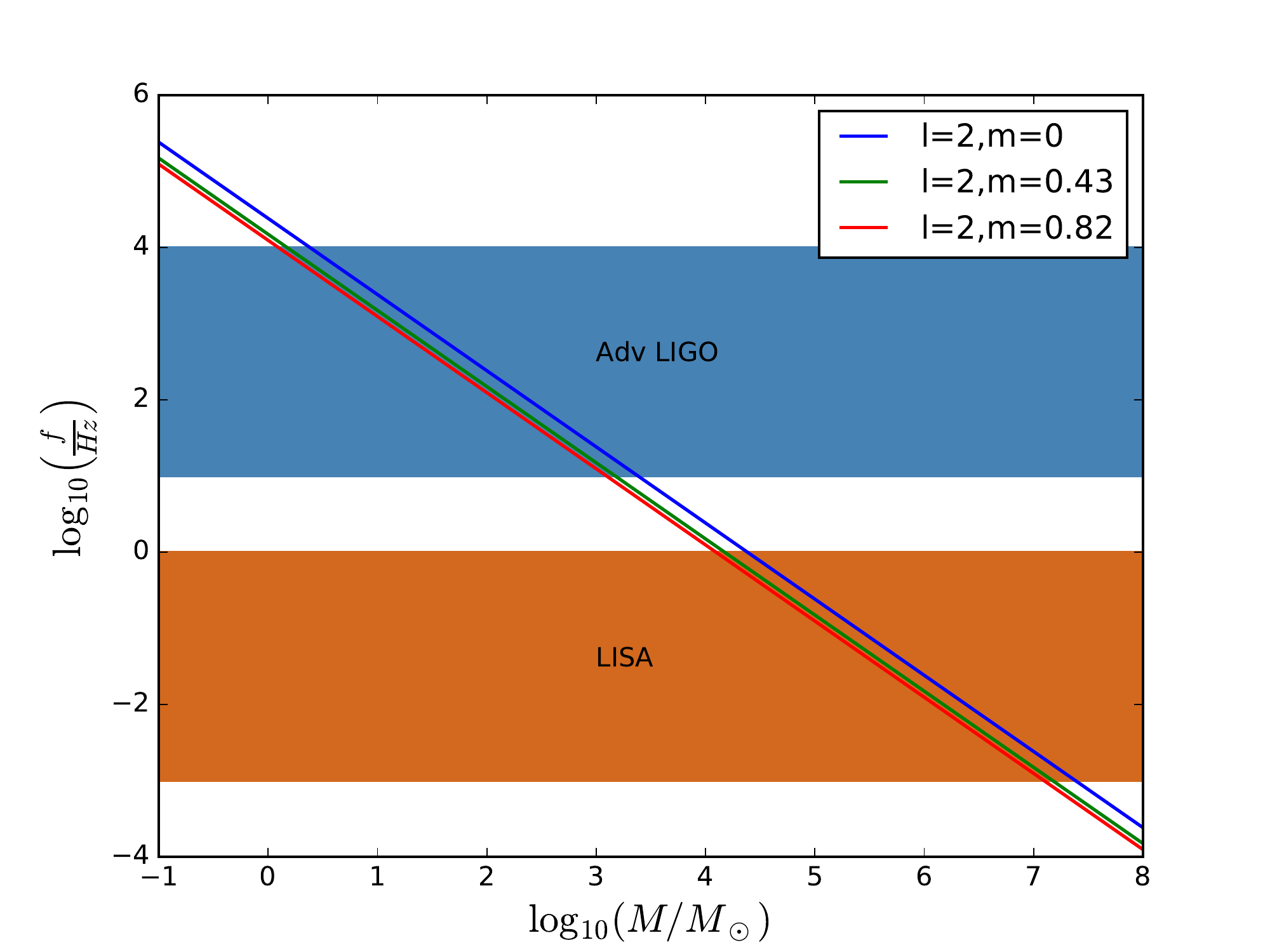}
\caption{This figure depicts how the oscillation frequency of the gravitational wave in the ring down phase changes as the mass of the black hole increases. For conveniences frequencies are plotted in Hz, while black hole mass is presented in solar units, but both on a Logarithmic scale. The oscillation frequencies have been plotted in the brane based scenarios for two lowest lying \KK mode masses with $l=2$. The same has been contrasted with the corresponding curve in \gr.  It is clear that for $M\sim 10^{3}M_{\odot}$ the frequencies associated with \gr\ are well within the aLIGO frequency band, however for the massive \KK modes they are outside. Since these massive modes have better chance of getting detected in the high mass regime it is most likely that they may become observable once LISA is operational. See text for more discussions.}
\label{Fig_Frequency}
\end{figure*}

Finally the time evolution of the gravitational perturbation can be obtained by either performing a \qnm analysis or by performing a fully numerical Cauchy evolution. We have performed both in this work and they are found to match very well with each other. This is expected as well as necessary for internal consistency of any gravitational perturbation computation. In particular from the \qnm analysis we have learned that the massive modes decay much slowly in contrast with the massless \gr\ modes (see e.g., \ref{Fig_02_GR_vs_KK_AL_123}), which is also confirmed by the Cauchy evolution (see e.g., \ref{Fig_QNM_Cauchy}). Thus keeping aside minute details overall behaviour of the time evolution of gravitational perturbation is identical whether one performs a \qnm analysis or complete Cauchy evolution. 

Having described the consistency of the time evolution obtained by using quasi-normal modes as well as a fully numerical Cauchy evolution of the perturbation equations, let us comment on possible detectability of the scenario presented above. For that purpose it is important to know the frequencies associated with the gravitational perturbation modes. The corresponding frequencies can broadly be divided into two classes, those originating from the real part of the quasi-normal modes of the gravitational perturbation and the universal one present in the very late time region \cite{Seahra:2004fg,Rosa:2011my} originating from the power law tail. As far as the possible detectability of the scenario presented here in aLIGO-like detectors using the real parts of the quasi-normal modes is concerned, one can safely say that most likely it is not a feasible option. This is mainly due to two reasons. First, the frequencies associated with these \KK modes are smaller compared to \gr\ (see \ref{Table_07} as well as 
\ref{Fig_Frequency}). Furthermore these \KK modes are supposed to be excited in the strong gravity regime, i.e., when mass of the black hole is large. On the other hand, as the mass of the black hole increases the frequency also decreases. This adds to the issue of detectability of these \KK modes. As evident from \ref{Table_07}, for black hole mass $\sim$ $10^{3}M_{\odot}$ the frequency of a mode in \gr\ is within the frequency band of aLIGO detectors. However the same is not true for the \KK modes where the frequencies are smaller and hence possibly outside the operational band of the aLIGO detectors. Nonetheless, all these frequencies pertaining to higher mass black holes are very well within the projected band of LISA and hence possibly detectable in near future (see \ref{Fig_Frequency}). The second point corresponds to the Signal-to-Noise ratio, since in order to detect the signal it is necessary to generate oscillations with a high Signal-to-Noise ratio. For this purpose as well we need collisions among heavier black holes (i.e., stronger gravity regime), so that higher order massive KK modes are excited. The frequency of these modes would correspondingly be lower and might get pushed out of the aLIGO frequency band but possibly be well within the LISA band.

\begin{figure*}[t!]
\centering
\includegraphics[scale=0.80]{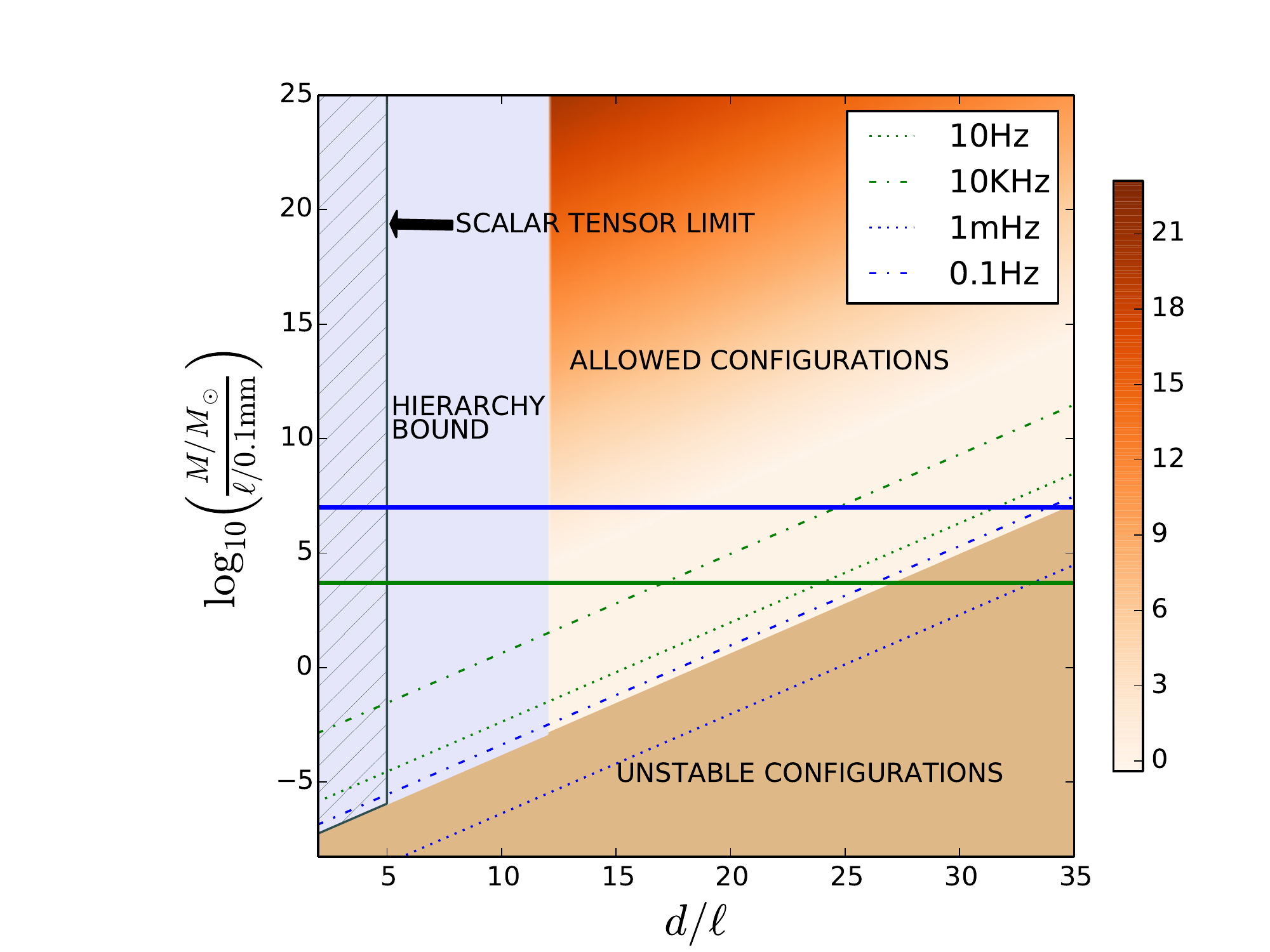}
\caption{This figure depicts the allowed region in the $(d/\ell,M/\ell)$ plane along with the frequency associated with the very late time behaviour of the perturbation modes for black holes. The accessible region in the plot is obtained by imposing a few restrictions on the $d/\ell$ as well as $M/\ell$ values. These correspond to --- (a) the Gregory-Laflamme instability, (b) the scalar-tensor limit and finally (c) the restriction on $d/\ell$ necessary to solve the hierarchy problem. The frequencies of the late time behaviour of the perturbation modes have been depicted in the accessible region by means of a colour coding. The labels on the right hand colour bar corresponds to a logarithm of frequencies (in Hz) to the base 10. The green dotted and dashed-dotted lines correspond to frequencies of 10 Hz and 10 KHz respectively, the extremities of aLIGO band. Similarly, the blue broken lines correspond to the extremities of the LISA band. The solid green and blue lines correspond to a configuration with $\ell=1\mathrm{\mu m}$ and $M$ of $50 M_{\odot}$ and $10^5 M_{\odot}$ respectively. For a given $d/\ell$ ratio, as $1/\ell$ increases the frequency also increases, while for a given $1/\ell$, as $d/\ell$ increases the frequency decreases. Hence the aLIGO band is completely within the accessible region while the LISA band has substantial overlap with the unstable regions. See text for more discussions.}
\label{Fig_Overall}
\end{figure*}

While, the late time behaviour of these massive gravitational perturbation modes corresponds to a universal frequency, associated with the power-law tail of the wave mode and proportional to the \KK mode mass \cite{Seahra:2004fg,Rosa:2011my}. Thus the frequency gets determined in terms of the $d/\ell$ ratio and the curvature length scale $\ell$ . Using the expression for $n$th \KK mode mass, $m_{n}=z_{n}(1/\ell)\exp(d/\ell)$, where $z_{n}$ are the zeros of the Bessel function $J_{\sqrt{13}/2}(x)$ one immediately arrives at the desired expression for the late time frequency as a function of $d/\ell$ and $1/\ell$ respectively. Given this universal late time frequency, one immediately observes that as $d/\ell$ and $1/\ell$ increase, the frequency also increases and hence for a given $d/\ell$ ratio, the frequency will be in aLIGO band for a larger value of $1/\ell$, but will fall within the LISA band for smaller values of $1/\ell$ (as evident from \ref{Fig_Overall} as well). This introduces additional 
complications in the detectability of these late time modes modulo the Gregory-Laflamme instability, which sets in for small $1/\ell$ values given a $d/\ell$ (see \ref{Fig_Overall}). At this point, it is interesting to note that given a black hole mass and a particular value of $\ell$, the frequency bands of aLIGO and LISA set natural observational bounds on $d/\ell$. In \ref{Fig_Overall}, we have considered two such scenarios, where the black hole masses are $50 M_{\odot}$ and $10^5 M_{\odot}$, while $\ell =1\mathrm{\mu m}$. The scenarios are depicted by the thick green and blue lines respectively. As is clearly evident from the plot the modes from the $50 M_{\odot}$ black hole can probe the range $17.0<d/\ell<23.9$. The  $10^5 M_{\odot}$ black hole has only a limited probe for $d/\ell$, because of the unstable configurations. The upper limit on $d/\ell$ in this case is therefore set by the boundary of the unstable region, so that $33.8<d/\ell<34.8$.  Thus as long as the universal frequency spectrum is 
concerned, the late time behaviour of the \KK modes has a better chance of detection in aLIGO rather than in LISA as clearly depicted in \ref{Fig_Overall}. The feasibility of the above detection, however is being determined by the Signal-to-Noise ratio, which will be much less for aLIGO while will be favourable for LISA. Hence even in this case there will be a tussle between the accessible regions in the $(d/\ell,1/\ell)$ space and the Signal-to-Noise ratio making the detectability difficult for aLIGO detectors for the late time behaviour of the massive quasi-normal modes as well.

The above exercise also opens up a few future avenues to explore. We have discussed the effect of higher dimensions on the quasi-normal modes in this work, however it is possible to address the nature of quasi-bound states and in particular how the presence of extra dimensions affect them. This may provide another observational test bed for detection of higher spatial dimensions. Besides whether one can obtain similar results for the quasi-bound states from Cauchy evolution as well remains to be verified. Also a through analysis of this allowed region in light of the recent detection of gravitational waves in aLIGO can lead to possible constraints on the extra-dimensional parameter space. Moreover the effect of higher dimensions on neutron star equation of state parameter, tidal love numbers associated with a brane black hole can lead to exciting results which we are currently pursuing and will report elsewhere.  
\section*{Acknowledgement}

Research of S.C. is supported by the SERB-NPDF grant (PDF/2016/001589) from SERB, Government of India and the research of S.S.G is partially supported by the SERB-Extra Mural Research grant (EMR/2017/001372), Government of India. This work was supported in part by NSF Grant PHY-1506497 and the Navajbai Ratan Tata Trust. The authors gratefully acknowledge discussions with David Hilditch, Emanuele Berti, Paolo Pani and Sanjeev Seahra. We also thank Shasvath Kapadia for carefully reading the manuscript and making useful suggestions. This work has been assigned the LIGO document number: LIGO-P1700295.
\appendix 
\labelformat{section}{Appendix #1} 
\labelformat{subsection}{Appendix #1}
\section{Deriving the perturbed gravitational field equations on the brane}\label{App_A}

In this appendix we present the detailed derivation of the perturbed gravitational field equations on the brane for completeness. We start with various geometrical quantities associated with spacetime curvature and express their perturbative expansion to leading order in $h_{AB}$. These correspond to,
\begin{align}
R^{(h)}_{ABCD}&=\frac{1}{2}\Bigg\lbrace \nabla _{C}\nabla _{D}h_{AB}+\nabla _{C}\nabla _{B}h_{AD}-\nabla _{C}\nabla _{A}h_{BD}
\nonumber
\\
&-\nabla _{D}\nabla _{C}h_{AB}-\nabla _{D}\nabla _{B}h_{AC}+\nabla _{D}\nabla _{A}h_{BC}\Bigg\rbrace~,
\label{GW_Eq10a}
\\
R^{(h)}_{AB}&=\frac{1}{2}\Bigg\lbrace -\nabla_{C}\nabla^{C}h_{AB}-2h^{C}_{D}R^{(g)~D}_{~~~~ACB}
\nonumber
\\
&+h^{D}_{A}R^{(g)}_{DB}+h^{D}_{B}R^{(g)}_{DA}\Bigg\rbrace~,
\label{GW_Eq10b}
\\
R^{(h)}&=\nabla_{C}\nabla_{B}h^{CB}-\nabla_{C}\nabla^{C}h=0~.
\label{GW_Eq10c}
\end{align}
In order to arrive at the last line the gauge conditions introduced in \ref{GW_Eq05} has been used. These gauge conditions further enables one to arrive at the following result,
\begin{align}\label{GW_Eq11}
\nabla_{C}\nabla_{B}h^{C}_{A}-\nabla_{B}\nabla_{C}h^{C}_{A}=h^{D}_{A}R^{(g)}_{DB}-h^{C}_{D}R^{(g)~D}_{~~~~ACB}~,
\end{align}
which has also been used in deriving \ref{GW_Eq10b}. Since we are ultimately interested in perturbation equations on the brane, it is instructive to rewrite all the bulk quantities in terms of brane variables. In particular conversion of the bulk covariant derivatives to the brane covariant derivative is a necessary and important step in that direction, since these are the terms appearing in \ref{GW_Eq10a}, \ref{GW_Eq10b} and \ref{GW_Eq10c} respectively. This can be done by elaborating all the bulk covariant derivatives into ordinary derivatives and Christoffel connections and then picking up all the terms involving brane covariant derivatives as well as terms depending on the bulk curvature and derivative with respect to $y$. Further since $x^{A}=(y,x^{\mu})$ it is clear that $e^{A}_{\mu}=\delta ^{A}_{\mu}$. Thus we obtain from \ref{GW_Eq10a},
\begin{align}\label{GW_Eq12}
2R^{(h)}_{ABCD}e^{A}_{\mu}n^{B}&e^{C}_{\nu}n^{D}=2\nabla _{\nu}\nabla _{y}h_{\mu y}-\nabla _{\nu}\nabla _{\mu}h_{yy}
\nonumber
\\
&-\nabla _{y}\nabla _{\nu}h_{\mu y}-\nabla _{y}\nabla _{y}h_{\mu \nu}+\nabla _{y}\nabla _{\mu}h_{y\nu}
\nonumber
\\
&=-\partial _{y}^{2}h_{\mu \nu}-2k\partial _{y}h_{\mu \nu}~.
\end{align}
Two other contractions in \ref{GW_Eq09} involving the perturbed Ricci tensor $R_{AB}^{(h)}$ results into,
\begin{align}\label{GW_Eq13}
R^{(h)}_{AC}e^{A}_{\mu}e^{C}_{\nu}&=-\frac{1}{2}\nabla _{C}\nabla ^{C}h_{\mu \nu}-h^{\alpha}_{\beta}R^{(g)\beta}_{~~~~\mu \alpha \nu}
\nonumber
\\
&+\frac{1}{2}h^{\alpha}_{\mu}R^{(g)}_{\alpha \nu}+\frac{1}{2}h^{\alpha}_{\nu}R^{(g)}_{\alpha \mu}
\nonumber
\\
&=-\frac{1}{2}~^{(4)}\square h_{\mu \nu}-\frac{1}{2}\partial _{y}^{2}h_{\mu \nu}+3k^{2}h_{\mu \nu}
\nonumber
\\
&-h^{\alpha}_{\beta}R^{(g)\beta}_{~~~~\mu \alpha \nu}+\frac{1}{2}h^{\alpha}_{\mu}R^{(g)}_{\alpha \nu}+\frac{1}{2}h^{\alpha}_{\nu}R^{(g)}_{\alpha \mu}~,
\end{align}
and,
\begin{align}\label{GW_Eq14}
R^{(h)}_{AC}n^{A}n^{C}=-\frac{1}{2}\nabla _{C}\nabla ^{C}h_{yy}-h^{\alpha}_{\beta}R^{(g)\beta}_{~~~y\alpha y}
=-h^{\alpha}_{\beta}R^{(g)\beta}_{~~~y\alpha y}~,
\end{align}
where \ref{GW_Eq10b} has been used. In order to arrive at the previous expressions the gauge conditions presented in \ref{GW_Eq05} have been used along with the following set of results,
\begin{align}\label{GW_Eq15}
g^{CD}\nabla _{C}\nabla _{D}h_{\alpha \beta}&=~^{(4)}\square h_{\alpha \beta}+\partial _{y}^{2}h_{\alpha \beta}
-6k^{2}h_{\alpha \beta}
\nonumber
\\
\nabla _{\nu}\nabla _{y}h_{\mu y}&=k\partial _{y}h_{\mu \nu}+3k^{2}h_{\mu \nu}
\nonumber
\\
-\nabla _{\nu}\nabla _{\mu}h_{yy}&=-2k^{2}h_{\mu \nu}
\nonumber
\\
-\nabla _{y}\nabla _{\nu}h_{y\mu}&=-k\partial _{y}h_{\mu \nu}-2k^{2}h_{\mu \nu}
\nonumber
\\
-\nabla _{y}\nabla _{y}h_{\mu \nu}&=-\partial _{y}^{2}h_{\mu \nu}-4k\partial _{y}h_{\mu \nu}-4k^{2}h_{\mu \nu}
\nonumber
\\
\nabla _{y}\nabla _{\mu}h_{y\nu}&=k\partial _{y}h_{\mu \nu}+2k^{2}h_{\mu \nu}~.
\end{align}
In arriving at the above relation we have also used the conditions that there exist only two non-vanishing connection components having the following forms: $\Gamma ^{y}_{\mu \nu}=kq_{\mu \nu}$ and $\Gamma ^{\mu}_{y\nu}=-k\delta ^{\mu}_{\nu}$. Use of these expressions for various projections of Riemann and Ricci tensor from \ref{GW_Eq12}, \ref{GW_Eq13} and \ref{GW_Eq14} in the perturbation equation for the bulk Weyl tensor as in \ref{GW_Eq09} leads to,
\begin{align}\label{GW_Eq16}
E_{\mu \nu}^{(h)}&=\frac{1}{2}\left(-\partial _{y}^{2}h_{\mu \nu}-2k\partial _{y}h_{\mu \nu} \right)
-\frac{1}{3}q_{\mu \nu}\left(-h^{\alpha}_{\beta}R^{(g)\beta}_{~~~y\alpha y}\right)
\nonumber
\\
&-\frac{1}{3}h_{\mu \nu}R^{(g)}_{yy}+\frac{1}{12}R^{(g)}h_{\mu \nu}
\nonumber
\\
&-\frac{1}{3}\Bigg(-\frac{1}{2}~^{(4)}\square h_{\mu \nu}-\frac{1}{2}\partial _{y}^{2}h_{\mu \nu}+3k^{2}h_{\mu \nu}
\nonumber
\\
&-h^{\alpha \beta}R^{(g)\beta}_{~~~~\mu \alpha \nu}+\frac{1}{2}h^{\alpha}_{\mu}R^{(g)}_{\alpha \nu}+\frac{1}{2}h^{\alpha}_{\nu}R^{(g)}_{\alpha \mu}\Bigg)~.
\end{align}
It is obvious that in order to separate out the perturbation of the bulk Weyl tensor into a four dimensional part and an additional part originating from extra dimensions, one needs to decompose all the quantities depending on the bulk metric $g_{AB}$ in terms of the four dimensional metric $q_{\alpha \beta}$. For that purpose we consider the following decompositions,
\begin{align}
R^{(g)\beta}_{~~~~\mu \alpha \nu}&=~^{(4)}R^{(q)\beta}_{~~~~\mu \alpha \nu}-k^{2}\left(q_{\mu \nu}\delta ^{\beta}_{\alpha}-q_{\mu \alpha}\delta ^{\beta}_{\nu}\right)
\nonumber
\\
R^{(g)\beta}_{~~~y\alpha y}&=-k^{2}\delta ^{\beta}_{\alpha}~,
\label{GW_Eq17a}
\\
R^{(g)}_{\alpha \mu}&=~^{(4)}R^{(q)}_{\alpha \mu}-4k^{2}q_{\alpha \mu}
\nonumber
\\
R^{(g)}_{yy}&=-4k^{2};\qquad R^{(g)}=~^{(4)}R^{(q)}-20k^{2}~.
\label{GW_Eq17b}
\end{align}
This eventually results into \ref{GW_Eq18}.
\section{Continued Fraction Method: Detailed Analysis}\label{App_B}

In this appendix, we provide a detailed and general derivation of the three term recursion relation pertaining to the continued fraction method, which we hope will be useful for the reader. Having derived \ref{Eq_qnm_02a} and \ref{Eq_qnm_02b} one normally makes an educated guess for $\psi_{n,l}$ and $\phi_{n,l}$ respectively. However as we will explicitly demonstrate  this is not necessary. One can start with an arbitrary choice for $\psi_{n,l}$ and $\phi_{n,l}$, but the structure of the differential equation itself will lead to the correct expressions for the master variables. Following this philosophy we decompose $\psi_{n,l}$ and $\phi_{n,l}$ as follows,
\begin{align}
\psi _{n,l}&=(r-2)^{\alpha}r^{\beta}\exp(\lambda r)f_{n,l}(r)~,
\\
\phi _{n,l}&=(r-2)^{\alpha}r^{\beta}\exp(\lambda r)g_{n,l}(r)~,
\end{align}
where $\alpha$, $\beta$ and $\lambda$ are arbitrary constants appearing in the master variables which we would like to uniquely determine using the structure of the differential equation. Substitution of these forms in \ref{Eq_qnm_02a} and \ref{Eq_qnm_02b} yields the following differential equations for $f_{n,l}$ and $g_{n,l}$ respectively,
\begin{widetext}
\begin{align}
r(r-2)\frac{d^{2}f_{n,l}(r)}{dr^{2}}&+\Big\lbrace 2\lambda r^{2}+\left(2\alpha+2\beta-4\lambda\right)r
+\left(2-4\beta\right) \Big\rbrace \frac{df_{n,l}(r)}{dr} 
\nonumber
\\
&+\Big\lbrace \left(\lambda ^{2}-m_{n}^{2}\right)r^{2}+\left(-2\lambda ^{2}+2\alpha \lambda +2\beta \lambda \right)r+\frac{\omega ^{2}r^{3}+2\alpha ^{2}}{r-2}-\frac{2\beta ^{2}-4\beta-6}{r}
\nonumber
\\
&+\left(2\lambda -4\beta \lambda+\beta (\beta -1)+\alpha (\alpha -1)-l(l+1)+2\alpha \beta \right)\Big\rbrace f_{n,l}(r)
-\frac{m_{n}^{2}}{r}g_{n,l}=0~,
\\
r(r-2)\frac{d^{2}g_{n,l}(r)}{dr^{2}}&+\Big\lbrace 2\lambda r^{2}+\left(2\alpha+2\beta-4\lambda\right)r
+\left(2-4\beta\right) \Big\rbrace \frac{dg_{n,l}(r)}{dr} 
\nonumber
\\
&+\Big\lbrace \left(\lambda ^{2}-m_{n}^{2}\right)r^{2}+\left(-2\lambda ^{2}+2\alpha \lambda +2\beta \lambda \right)r+\frac{\omega ^{2}r^{3}+2\alpha ^{2}}{r-2}-\frac{2\beta ^{2}-4\beta}{r}
\nonumber
\\
&+\left(2\lambda -4\beta \lambda+\beta (\beta -1)+\alpha (\alpha -1)-l(l+1)+2\alpha \beta \right)\Big\rbrace f_{n,l}(r)
-4f_{n,l}=0~.
\end{align}
\end{widetext}
Given these differential equations one changes the variable from $r$ to $\xi$, such that $r=2/(1-\xi)$. In order for these differential equations to have regular singular points after the variable change it is necessary that the terms behaving as $r^{2}$, $r$ and $1/(r-2)$ in the above should vanish, which would require, at the first level, $\alpha =-2i\omega$. In which case one can use the following relation, $r^{3}-8=(r-2)(r^{2}+2r+4)$, such that the other two parameters $\lambda$ and $\beta$ are determined as: $\lambda = \sqrt{m_{n}^{2}-\omega ^{2}}$, and $\beta = 2i\omega+(1/\lambda)(-\omega ^{2}+\lambda ^{2})=-(1/\lambda)(\omega -i\lambda)^{2}$. Hence the substitutions of these three constants leads to the following ansatz for the master variable suited with the above problem
\begin{align}
\psi _{n,l}&=\Big(\frac{r-2}{r}\Big)^{-2i\omega}r^{-b}e^{\lambda r}f_{n,l}~,
\nonumber
\\
\lambda &= \sqrt{m_{n}^{2}-\omega ^{2}};\qquad b =\frac{\omega ^{2}-\lambda ^{2}}{\lambda}~,
\label{Eq_qnm_03a}
\\
\phi _{n,l}&=\Big(\frac{r-2}{r}\Big)^{-2i\omega}r^{-b}e^{\lambda r}g_{n,l}~.
\label{Eq_qnm_03b}
\end{align}
Using \ref{Eq_qnm_03a} and \ref{Eq_qnm_03b} respectively, both the differential equations can be casted in the following form,
\begin{align}
r(r-2)\frac{d^{2}y}{dr^{2}}&+\Big\lbrace A r^{2}+Br+C \Big\rbrace  \frac{dy}{dr} 
\nonumber
\\
&+\Big\lbrace D+\frac{E}{r} \Big \rbrace y +\Big\lbrace F+\frac{G}{r} \Big\rbrace z=0~,
\label{Eq_qnm_04}
\end{align}
where $y$ stands for $f_{n,l}$ and $z$ stands for $g_{n,l}$ or vice versa and the constants appearing in the above differential equation will depend on the parameters introduced above in \ref{Eq_qnm_03a} and \ref{Eq_qnm_03b}. As mentioned earlier it is advantageous to introduce a new variable $\xi=(r-2)/r$ in lieu of $r$, such that, $r=2/(1-\xi)$. Eliminating the variable $r$ appearing in \ref{Eq_qnm_04} one can rewrite the differential equation in terms of the new variable $\xi$. Simplifying the resulting differential equation further we obtain,
\begin{align}
\xi (1-\xi)^{2}\frac{d^{2}y}{d\xi ^{2}}&+\Big\lbrace \left(2A+B+\frac{C}{2}\right)
\nonumber
\\
&-\left(2+B+C\right)\xi+\left(2+\frac{C}{2}\right)\xi ^{2}\Big\rbrace \frac{dy}{d\xi}
\nonumber
\\
&+\Big\lbrace \left(D+\frac{E}{2}\right)-\frac{E}{2}\xi \Big \rbrace y 
\nonumber
\\
&+\Big\lbrace \left(F+\frac{G}{2}\right)-\frac{G}{2}\xi \Big\rbrace z=0~.
\label{Eq_qnm_05}
\end{align}
The above differential equation can be solved using the power series technique. Keeping this in mind let us assume the following series expansion for $y$ in powers of $(r-2)/r$, or in terms of $\xi$ as
\begin{align}\label{Eq_qnm_06}
y\equiv \sum _{j=0}c_{j}\left(\frac{r-2}{r}\right)^{j}=\sum _{j=0}c_{j}\xi ^{j};\qquad z=\sum _{j=0}d_{j}\xi^{j}~,
\end{align}
where $c_{j}$ and $d_{j}$ are arbitrary coefficients that needs to be determined. Substitution of the above power series in the differential equation given by \ref{Eq_qnm_05} results into an equation involving various powers of $\xi$. Simplifying the above algebraic equation further and writing it in such a manner that all the powers of $\xi$ coincides, we finally obtain the following three term recursion relation between three coefficients $c_{j-1}$, $c_{j}$ and $c_{j+1}$ as well as $d_{j}$ and $d_{j-1}$ appearing in the series expansion in \ref{Eq_qnm_06} as,
\begin{align}
(j+1)&\left(j+2A+B+\frac{C}{2}\right)c_{j+1}-\Big\lbrace 2j(j-1)
\nonumber
\\
&+n \left(2+B+C\right)-\left(D+\frac{E}{2}\right) \Big\rbrace c_{j}
\nonumber
\\
&+\Big\{(j-1)(j-2)+\left(2+\frac{C}{2}\right)(j-1)-\frac{E}{2} \Big\}c_{j-1}
\nonumber
\\
&+d_{j}\left(F+\frac{G}{2}\right)-\frac{G}{2}d_{j-1}=0~.
\end{align}
As we have mentioned earlier the same recursion relation holds for expansion coefficients of $f_{n,l}$ and $g_{n,l}$ both, but the constants appearing in the recursion relation will have distinct values for the two situations. In the case of $f_{n,l}$, we have, the following expressions for the constants,
\begin{align}
A&=2\lambda;\qquad B=2\alpha +2\beta -4\lambda~,
\nonumber
\\
C&=2-4\beta;\qquad E=-2\beta ^{2}+4\beta +6~,
\nonumber
\\
D&=2\lambda -4\beta \lambda+\beta (\beta -1)+\alpha (\alpha -1)
\nonumber
\\
&-l(l+1)+2\alpha \beta +4\omega ^{2}~,
\nonumber
\\
F&=0;\qquad G=-m_{n}^{2}~,
\end{align}
while, the corresponding values associated with the differential equation for $g_{n,l}$ becomes,
\begin{align}
A&=2\lambda;\qquad B=2\alpha +2\beta -4\lambda~,
\nonumber
\\
C&=2-4\beta;\qquad E=-2\beta ^{2}+4\beta~,
\nonumber
\\
D&=2\lambda -4\beta \lambda+\beta (\beta -1)+\alpha (\alpha -1)
\nonumber
\\
&-l(l+1)+2\alpha \beta +4\omega ^{2}~,
\nonumber
\\
F&=-4;\qquad G=0~.
\end{align}
Using these constants one can write down the three term recursion relation for both the master variables, which we have used to arrive at \ref{Eq_qnm_07}.
\bibliography{GW_References}

\bibliographystyle{./utphys1}
\end{document}